\documentclass[fleqn,aps,prb,twocolumn,longbibliography]{revtex4-2}
\usepackage{amsmath,amssymb,amsfonts,float,graphics,epsfig,epstopdf,color,verbatim,tabularx,bm,multirow,appendix,hyperref}
	\usepackage{amsmath}
	\usepackage{amssymb}
        \usepackage{ulem}
	\usepackage{mathtools}
	\usepackage{graphicx}
	\usepackage{lipsum}
        \usepackage[german,english]{babel}
	\usepackage{bm}
	\usepackage{hyperref}
	\usepackage{url}
	\usepackage[utf8]{inputenc}
	\usepackage{subfigure}
	\usepackage{slashed,bbm}
	\usepackage{graphics,psfrag,epsfig}
	\usepackage{dsfont}
	\usepackage{setspace}
	\usepackage{wasysym}
	\usepackage{slashed}
	\usepackage{lipsum}
	\usepackage{braket}
	\usepackage{physics}
	\usepackage{enumitem}
	\usepackage{xcolor}
	\definecolor{dark-red}{rgb}{0.4,0.15,0.15}
	\definecolor{dark-blue}{rgb}{0.15,0.15,0.4}
	\definecolor{medium-blue}{rgb}{0,0,0.5}
	\hypersetup{
		colorlinks, linkcolor={dark-blue},
		citecolor={dark-blue}, urlcolor={medium-blue}
	}
	\usepackage{ulem}
	\newcommand{\be}{\begin{equation}}
		\newcommand{\ee}{\end{equation}}
	\newcommand{\bea}{\begin{eqnarray}}
		\newcommand{\eea}{\end{eqnarray}}

\begin{document}
		
		\title{Zipping many-body quantum states: a scalable approach to diagonal entropy}

		\author{Yu-Hsueh Chen}
		\affiliation{Department of Physics, University of California at San Diego, La Jolla, California 92093, USA}
		\author{Tarun Grover}
		\affiliation{Department of Physics, University of California at San Diego, La Jolla, California 92093, USA}

\begin{abstract}
The outcomes of projective measurements on a quantum many-body system in a chosen basis are inherently probabilistic. The Shannon entropy of this probability distribution (the ``diagonal entropy'') often reveals universal features, such as the existence of a quantum phase transition. A brute-force tomographic approach to estimating this entropy scales exponentially with the system size. Here, we explore using the Lempel-Ziv lossless image compression algorithm as an efficient, scalable alternative, readily implementable in a quantum gas microscope or programmable quantum devices. We test this approach on several examples: one-dimensional quantum Ising model, and two-dimensional states that display conventional symmetry breaking due to quantum fluctuations, or strong-to-weak symmetry-breaking due to local decoherence. We also employ the diagonal mixed state to put constraints on the phase boundaries of our models.  In all examples, the compression method accurately recovers the entropy density while requiring at most polynomially many images. We also analyze the singular part of the diagonal entropy density using renormalization group on a replicated action. In the 1+1-D quantum Ising model, we find that it scales as $|t| \log|t|$, where $t$ is the deviation from the critical point, while in a 2+1-D state with amplitudes proportional to the Boltzmann weight of the 2D Ising model, it follows a $t^2 \log|t|$ scaling.

\end{abstract}
		
\maketitle
\section{Introduction}
The entropy of a many-body state $\rho$, defined as $S = -\tr\left(\rho \log \rho\right)$, is a remarkably versatile measure of a system's  effective degrees of freedom \cite{boltzmann1877beziehung,gibbs1902elementary,Neumann1927,Holzhey94,Calabrese04}. Entropy remains well-defined even in out-of-equilibrium systems while retaining a straightforward interpretation: it quantifies the von Neumann entanglement between the system and its environment. Despite its appeal, measuring entropy experimentally is challenging. One reason is that $S$ is a non-linear function of the density matrix, and can be interpreted as the expectation value of an unwieldy state-dependent operator, namely, $-\log \rho$. Here, we focus on a class of problems where a different non-linear observable, the Shannon entropy \cite{shannon1948mathematical} of measurement outcomes in a chosen basis --- denoted as $S_d$ (``diagonal entropy'') --- is a useful quantity that can be measured efficiently. Specifically, we demonstrate that for a class of problems, the volume-law coefficient of $S_d$ can be estimated using a well-known lossless data compression algorithm \cite{ziv1977universal}. We focus on two distinct class of problems: quantum phase transitions driven by a non-thermal parameter, and decoherence induced transitions in many-body mixed states, such as strong-to-weak symmetry breaking, and decodability transitions in error-correcting code. We  also explore the universal aspects of diagonal entropy using a renormalization group approach.

Our setup is as follows. Let's subject a quantum many-body state $\rho$ to a projective measurement in a preferred local basis $\{|x_j\rangle\}$, producing a snapshot $x_\mathbf{j}$ with Born probability $\rho_{x_\mathbf{j}} := \langle x_\mathbf{j} | \rho | x_\mathbf{j} \rangle$. Such an experiment can be performed with a quantum gas microscope \cite{schlosser2001sub,bakr2009quantum,sherson2010single,cheuk2015quantum,haller2015single,parsons2015site,edge2015imaging,gross2021quantum,browaeys2020many}, or in various programmable quantum devices ~\cite{kjaergaard2020superconducting,blatt2012quantum}. We assume that the local Hilbert space is finite-dimensional, so that $x_j$ can take $\mathcal{D}$ distinct values (in our examples, $\mathcal{D} = 2$). Using $\{\rho_{x_\mathbf{j}}\}$, one may define the diagonal entropy as \be S_d= -\sum_{x_\mathbf{j}}  \rho_{x_\mathbf{j}} \log( \rho_{x_\mathbf{j}}).\ee  $S_d$ provides an upper bound to the von Neumann entropy $S_d \geq S$. A salient feature of $S_d$ is that, even for the ground states of local Hamiltonians, it generically exhibits a volume-law scaling with the system size~\cite{stephan2009shannon,zaletel2011logarithmic,alcaraz2013universal,luitz2014participation}. Our goals will be twofold. First, we will explore problems where $S_d$ is a useful theoretical tool --- for example, it may constrain the phase boundaries where von Neumann entropy $S$ becomes singular. More importantly, we will demonstrate that for the class of problems we study, the volume-law coefficient of $S_d$ can be accurately estimated using only polynomially many images. This is surprising because for a generic quantum state, estimating $S_d$ is expected to require exponentially many images. We conjecture that this enormous reduction in complexity is a consequence of the fact that the probabilities $\{\rho_{x_\mathbf{j}}\}$ are related to  correlation functions with respect to a local effective action.

Let us first discuss the theoretical utility of the diagonal entropy {\cite{stephan2009shannon,zaletel2011logarithmic,alcaraz2013universal,luitz2014participation,PhysRevLett.128.130605,PhysRevB.106.214316, Liu_2025, PhysRevLett.132.140401, Turkeshi_2024,tirrito2024anticoncentrationmagicspreadingergodic, sierant2025many, wang2025decoherenceinducedselfdualcriticalitytopological}}. The leading-order contribution to $S_d$  generically scales with system volume, i.e., $S_d \sim s^{\infty}_d V + ...$, where the subscript $\infty$ indicates $s^{\infty}_d = \lim_{V\to \infty} S_d/V$. Much attention has been devoted to the subleading terms that encode universal data of critical ground states \cite{stephan2009shannon,zaletel2011logarithmic,alcaraz2013universal,luitz2014participation,PhysRevLett.128.130605,PhysRevB.106.214316}. For ground states, we focus primarily on the volume-law coefficient $s^{\infty}_d$, both because it is readily accessible using our method and because it has received less attention than subleading terms. We relate the singular behavior of $s^{\infty}_d$ to universal aspects of the phase transitions we encounter. For example, in the ground state of 1+1-D quantum Ising model, we show that the singular part of $s^{\infty}_d$ close to the quantum phase transition relates to the surface free energy of the classical 2D Ising model~\cite{fisher1967interfacial}, and scales as  $|t| \log(1/|t|)$ where $t$ is the non-dimensional deviation from the critical point. Our numerical calculations will of course be done on finite systems, and we will denote the entropy density $S_d/V$ simply as $s_d$.

Another conceptual motivation for studying $S_d$ is its ability to reveal universal many-body phenomena visible only in non-linear functions of the density matrix $\rho$. A prominent example concerns subjecting a topologically ordered state, such as the toric code, to a finite-depth local channel \cite{dennis2002,wang2003confinement, lee2023quantum, fan2023diagnostics,bao2023mixed,chen2023separability,sang2023mixed}. 
Since the channel is finite-depth, any correlation function of the form $\tr(\rho O_1(x) O_2(0))$, where $O_1, O_2$ are state-independent operators, remains  qualitatively unchanged and is an analytic function of the decoherence rate. Nevertheless, beyond a certain threshold, the system loses quantum memory, i.e., error-correction becomes impossible. This is an intrinsic mixed-state transition and it is visible in various diagnostics that are non-linear functions of the density matrix, such as the von Neumann entropy $S = - \tr \left(\rho \log \rho\right)$. Intuitively, across the transition, information is irreversibly lost to the environment,  making entropic quantities such as $S$ sensitive to the transition. 

Unlike von Neumann entropy $S$, the diagonal entropy $S_d$ is clearly basis dependent. Applying the Kramers-Wannier duality transformation on the aforementioned decohered toric code illustrates why $S_d$ in a specific basis may also detect a conceptually similar transition. In the dual formulation, one subjects a paramagnet state $|x_\mathbf{j} = 1\rangle$ to an Ising-symmmetric channel with local Kraus operators $Z_i Z_j$ on each link $\langle i, j\rangle$. Beyond a threshold decoherence rate, the density matrix can be expressed as a classical mixture of GHZ-like states~\cite{chen2023separability}. 
This is the strong-to-weak spontaneous symmetry breaking (SW-SSB) phenomena that has recently been discussed from various different perspectives \cite{lee2023quantum,ma2023topological,lessa2024strong,sala2024spontaneous,gu2024spontaneoussymmetrybreakingopen, huang2024hydrodynamicseffectivefieldtheory,kuno2024strong,zhang2024strongtoweakspontaneousbreaking1form,zhang2024fluctuationdissipationtheoreminformationgeometry, liu2024diagnosingstrongtoweaksymmetrybreaking,shah2024instabilitysteadystatemixedstatesymmetryprotected,guo2024strongtoweakspontaneoussymmetrybreaking,kim2024errorthresholdsykcodes,ando2024gaugetheorymixedstate,chen2024strongtoweaksymmetrybreakingentanglement,orito2025strongweaksymmetriesspontaneous,sun2025schemedetectstrongtoweaksymmetry} {and has been realized experimentally \cite{chen2025nishimori}}
. 
Since symmetry plays an important role in the dual picture, one may ask if the diagonal entropy in the \textit{Ising symmetric basis} has any special significance. Indeed, it is easy to show that the decohered density matrix is already diagonal in the Ising symmetric basis, and therefore, $S_d = S$, which implies that $S_d$ also sees the same phase transition as $S$. More interestingly, since $\rho_d$ can be obtained by applying a strongly symmetric, short-depth channel to $\rho$, the existence of an SW-SSB phase for $\rho$ implies the existence of an SW-SSB phase for $\rho_d$
\cite{lessa2024strong}. 
Further, by continuity, if the initial state or the Kraus operators are slightly perturbed while preserving Ising symmetry, an SW-SSB phase transition persists for both $\rho$ and $\rho_d$, although their phase boundaries need not coincide.

Having established that the diagonal entropy $S_d$ is conceptually useful, we now discuss its measurement in experiments. A brute-force tomographic approach to measuring $S_d$ records measurement outcomes (bit strings) and constructs a histogram of these results. Since the number of possible measurement outcomes scales linearly with the Hilbert space size, this scheme scales exponentially with the system size and is therefore not practical. We note that remarkable progress has been made recently in estimating observables, including Renyi entropies $S_n = \frac{1}{1-n}\log \tr \rho^n$ using shadow tomography/randomized measurement methods \cite{da2011practical, van2012measuring, aaronson2018shadow, elben2018renyi, huang2020predicting, elben2023randomized}. However, these approaches generically also scale exponentially with the support of the density matrix $\rho$ when estimating $S_n/V$. Significant progress has also been made in numerically computing Renyi entropies for variational wavefunctions and for sign-problem-free Hamiltonians~\cite{Hastings10,Melko_Hubbard,zhang_criticalee,Jiang12,nielsen2012laughlin,Grover13,Assaad13a,drut2016entanglement,d2020entanglement,zhao2022measuring,zhou2024incremental}. However, without explicit knowledge of the entanglement Hamiltonian $\log(\rho)$~\cite{mendes2020measuring}, developing scalable algorithms to estimate either von Neumann entropy or the diagonal entropy remains challenging.

In this work we will exploit the lossless data compression methods discovered by Lempel and Ziv \cite{ziv1977universal} to estimate $S_d$ using images generated from making a projective measurement on the underlying state.  We are particularly inspired by the results in Ref.~\cite{stefano2019quantifying,martiniani2020correlation}, where a variant of this compression algorithm was used to decipher universal aspects of certain non-equilibrium classical statistical mechanics models. Given a bit string $x_{\mathbf{j}}$ of length $N$ generated from an unknown probability distribution, this algorithm provides a lossless compression scheme to obtain a compressed file whose size is $\mathcal{N}(x_\mathbf{j})$. One can then show that the expectation value of the ratio $\mathcal{N}(x_\mathbf{j})/\mathcal{N}_\text{shuffle}$, where $\mathcal{N}_\text{shuffle}$ is the size of the compressed file for a random binary
sequence, approaches the Shannon entropy density $S_d/N$ for the unknown probability distribution from which the bit string was generated. Although this algorithm has been applied to a variety of problems in classical statistical mechanics, both in and out of equilibrium \cite{sheinwald1990two,melchert2015analysis,stefano2019quantifying, martiniani2020correlation}, we are not aware of any application to estimating diagonal entropy of many-body quantum systems. We will apply this algorithm to the problems of our interest hinted above. Our main results are:

\begin{enumerate}

    \item We study diagonal entropy for the following three different models in the Ising symmetric basis (which corresponds to the Pauli-$X$ basis in our notation): 
    
    (i) Ground state of the 1+1-D quantum Ising model. 
    
    (ii) The Kramers-Wannier dual of the Castelnovo-Chamon state (Ref.~\cite{castelnovo2008quantum}) $|\Psi(q)\rangle = \prod_{\langle i,j\rangle}[(1-q)I + q Z_i Z_j] |x_\mathbf{j} = 1\rangle$, subjected to Kraus operators $Z_i Z_j$ on each bond $\langle i j \rangle$ with probability $p$. Note that the phase diagram is a function of two variables ($p,q$).
    
    (iii) The mixed state obtained by subjecting the paramagnetic state $|x_{\mathbf{j}} = 1\rangle$  to `coherent errors' of the form $e^{i \phi \sum_{\langle i,j\rangle} Z_i Z_j}$. \

We provide evidence that, in each of these models, the diagonal entropy density $S_d/V$ can be accurately estimated using the Lempel-Ziv data compression method applied to images derived from projective measurements on the state of interest. The number of required images scales at most polynomially with system size. 

Notably, our approach is limited only by the ability to generate images with the correct Born distribution.  Once these images are available (e.g., using quantum gas microscopy~\cite{schlosser2001sub,bakr2009quantum,sherson2010single,cheuk2015quantum,haller2015single,parsons2015site,edge2015imaging,gross2021quantum}), there is essentially no limit on the system size. For example, in the 2+1-D system described in (ii) above, we studied diagonal entropy in systems as large as $256 \times 256$ sites, enabled by an efficient classical algorithm to generate images. This stands in marked contrast to Monte Carlo-based methods for Renyi entropies $S_n$ ~\cite{Hastings10,Melko_Hubbard,zhang_criticalee,Jiang12,nielsen2012laughlin,Grover13,Assaad13a,drut2016entanglement,d2020entanglement,zhao2022measuring,zhou2024incremental,torlai2018neural,torlai2019integrating}, where the sampled quantity is exponentially small in the support of the density matrix, making scaling difficult even in the absence of the sign problem. However, we again note that our method applies only to the diagonal entropy density, which, unlike Renyi entropies, is a basis-dependent quantity.

    \item As hinted above, we study the phase diagram of the dual of Castelnovo-Chamon state (Ref.\cite{castelnovo2008quantum}), $|\Psi(q)\rangle = \prod_{\langle i,j\rangle}[(1-q)I + q Z_i Z_j] |x_\mathbf{j} = 1\rangle$, subjected to Kraus operators $Z_i Z_j$ on each bond $\langle i j \rangle$ with probability $p$, as well as Kraus operators $X_j$ on each site with probability $r$ (measurement in the Pauli-$X$ basis corresponds to the special case of maximal X dephasing). The three-dimensional phase diagram of the corresponding mixed state as a function of $(p,q,r)$ shows three distinct phases: a symmetric phase, a conventional symmetry-broken phase, and an SW-SSB phase. Notably, along the $p$ and $q$ axes, the critical point associated with the actual state coincides with singularities of the diagonal density matrix in the Pauli-$X$ basis. Along the $p$ axis, the phase transition corresponds to the SW-SSB transition \cite{lee2023quantum,ma2023topological,lessa2024strong,sala2024spontaneous,gu2024spontaneoussymmetrybreakingopen, huang2024hydrodynamicseffectivefieldtheory,kuno2024strong,zhang2024strongtoweakspontaneousbreaking1form,zhang2024fluctuationdissipationtheoreminformationgeometry, liu2024diagnosingstrongtoweaksymmetrybreaking,shah2024instabilitysteadystatemixedstatesymmetryprotected,guo2024strongtoweakspontaneoussymmetrybreaking,kim2024errorthresholdsykcodes,ando2024gaugetheorymixedstate,chen2024strongtoweaksymmetrybreakingentanglement,orito2025strongweaksymmetriesspontaneous,sun2025schemedetectstrongtoweaksymmetry}, whose singular behavior is related to the Nishimori critical point \cite{dennis2002,wang2003confinement, lee2023quantum, fan2023diagnostics,bao2023mixed,chen2023separability,sang2023mixed}. We are unable to detect any singularity in the entropy density for this transition (see Sec.\ref{sec:Nishimori} for a discussion). As expected, subleading terms in the entropy density can detect this transition, but they do not appear amenable to the Lempel-Ziv compression algorithm.

    \item We study the universal scaling behavior of the diagonal entropy density using the replica trick and renormalization group arguments. Specifically, we argue that the diagonal entropy density $s_d$ for the 1+1-D transverse-field Ising model in the Ising symmetric basis is proportional to the boundary contribution to the free energy in a classical 2D Ising model~\cite{fisher1967interfacial}, and therefore its singular part scales as $s^{\text{sing}}_{d} \sim |t| \log(1/|t|)$, where $t$ is the deviation from the critical point. In contrast, in the ground state of the Castelnovo-Chamon model, we find that $s_d$ maps to the \textit{bulk} free energy of the 2D classical Ising model, and consequently its singular part scales as $s^{\rm{sing}}_{d} \sim t^2 \log(1/|t|)$, where  $t$  again denotes the deviation from the critical point.
    Our numerical results for the diagonal entropy are consistent with these predictions.
    
\end{enumerate}

    The paper is organized as follows: In Sec.\ref{sec:diagonal_entropy}, we discuss the virtues of diagonal entropy in a few different contexts, and provide an introduction to the Lempel-Ziv algorithm for its estimation. In Sec.\ref{sec:tfim}, we discuss diagonal entropy in the maximally dephased 1+1-D quantum Ising model, both numerically and analytically. Sec.\ref{sec:chamon} is devoted to the aforementioned 2+1-D paramagnet state subjected to decoherence. We discuss the phase diagram of this state and as well as the state obtained from subjecting it to Pauli-$X$ dephasing. We also study the diagonal entropy in the maximally dephased ground state using renormalization group applied to a replicated model. We then study its diagonal entropy numerically using Lempel-Ziv compression and a tensor network method. Sec.\ref{sec:coherent} studies the diagonal entropy of a mixed state obtained by subjecting a paramagnet to an Ising symmetric unitary, followed by maximal Pauli-$X$ depasing. Sec.\ref{sec:complexity} is devoted to analyzing the sample complexity of estimating diagonal entropy. We conclude in Sec.\ref{sec:discuss} by discussing directions for future work.

\section{Diagonal entropy: uses and  scalable estimation}
\label{sec:diagonal_entropy}

\subsection{Diagonal entropy as a diagnostic for certain phase transitions}
\label{sec:uses}

The main object of our study is the diagonal entropy $S_d$ of a many-body density matrix $\rho$ in a chosen product basis, which we denote as $\{ |x_\mathbf{j}\rangle\}$:
\begin{equation}
\label{Eq:diagonal_entropy}
\begin{aligned}
        S_d &=- \sum_{x_\mathbf{j}} \langle x_\mathbf{j} |\rho | x_\mathbf{j}\rangle \log( \langle x_\mathbf{j} |\rho | x_\mathbf{j}\rangle) \\
        & =- \sum_{x_\mathbf{j}} \rho_{x_\mathbf{j}} \log \rho_{x_\mathbf{j}}
\end{aligned}
\end{equation}
All logarithms (log($\cdot$)) are base 2 unless stated otherwise. When using the natural logarithm (base $e$), we denote it as $\ln(\cdot)$.

Since $S_d$ is clearly a basis-dependent object, an immediate question is: what, if any, information about the state $\rho$ can be inferred from $S_d$? As discussed in previous works~\cite{stephan2009shannon,zaletel2011logarithmic,alcaraz2013universal,luitz2014universal,luitz2014participation}, the sub-volume-law contributions to $S_d$ can encode universal information about critical states. Partly motivated by recent progress in many-body open quantum systems~\cite{dennis2002,wang2003confinement, lee2023quantum, fan2023diagnostics,bao2023mixed,chen2023separability,sang2023mixed,lee2023quantum,ma2023topological,lessa2024strong,sala2024spontaneous,gu2024spontaneoussymmetrybreakingopen, huang2024hydrodynamicseffectivefieldtheory,kuno2024strong,zhang2024strongtoweakspontaneousbreaking1form,zhang2024fluctuationdissipationtheoreminformationgeometry, liu2024diagnosingstrongtoweaksymmetrybreaking,shah2024instabilitysteadystatemixedstatesymmetryprotected,guo2024strongtoweakspontaneoussymmetrybreaking,kim2024errorthresholdsykcodes,ando2024gaugetheorymixedstate,chen2024strongtoweaksymmetrybreakingentanglement,orito2025strongweaksymmetriesspontaneous,sun2025schemedetectstrongtoweaksymmetry}, here we provide a few additional perspectives on this question, which will guide our subsequent discussion. A key insight is that if the state $\rho$ has a certain symmetry, and the measurement basis $\{ |x_\mathbf{j}\rangle\}$ is also invariant under the same symmetry, then the diagonal entropy can capture certain universal features where the symmetry plays an important role (e.g., spontaneous symmetry breaking or symmetry-enforced separability transitions).
Let us elaborate this further in a few different contexts:

(i) \underline{Strong-to-weak SSB}:

Let us recall that a density matrix is said to have a ``strong symmetry'' if $U \rho = e^{i \theta} \rho$ and a ``weak symmetry'' if $U \rho U^{\dagger} = \rho$, where $U$ is the generator of the symmetry \cite{de2022symmetry, ma2022average}.  
We will focus on examples where the mixed state of interest respects a strong $\mathbb{Z}_2$ symmetry: $U \rho = \pm \rho$, where $ U = \prod_{j = 1}^N X_j $, with $N$ the total system size. As a function of the decoherence rate, the mixed state can undergo a phase transition in which the strong $\mathbb{Z}_2$ symmetry is spontaneously broken down to a weak $\mathbb{Z}_2$ symmetry~\cite{lee2023quantum,ma2023topological,lessa2024strong,sala2024spontaneous,gu2024spontaneoussymmetrybreakingopen, huang2024hydrodynamicseffectivefieldtheory,kuno2024strong,zhang2024strongtoweakspontaneousbreaking1form,zhang2024fluctuationdissipationtheoreminformationgeometry, liu2024diagnosingstrongtoweaksymmetrybreaking,shah2024instabilitysteadystatemixedstatesymmetryprotected,guo2024strongtoweakspontaneoussymmetrybreaking,kim2024errorthresholdsykcodes,ando2024gaugetheorymixedstate,chen2024strongtoweaksymmetrybreakingentanglement,orito2025strongweaksymmetriesspontaneous,sun2025schemedetectstrongtoweaksymmetry}. We will use the acronym ``SW-SSB'' for the corresponding symmetry-broken phase. As previously discussed, detecting such a transition typically requires access to quantities that are nonlinear functions of the density matrix, such as Rényi entropies $S_n = \frac{1}{1-n} \log \left(\tr \rho^n\right)$ or the von Neumann entropy $S = -\tr \rho \log \rho$, which is rather challenging.

Let us now consider making a projective measurement on $\rho$ in the $X$-basis.  
$\rho_{x_{\mathbf{j}}} = \langle x_\mathbf{j} |\rho | x_\mathbf{j}\rangle$ represents the probability of obtaining a measurement outcome $x_\mathbf{j} = (x_1, \cdots, x_N)$, and the diagonal entropy $S_d$ is the corresponding classical Shannon entropy. The process of preparing $\rho$ and subsequently measuring it in the Pauli-$X$ basis can also be understood as preparing the \textit{diagonal mixed state} $\rho_d = \sum_{x_\mathbf{j}} \rho_{x_\mathbf{j}} |x_\mathbf{j}\rangle \langle x_\mathbf{j}|$ by applying a maximal dephasing channel in the Pauli-$X$ basis to $\rho$: %
$\rho_d = \mathcal{E}_d[\rho]$, where $\mathcal{E}_d[\cdot] = \prod_j \mathcal{E}_j[\cdot]$, with $\mathcal{E}_j[\cdot] = (\rho + X_j \rho X_j)/2$.  
The diagonal entropy of $\rho$ then corresponds to the von Neumann entropy of $\rho_d$.  
Since $\mathcal{E}_d[\cdot]$ is a strongly symmetric, finite-depth local channel, several intrinsic properties of $\rho$ cannot be altered by $\mathcal{E}_d[\cdot]$. In particular, as shown in Ref.~\cite{lessa2024strong}, if $\rho$ is in the SW-SSB phase, then $\rho_d$ will be as well. We will further show in later sections that several information-theoretic quantities of $\rho_d$ that detect SW-SSB provide bounds for the corresponding quantities of $\rho$. Pertinently, we will construct examples where the phase diagram for the density matrix $\rho_d$ has the same qualitative structure as that for $\rho$, and along certain axes of the tuning parameter, $S_d$ in fact equals $S$, thereby inheriting the singularities of $S$ across transitions.

(ii) \uline{Average strange-correlator/temporal-boundary correlator:}

Using $|x_\mathbf{j}\rangle = \prod_j Z_j^{(1-x_j)/2}|x_\mathbf{j} = 1\rangle$, let us rewrite $\rho_{x_\mathbf{j}} $ as
\begin{equation}
\label{Eq:boundary_correlator}
    \rho_{x_\mathbf{j}} = {\langle  x_\mathbf{j} = 1 | \prod_j Z_j^{(1-x_j)/2} \rho \prod_j  Z_j^{(1-x_j)/2}|x_\mathbf{j} = 1 \rangle }.
\end{equation}
We now use Eq.~\eqref{Eq:boundary_correlator} to heuristically interpret the diagonal entropy $S_d$ as an `average temporal boundary correlator' associated with a bulk action. To see this, let us assume that $\rho$ is obtained by applying a local, finite-depth quantum channel $\mathcal{E}$ to the ground state $|\Psi_0\rangle$ of a local Hamiltonian $H$, i.e., $\rho = \mathcal{E}[|\Psi_0\rangle \langle \Psi_0 |]$. In this setup, one can express the coarse-grained version of the state $\rho$ as a path integral in a $d+1$-dimensional spacetime:
\begin{equation}
\begin{aligned}
\rho \sim \int \mathcal{D}\varphi(r, \tau) & 
|\varphi(r,\infty) \rangle  \langle  \varphi(r,-\infty) |  \\
    & 
        e^{ - S_H[\varphi(r,\tau)] - S_\mathcal{E}[\varphi(r,\infty),\varphi(r,-\infty)]} ,
\end{aligned}
\end{equation}
where $S_H[\varphi(r,\tau)] = \int_{-\infty}^{\infty} d\tau \int d^d r \mathcal{L}_H[\varphi(r,\tau)] $ denotes the effective imaginary-time action associated with the Hamiltonian $H$ while $S_\mathcal{E}[\varphi(r,\infty),\varphi(r,-\infty)]] = \int  d^dr \mathcal{L}_\mathcal{E}[\varphi(r,\infty),\varphi(r,-\infty)] $ represents the effect of the channel $\mathcal{E}[\cdot]$.
We note that there is no integral over time in $S_\mathcal{E}$ since $\mathcal{E}$ is a finite-depth local channel.
Using $| x_\mathbf{j} = 1 \rangle \sim \int \mathcal{D}\varphi (r, -\infty )  |\varphi (r, -\infty ) \rangle $ and $\langle  x_\mathbf{j} = 1 |  \sim \int \mathcal{D}\varphi (r, \infty )  \langle \varphi (r, \infty ) | $, one can rewrite Eq.\eqref{Eq:boundary_correlator} as 
\begin{equation}
\label{Eq:boundary_continuum}
    \rho_{x_\mathbf{j}} \sim    \int \mathcal{D}\varphi(r, \tau)      e^{ - S_H - S_\mathcal{E}} [\cdots \varphi(r,-\infty) \varphi(r, \infty) \cdots],
\end{equation}
which is nothing but the temporal boundary correlation between $\rho$ and the symmetric product state $|x_\mathbf{j} = 1\rangle$ at $\tau = \infty, -\infty$.
The insertion of fields in ``$\cdots$'' depends on the measurement outcome $x_\mathbf{j}$. For example, $\rho_{x_\mathbf{j}}$ in a configuration where $x_\mathbf{j}$ satisfies $x_j = -1$ for $j = k,l$ and $x_j = 1$ otherwise corresponds to the two-point correlation function $\rho_{x_\mathbf{j}} \sim \langle [\varphi(k,-\infty) \varphi(k, \infty)]  [\varphi(l,-\infty) \varphi(l, \infty)] \rangle$, and it is precisely the type-II strange correlator introduced in Ref.~\cite{lee2022symmetry} to detect decoherence-induced transitions. It has been argued that the boundary correlator is short-ranged if $\rho$ is smoothly connected to the symmetric product state, whereas it can be long-ranged (i.e., it saturates to a constant as $|j-k| \rightarrow \infty$ or decays as a power law) in an SPT phase or a long-range entangled state \cite{you2014wave, lee2022symmetry, zhang2022strange}. Therefore,  
$S_d = - \sum_{x_\mathbf{j}} \rho_{x_\mathbf{j}} \log \rho_{x_\mathbf{j}}$ represents the average of the logarithm of the boundary correlator $\rho_{x_\mathbf{j}}$.  
Later, we will use the above connection between $\rho_{x_\mathbf{j}}$ and the boundary correlator to support the validity of our approach in efficiently estimating the diagonal entropy density.

(iii) \uline{Detecting Transitions invisible in von Neumann entropy:}

Finally, the diagonal entropy $S_d$ can be used to probe certain transitions of $\rho$ that are invisible even to the von Neumann entropy $S(\rho) = -\tr(\rho \log 
\rho)$. One example is subjecting a 2+1-D cluster state to phase-flip errors on all edges, described by $\mathcal{E}_e[\cdot] = (1-p)(\cdot) + p Z_e (\cdot) Z_e$.  
As a function of $p$, the mixed state undergoes a ``symmetry-enforced separability transition,'' where for $p > p_c \approx 0.109$ ($p < p_c$), the mixed state can (cannot) be expressed as a convex sum of symmetric states that spontaneously break a one-form symmetry \cite{chen2023symmetry}.  
One can show that the von Neumann entropy $S(\rho)$ is a smooth function of $p$ and thus cannot be used to detect the transition. Instead, the transition can be understood using the following convex decomposition ansatz \cite{chen2023symmetry}:
\begin{equation}
\begin{aligned}
\rho & = \sum_{x_\mathbf{j}} \rho^{1/2} |x_\mathbf{j}\rangle \langle x_\mathbf{j}| \rho^{1/2} = \sum_{x_\mathbf{j}} 
 |\psi_{x_\mathbf{j}}\rangle \langle \psi_{x_\mathbf{j}}| \\
& = \sum_{x_\mathbf{j}} \rho_{x_\mathbf{j}} | \tilde{\psi}_{x_\mathbf{j}}\rangle \langle \tilde{\psi}_{x_\mathbf{j}}|.
\end{aligned}
\end{equation}
Here $|\psi_{x_\mathbf{j}}\rangle = \rho^{1/2}|x_\mathbf{j}\rangle$ and $|\tilde{\psi}_{x_\mathbf{j}} \rangle = |\psi_{x_\mathbf{j}}\rangle/\sqrt{\langle \psi_{x_\mathbf{j}}| \psi_{x_\mathbf{j}}\rangle} = |\psi_{x_\mathbf{j}}\rangle/\sqrt{\rho_{x_\mathbf{j}}}$ is the normalized wavefunction.
It has been shown in Ref.~\cite{chen2023symmetry} that the Shannon entropy of this decomposition, which is equivalent to the diagonal entropy $S_d$, can be mapped to the disorder-averaged free energy of the random-bond Ising model (RBIM) along the Nishimori line and is therefore singular at $p = p_c \approx 0.109$. Therefore, the diagonal entropy of $\rho$ can be used to probe the symmetry-enforced separability transition, which is sometimes insensitive to the von Neumann entropy of $\rho$.

\subsection{Estimating diagonal entropy density using image compression} \label{sec:LZ}

We now discuss the efficient estimation of the diagonal entropy density $s_d$ in experiments, for example, in a quantum gas microscope~\cite{schlosser2001sub,bakr2009quantum,sherson2010single,cheuk2015quantum,haller2015single,parsons2015site,edge2015imaging,gross2021quantum}, which allows for the simultaneous measurement of all degrees of freedom in a preferred basis. Our method is based on lossless image compression, in particular, the Lempel-Ziv algorithm~\cite{ziv1977universal}. This algorithm has previously been applied to estimate the entropy of classical statistical mechanics models, as discussed in Ref.~\cite{sheinwald1990two,melchert2015analysis,stefano2019quantifying,martiniani2020correlation}.

We are interested in the diagonal entropy of a quantum state $\rho$ in a desired basis, which we choose as the Pauli-$X$ basis. The first step is to prepare the state $\rho$, followed by a projective measurement in the Pauli-$X$ basis. The observer then obtains a specific measurement outcome, a bit string $x_\mathbf{j} = (x_1, \cdots, x_N)$, with probability $\rho_{x_\mathbf{j}} = \langle x_\mathbf{j}|\rho|x_\mathbf{j}\rangle$. A straightforward way to obtain $S_d$ is to repeat the above process $N_s$ times and record the number of times each outcome $x_\mathbf{j}$ appears, which we denote as $N_{x_\mathbf{j}}$. The probability of a measurement outcome $x_\mathbf{j}$ can  be estimated as $\rho_{x_\mathbf{j}} \approx {N_{x_\mathbf{j}}}/ {N_s}$, and one may then compute the diagonal entropy using Eq.~\eqref{Eq:diagonal_entropy}. However, such an approach is unrealistic, as the number of measurements needed to estimate the diagonal entropy density at a fixed tolerance or error will generally scale exponentially with the system size.

To make progress, we note that if the probability $\rho_{{x}_{\mathbf{j}}} $ can be related to correlators with respect to some local Hamiltonian (see Eq.~\eqref{Eq:boundary_continuum}), then the \textit{image} of a typical measurement outcome $x_\mathbf{j}$ is expected to encode information about the Shannon entropy. For example, if $\rho$ has high fidelity with the symmetric product state $|x_\mathbf{j} = 1\rangle$, one typically observes that most of the sites point in the positive $x$-direction. The resulting image appears highly ordered, indicating that the system has low diagonal entropy. On the other hand, if $\rho$ is a completely random mixed state, the observer typically obtains a random bit string. The resulting image is highly disordered, suggesting that the system has large diagonal entropy. Following Refs.~\cite{stefano2019quantifying,martiniani2020correlation}, this intuition can be \textit{quantitatively} captured by the size of the losslessly compressed data file.
Specifically, consider the ``computable information density'' (CID) defined as
\begin{equation}
\label{Eq:CID}
    \text{CID}(x_\mathbf{j}) = \frac{\mathcal{N}(x_\mathbf{j})}{\mathcal{N}_\text{shuffle}},
\end{equation}
where $\mathcal{N}(x_\mathbf{j})$ is the size of the compressed file with the measurement outcome $x_\mathbf{j}$ and $\mathcal{N}_\text{shuffle}$ is the size of the compressed file for a random binary
sequence.  
The reason for dividing \( \mathcal{N}(x_\mathbf{j}) \) by \( \mathcal{N}_\text{shuffle} \) is simply to normalize the CID to unity for random binary sequences. It is known that if $x_\mathbf{j}$ is a sequence sampled from a stationary and ergodic process, the expectation value of CID, given by \be \mathbb{E}[\text{CID}] = \sum_{x_\mathbf{j}} p_{x_\mathbf{j}} \text{CID}[(x_\mathbf{j})],\ee approaches the Shannon entropy per site --- in our case, the diagonal entropy density $s_d = S_d/N$ --- in the limit $N \rightarrow \infty$ \cite{Wyner1989asymptotic,Nobel1992recurrence,ornstein2002entropy,wyner1991fixed,wyner2002sliding,ornstein1990universal}. We will numerically demonstrate that $\mathbb{E}[\text{CID}]$ closely approximates $s_d$ in all the examples we consider. We attribute this to the inherent locality in the probability distribution $\rho_{x_\mathbf{j}}$ hinted above.

We now summarize the protocol to estimate the diagonal entropy density:
\begin{enumerate}
  \item Prepare the target mixed state $\rho$ and then measure it in the Pauli-$X$ basis on all sites. The observer obtains an outcome $x_\mathbf{j} = (x_1, \cdots, x_N)$ with probability $p_{x_\mathbf{j}} = \rho_{x_\mathbf{j}}$.
  
  \item Compress the measurement outcome $x_\mathbf{j}$ using a lossless data compression algorithm, and calculate the size of the compressed file $\mathcal{N}(x_\mathbf{j})$.  
  We use the Lempel-Ziv 77 (LZ77) coding algorithm~\cite{ziv1977universal} throughout the paper, although we do not expect the choice of a specific variant of the Lempel-Ziv algorithm to be crucial. See App.~\ref{sec:LZ77} for a brief introduction.
  
  \item Compute $\text{CID}(x_\mathbf{j}) = \mathcal{N}(x_\mathbf{j})/N_\text{shuffle}$.

  \item Repeat steps 1–3 and take the average to obtain $\mathbb{E}[\text{CID}] = \sum_{x_\mathbf{j}} p_{x_\mathbf{j}} \text{CID}(x_\mathbf{j}) \approx \sum_{x_\mathbf{j} \in \text{samples}} \text{CID}(x_\mathbf{j})/N_s$, where $N_s$ is the number of samples.
  Here, $\sum_{x_\mathbf{j} \in \text{samples}}$ denotes summation over all samples generated in the first step. In Sec.~\ref{sec:complexity}, we discuss the scaling of the error with the number of samples.
\end{enumerate}

\section{1+1-D Transverse Field Ising model} \label{sec:tfim}

In this section, we will illustrate our approach to the diagonal entropy density for the ground state of the (1+1)-D transverse field Ising model (TFIM).
The Hamiltonian is:
\begin{equation}
\label{Eq:ham_TFIM}
H = - (1-J)\sum_i X_i - J \sum_{\langle i,j \rangle } Z_i Z_{j},
\end{equation}
where $\langle i,j\rangle$ denotes the nearest-neighbor pair.
The ground state density matrix $\rho(J) = |\Psi (J) \rangle \langle \Psi(J)|$ is in the symmetric phase for $J < J_c$ and in the symmetry-breaking phase for $J > J_c$, where $J_c = 0.5$. The universal, subleading terms in the diagonal entropy were discussed in Ref.~\cite{stephan2009shannon}, and as we discuss below, the volume-law coefficient of the diagonal entropy $S_d$ in the Ising-symmetric basis can also detect this transition (due to Kramers-Wannier duality, the diagonal entropy in the Pauli-$Z$ basis can be related to the one in the Pauli-$X$ basis~\cite{stephan2009shannon}).  

Of course, this transition can simply be detected via the connected two-point correlator $\langle Z_i Z_j\rangle \overset{\text{def}}{=} \tr(\rho Z_i Z_j) - \tr(\rho Z_i) \tr(\rho Z_j)$; thus, the diagonal entropy offers no apparent advantage. However, let us subject the ground state to the maximal dephasing channel $\rho(J) \rightarrow \mathcal{E}_d [\rho(J)]$, where $\mathcal{E}_d [\cdot] = \prod_j \mathcal{E}_j[\cdot]$ with $\mathcal{E}_j[\cdot] = (\rho + X_j \rho X_j)/2$. The correlator $\langle Z_i Z_j\rangle$ now vanishes identically due to symmetry constraints, whereas $S_d$ continues to serve as a diagnostic for what is now an \textit{SW-SSB transition}.  
We note that this particular SW-SSB transition can also be detected by the expectation value of Ising-symmetric operators, such as $\tr(\rho \sum_i X_i)/L$. However, as discussed below, the diagonal entropy is sensitive even to those SW-SSB transitions that are undetectable via such linear observables. In addition to serving as a testbed for our methodology, we will later use this example to analyze the complexity of estimating $S_d$ in experiments (Sec.~\ref{sec:complexity}).

\subsection{Structure of the diagonal density matrix and its relation to SW-SSB}

Let us elaborate on the connection between SW-SSB and the singularity of the diagonal entropy in the Ising-symmetric basis. $S_d$ is simply the von Neumann entropy of the diagonal mixed state $\rho_d$, which is obtained by applying the maximal dephasing channel $\mathcal{E}_d [\cdot] = \prod_j \mathcal{E}_j[\cdot]$ with $\mathcal{E}_j[\cdot] = [(\cdot) + X_j (\cdot) X_j]/2$ to $\rho$. This channel leaves all observables involving only Pauli-$X$ matrices unchanged.  
Therefore, the two-point connected correlation $\tr(\rho_d X_i X_j) - \tr(\rho_d X_i) \tr(\rho_d X_j)$ decays exponentially as a function of $|i-j|$ for $J \neq J_c$ and polynomially at $J = J_c$, since $|\Psi(J=J_c)\rangle$ is critical. It is then natural to expect that the von Neumann entropy of $\rho_d$, which equals the diagonal entropy $S_d$ of $\rho$ in the Ising-symmetric basis, is non-analytic at $J = J_c$. Later, we will use effective field theory to analyze its universal scaling behavior.

We claim that $\rho_{d}(J > J_c)$ is, in fact, an SW-SSB state \cite{lee2023quantum, ma2023topological, lessa2024strong}. By definition, an SW-SSB state $\rho$ satisfies the following two properties \cite{lessa2024strong}:   

\begin{enumerate}[label=(\alph*)]
    \item There is no long-range order in the two-point correlation function, i.e., $\lim_{|i-j| \rightarrow \infty } \tr(\rho Z_i Z_j) = 0$.  

    \item The fidelity between $\rho$ and $Z_i Z_j \rho Z_i Z_j$ saturates to a finite constant as $|i-j| \to \infty$, i.e., $\lim_{|i-j| \rightarrow \infty } F(\rho_d, Z_i Z_j \rho_d Z_i Z_j) = c > 0$.
\end{enumerate}

$\rho_d$ satisfies (a) because $\tr(\rho_d Z_i Z_j) = \tr( \mathcal{E}_d[\rho] Z_i Z_j) = \tr(\rho \mathcal{E}_d[Z_i Z_j]) = 0$, where, in the last equality, we have used $\mathcal{E}_d[Z_i Z_j] = 0\ \forall i,j$ due to Ising symmetry.  On the other hand, $\rho_d$ satisfies (b) due to the data-processing inequality:  
\bea 
\label{Eq:data_process}
F(\rho, Z_i Z_j \rho Z_i Z_j) & \leq & F(\mathcal{E}_d[\rho], \mathcal{E}_d[Z_i Z_j \rho Z_i Z_j]) \nonumber  \\
& = & F(\rho_d, Z_i Z_j \rho_d Z_i Z_j)
\eea 
where we have used $\mathcal{E}_d[Z_i Z_j (\cdot) Z_i Z_j] = Z_i Z_j\mathcal{E}_d[ \cdot ]Z_i Z_j$ in the second step.  Since $\rho(J) = |\Psi(J)\rangle \langle \Psi(J)|$ is a pure state, the left-hand side of Eq.~\eqref{Eq:data_process} satisfies $\lim_{|i-j|\rightarrow \infty} F(\rho(J), Z_i Z_j \rho(J) Z_i Z_j) = \lim_{|i-j|\rightarrow \infty} |\langle \psi(J)| Z_i Z_j |\psi(J)\rangle |^2 \neq 0$ for $J > J_c$. Therefore, the diagonal density matrix $\rho_d$ is an SW-SSB state for $J > J_c$.

		\begin{figure}
			\centering
			\includegraphics[width=\linewidth]{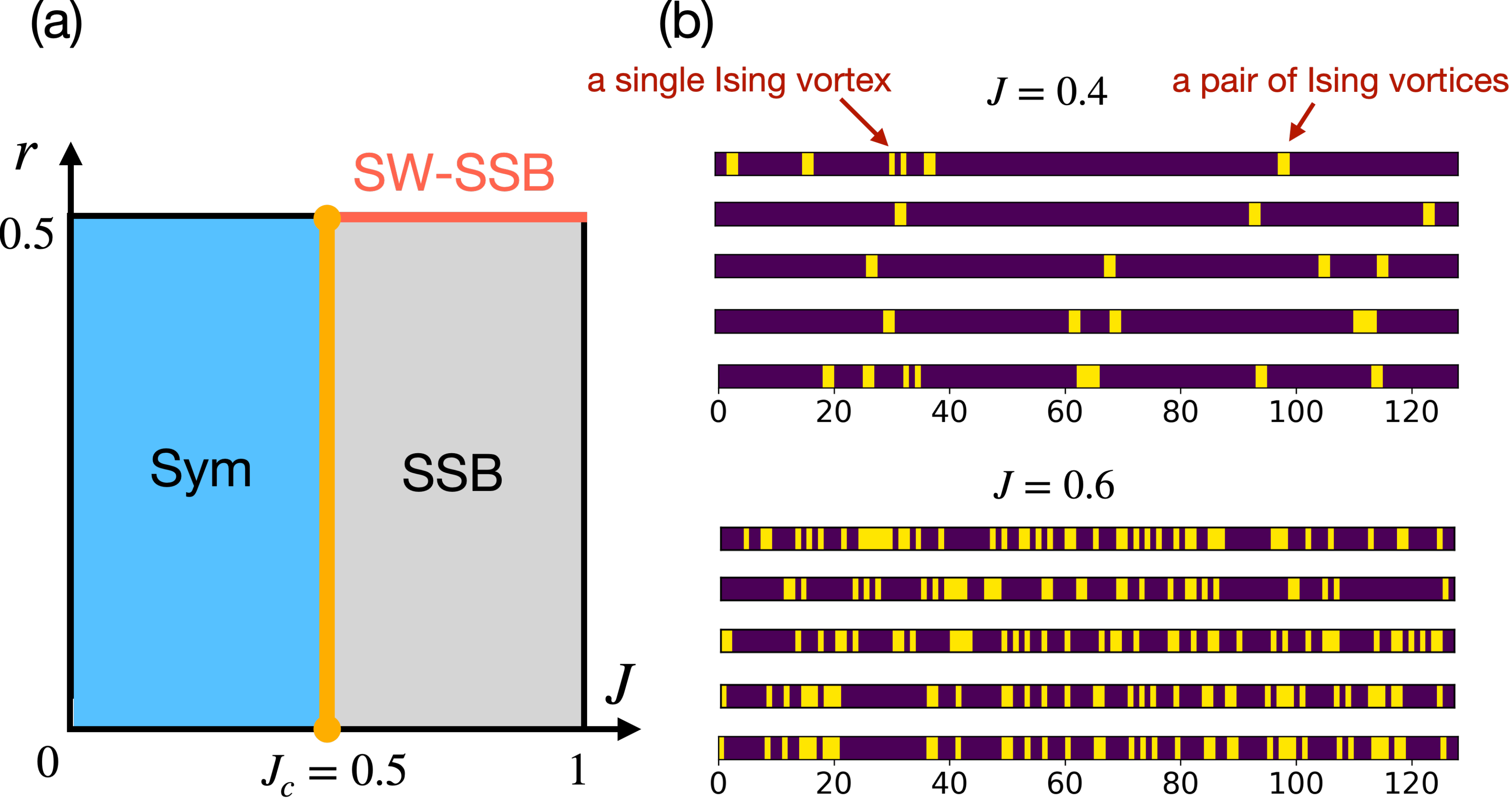}
			\caption{ 
				(a) The phase diagram $\rho(J,r)$ of the ground state of (1+1)-D TFIM [Eq\eqref{Eq:ham_TFIM}] $|\Psi(J)\rangle$ subjected to the channel  $\mathcal{E}_{r}[\cdot] = \prod_j \mathcal{E}_{j,r}[\cdot]$, $\mathcal{E}_{j,r}[\cdot] = (1-r)[\cdot] + r X_j [\cdot] X_j$.
				(b) 
				Typical images $x_\mathbf{j}$ based on the probability distribution $\rho_{x_\mathbf{j}}(J)=\langle x_\mathbf{j} |\rho(J,r) |x_\mathbf{j}\rangle  = |\langle x_\mathbf{j}|\Psi(J)\rangle|^2$ at $J = 0.4$ and $J = 0.6$. The total system size is $L = 128$.
			}
			\label{Fig:tfim_image}
		\end{figure}

More generally, one can apply the following channel $\mathcal{E}_{r}$ to the ground state $\rho(J)$ of TFIM, which interpolates between no dephasing and maximal dephasing as $r$ varies from 0 to 1/2:  
\be  
\label{Eq:dephasing}  
\mathcal{E}_{r}[\cdot] = \prod_j \mathcal{E}_{j,r}[\cdot], \quad \mathcal{E}_{j,r}[\cdot] = (1-r)[\cdot] + r X_j [\cdot] X_j.  
\ee  
The resulting density matrix, $\rho(J,r) = \mathcal{E}_{r}[\rho(J)]$, satisfies
\be 
\tr [\rho(J,r)Z_i Z_j ]= (1-2r)^2 \tr [\rho(J,r=0)Z_i Z_j ].
\ee 
Following the aforementioned criteria for SW-SSB, only the $r = 1/2$ line in the $(J,r)$ plane corresponds to an SW-SSB transition, whereas for any $r < 1/2$, the transition is a standard SSB transition. The phase diagram is shown in Fig.~\ref{Fig:tfim_image}(a).

\subsection{Universal scaling behavior of diagonal entropy}
\label{sec:tfim_analytical}

We now discuss the universal behavior of the diagonal entropy density. Our main conclusion is that the singular part of $\lim_{V\to \infty} S_d/V$ scales as $|(J - J_c) \log |J - J_c||$. This implies that the first derivative of $S_d$, i.e., $dS_d / dJ$, diverges at $J = J_c$.

\begin{figure}
	\centering
\includegraphics[width=\linewidth]{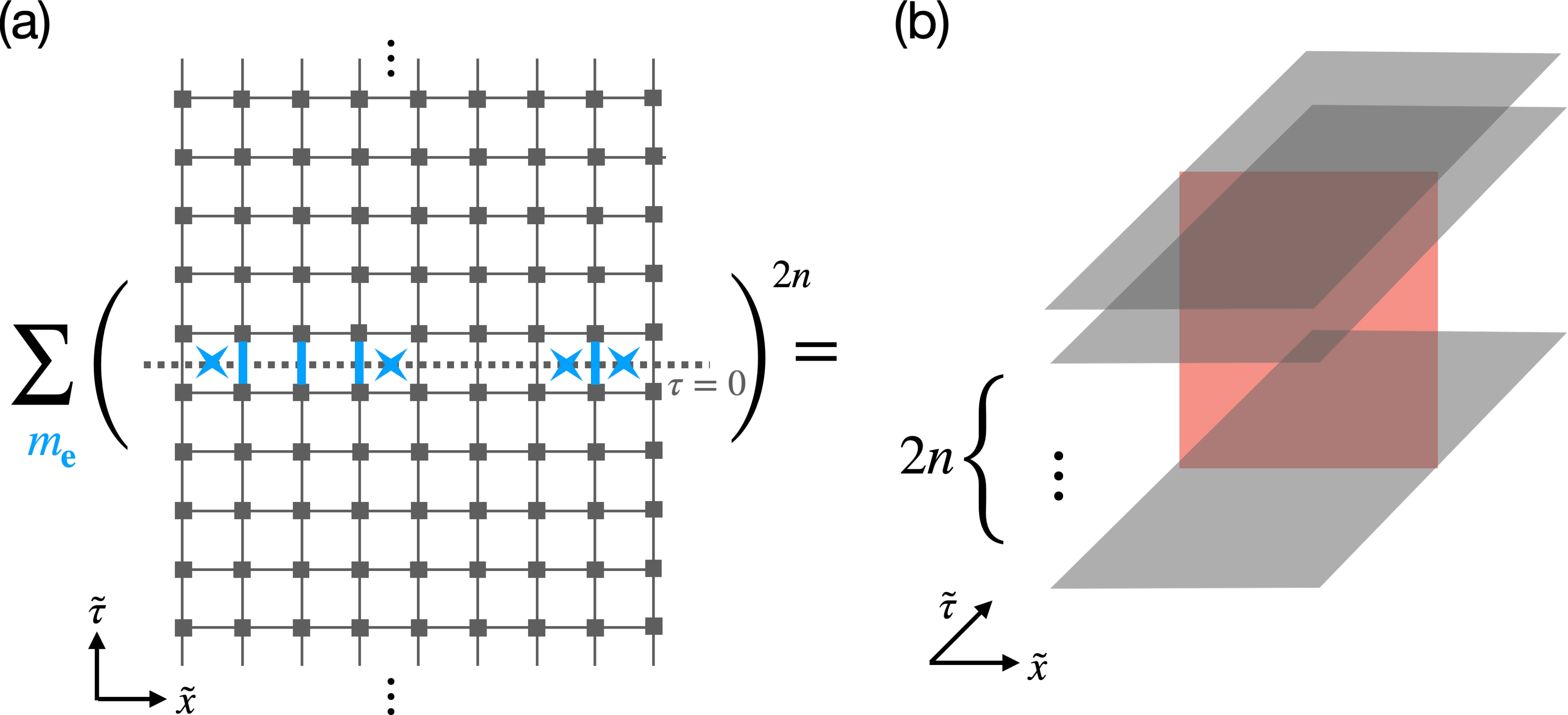}
	\caption{ Schematic representation of Eq.~\eqref{Eq:sum_me}. Fig.~(a) illustrates a term in the summation, representing $\mathcal{Z}_{m_\mathbf{e}}$, the partition function of a 2D classical Ising model with disorder $m_{\mathbf{e}}$ restricted to the $\tau = 0$ line. The blue bonds label the positions where $ m_e = -1$ and the blue crosses indicate the creation of an Ising vortex (corresponding to the location where $x_j = -1$). 
    Fig.~(b) depicts the right-hand side of Eq.~\eqref{Eq:sum_me}, which corresponds to the partition function of $2n$ copies of Ising models on an infinite plane, interacting with each other through translationally invariant interactions along the $\tau = 0$ line.
	}
\label{Fig:tfim_replica}
\end{figure}

Let us first examine the physical meaning of the probability distribution $\rho_{x_\mathbf{j}} = |\langle x_\mathbf{j} | \Psi\rangle|^2$. Following Eqs.\eqref{Eq:boundary_correlator}-\eqref{Eq:boundary_continuum}, $   \rho_{x_\mathbf{j}} = |\langle x_\mathbf{j} = 1| \prod_j Z_j^{(1-x_j)/2} |\Psi\rangle |^2 $.
In the continuum limit, $| x_\mathbf{j} = 1 \rangle \sim \int \mathcal{D}\varphi (r, 0)  |\varphi (r, 0) \rangle $ and 
$|\Psi\rangle \sim \int \mathcal{D} \varphi (r, \tau) e^{-S[\varphi(r,\tau)]} |\varphi(r, 0) \rangle$, where $S[\varphi(r,\tau)] = \int_{\tau = -\infty}^0 d\tau \int d r [\varphi (\partial_\tau^2 + \nabla_r^2 + m^2 ) \varphi + u \varphi^4$].
Therefore,
\begin{equation}
\label{Eq:rhod_tfim_corr}
    \rho_{x_\mathbf{j}} \sim |\int \mathcal{D}\varphi(r,\tau) [\cdots \varphi(r_j,0) \cdots ]  e^{-S[\varphi(r,\tau)]} |^2,
\end{equation}
which is the square of the multi-point correlator in the scalar $\phi^4$ theory at the imaginary time $\tau = 0$. 
It proves beneficial to examine the lattice-spacetime formulation of Eq.\eqref{Eq:rhod_tfim_corr}:
\begin{equation}
\label{Eq:tfim_corr_discrete}
\rho_{x_{\mathbf{j}}} \sim |\sum_{z_{\mathbf{j}, \boldsymbol{\tau}}} (\prod_{j = 0}^L z^{\frac{1-x_j}{2}}_{j, \tau = 0}) e^{\beta \sum_{\langle (j, \tau), (j', \tau') \rangle} z_{j, \tau} z_{j', \tau'}}|^2.
\end{equation}
Using the standard Kramers-Wannier duality, one can express the multi-point correlator as the following disorder model on the dual lattice:
\begin{equation}
\label{Eq:rhox_dual}
\begin{aligned}
\rho_{x_{\mathbf{j}}} & \sim |\sum_{s_{\mathbf{\tilde{
i}}}} e^{K (\sum_{\langle \tilde{i}, \tilde{j} \rangle  \in \tau \neq 0} s_{\tilde{i}} s_{\tilde{j}} +  \sum_{\langle \tilde{i}, \tilde{j} \rangle \in \tau  = 0} m_{\langle \tilde{i}, \tilde{j} \rangle} s_{\tilde{i}} s_{\tilde{j}}) }|^2 \\
& = \mathcal{Z}_{m_{\mathbf{e}}}^2.
\end{aligned}
\end{equation}
Here, $\tanh(\beta) = e^{-2K}$, and $m_{\mathbf{e}} = \{ m_{\langle \tilde{i}, \tilde{j}\rangle} = \pm 1\}$ is any bond configuration satisfying $\prod_{ \langle \tilde{i}, \tilde{j} \rangle \in j} m_{\langle \tilde{i}, \tilde{j} \rangle} = x_j$ (note that we sometimes write $e = \langle \tilde{i}, \tilde{j} \rangle$ for notational simplicity).  
Eq.~\eqref{Eq:rhox_dual} represents the partition function of a classical Ising model with disorder $m_{\mathbf{e}}$ restricted to the $\tau = 0$ line [see Fig.~\ref{Fig:tfim_replica}(a)]. In this dual picture, $x_j = -1$ corresponds to the presence of an Ising vortex, defined via $\prod_{ \langle \tilde{i}, \tilde{j} \rangle \in j} m_{\langle \tilde{i}, \tilde{j} \rangle} = -1$.  
In the symmetric phase, configurations in which a single Ising vortex is far from the others are exponentially suppressed, and the typical measurement outcomes correspond to situations where Ising vortices are always bound in pairs. On the other hand, in the symmetry-breaking phase, the two-point correlation function saturates to a finite constant, corresponding to a situation where Ising vortices are deconfined.

To formalize this intuition and understand the universal behavior of the diagonal entropy, $S_d = -\sum_{x_{\mathbf{j}}} \rho_{x_{\mathbf{j}}} \log \rho_{x_\mathbf{j}}$, we employ the replica trick to explicitly derive the statistical mechanical model for $S^{(n)}_d = {\log \sum_{x_\mathbf{j}} \rho^n_{x_\mathbf{j}}}/{(1-n)}$, and then take the replica limit $n \to 1$.
The central quantity associated with $S^{(n)}_d$ is obtained by summing over the disorder $m_\mathbf{e}$ of $\mathcal{Z}^{2n}_{m_\mathbf{e}}$, which can be computed as 
\begin{equation}
\label{Eq:sum_me}
\begin{aligned}
& \sum_{m_\mathbf{e}}\mathcal{Z}_{m_\mathbf{e}}^{2n} \\
& =  \sum_{s^{(\alpha)}_{\mathbf{\tilde{i}}}}  \Bigg[ e^{\beta  \sum_{\substack{\langle \tilde{i}, \tilde{j} \rangle  \in \tau \neq 0} }  \sum\limits_{\alpha=1}^{2n} s^{(\alpha)}_{\tilde{i}} s^{(\alpha)}_{\tilde{j}}} \prod_{\substack{\langle \tilde{i}, \tilde{j} \rangle \\ \in \tau = 0}} \cosh  (\beta \sum_{\alpha =1}^{2n} s^{(\alpha)}_{\tilde{i}} s^{(\alpha)}_{\tilde{j}}) \Bigg]\\
& = \sum_{s^{(\alpha)}_{\mathbf{\tilde{i}}}}  e^{-H[\beta; s^{(\alpha)}_{\mathbf{\tilde{i}}}]} = \sum_{s^{(\alpha)}_{\mathbf{\tilde{i}}}}  e^{ - ( \sum\limits_{\alpha=1}^{2n} 
 (H_{\text{Ising}, \tau>0 }^{\alpha}+H_{\text{Ising}, \tau>0}^{\alpha})  +  H_{\text{int}})}.
\end{aligned}
\end{equation}
Here, $H_{\text{Ising}, \tau > 0 }^{\alpha} = -\beta \sum_{\langle \tilde{i}, \tilde{j}\rangle \in \tau > 0} s^{(\alpha)}_{\tilde{i}} s^{(\alpha)}_{\tilde{j}}$ represents the $\alpha$-th copy of the Ising model on the upper half-infinite plane. A similar relation holds for $H_{\text{Ising}, \tau < 0 }^{\alpha}$ on the lower half-infinite plane. On the other hand, {$H_{\text{int}} = -\sum_{{x}} \ln \cosh( \beta \sum_{\alpha=1}^{2n} s^{(\alpha)}_{{x}, 0^+} s^{(\alpha)}_{{x}, 0^-})$} represents the interaction terms that couple the $2n$ copies of Ising models through the boundary spins $s^{(\alpha)}_{{x}, 0^+}$ and $s^{(\alpha)}_{{x}, 0^-}$ [see Fig.~\ref{Fig:tfim_replica}(b) for a schematic representation].  

Eq.~\eqref{Eq:sum_me} implies that $F(2n) = -\log(\sum_{m_{\mathbf{e}}}\mathcal{Z}_{ m_\mathbf{e}}^{2n})$ represents the free energy of $2n$ copies of Ising models on the infinite plane (or equivalently, $4n$ copies of Ising models on the half-infinite plane), interacting with one another through translationally invariant interactions along the $\tau = 0$ line. Crucially, since $\cosh(x)$ is an even function, $H_{\text{int}}$ is invariant under the spin-flip operation applied to all copies exclusively on the upper (or lower) half-infinite plane: $s^{(\alpha)}_{x, \tau > 0} \rightarrow -s^{(\alpha)}_{x, \tau > 0}$ for all $\alpha$. This fact will play an important role in the relevance/irrelevance of interactions between different copies of the Ising model. Meanwhile, let us first relate $S_d^{(n)}$ to the difference in boundary free energy between a single $2n$-coupled Ising model and $n$ copies of a $2$-coupled Ising model. Denoting the free energy of a single Ising model on the infinite plane as $F_{\text{Ising}}(1)$, the boundary (or interfacial) free energy of a $2n$-coupled Ising model is defined as $F_b(2n) \equiv F(2n) - F_{\text{Ising}}(2n)$.
It follows that 
\begin{equation}
\label{Eq:Sd_Fb}
\begin{aligned}
S_d^{(n)} & =\frac{1}{1-n} \log( \frac{\sum_{m_\mathbf{e}}\mathcal{Z}_{m_\mathbf{e}}^{2n}}{(\sum_{m_\mathbf{e}}\mathcal{Z}_{m_\mathbf{e}}^{2})^n})
 = \frac{F(2n) - n F(2)}{n-1} \\
& = \frac{[F(2n) - F_{\text{Ising}}(2n)] - n [F(2) - F_{\text{Ising}}(2)]}{n-1} \\
& = \frac{F_b(2n) - n F_b(2)}{n-1},
\end{aligned}
\end{equation}
where we have used the additivity of free energies for the decoupled models, i.e., $F_{\text{Ising}}(2n) = n F_{\text{Ising}}(2)$, in the second line. Therefore, the universal behavior of $S_d$ can be understood by studying the boundary free energy $F_b(2n)$ and taking the replica limit $n \to 1$ in Eq.~\eqref{Eq:Sd_Fb}.

We now study the effect of $H_{\text{int}}$ in Eq.~\eqref{Eq:sum_me} close to the decoupled Ising conformal critical point on the half-infinite planes. The most relevant term allowed by symmetry is described by the following action  
\begin{equation}  
\mathcal{S}_{\text{int}} = \lambda \int dx  \sum_{\alpha \neq \beta} \sigma^{(\alpha)}_{(x, 0^+)} \sigma^{(\alpha)}_{(x, 0^-)}  \sigma^{(\beta)}_{(x, 0^+)}  \sigma^{(\beta)}_{(x, 0^-)},  
\end{equation}  
where $\sigma^{(\alpha)}_{(x, \tau)}$ denotes the spin operators in the continuum limit. We emphasize that the integral is only over space, as $H_{\text{int}}$ only involves interactions along the $\tau = 0$ line. Furthermore, the more relevant term $\int dx \sum_{\alpha} \sigma^{(\alpha)}_{(x, 0^+)} \sigma^{(\alpha)}_{(x, 0^-)}$ is prohibited by the aforementioned symmetry, which flips the spins for all copies exclusively on the upper (or lower) half-plane, simultaneously sending $\sigma^{(\alpha)}_{(x, 0^+)}$ to $-\sigma^{(\alpha)}_{(x, 0^-)}$ for all $\alpha = 1, \ldots, 2n$. Since the scaling dimension of $\sigma^{(\alpha)}_{(x, 0^\pm)}$ is $\text{dim}[\sigma^{(\alpha)}_{(x, 0^\pm)}] = 1/2$ \cite{CFT1997}, one finds that $\text{dim}[\sigma^{(\alpha)}_{(x, 0^+)} \sigma^{(\alpha)}_{(x, 0^-)}  \sigma^{(\beta)}_{(x, 0^+)}  \sigma^{(\beta)}_{(x, 0^-)}] = 2 <1$, making it an \textit{irrelevant} perturbation. Therefore, the singular part of $F_b(2n)$ takes the same scaling form as the singular part of the boundary free energy of a single Ising model, $F_b(2n = 1)$, which is given by $F^{\text{sing}}_b = c_{2n=1} |(J-J_c)\log(|J-J_c|)|$ \cite{fisher1967interfacial}. Here, the coefficient $c_{2n=1}$ is non-universal and generally depends on microscopic details, including irrelevant terms. Consequently, in general, $F^{\text{sing}}_b(2n) = 2n c_{2n} |(J-J_c)\log(|J-J_c|)|$. Expanding $c_{2n}$ in a Taylor series as $c_{2n} \approx c_{2} + (2n-2) c'_2$ for $n \approx 1$, the singular part of $s^{\infty}_d = \lim_{V\to \infty} S_d/V$ can be derived by taking the replica limit $n \to 1$ in Eq.~\eqref{Eq:Sd_Fb}:  
\begin{equation}  
\begin{aligned}  
\label{Eq:sd_replicalimit}  
s^{\infty,\text{sing}}_d & = \lim_{n \to 1}\Big( 2n \frac{ (2n-2) c'_2}{n-1} \Big) |(J-J_c)\log(|J-J_c)|)|  \\  
& = 2 c'_2 |(J-J_c)\log(|J-J_c)|)|.  
\end{aligned}  
\end{equation}  
Therefore, we expect that the first derivative of $s_d$ will show divergent behavior at the critical point --- this feature will later be confirmed through numerical results.

\subsection{Numerical Estimation of diaogonal entropy using Lempel-Ziv compression in 1+1-D TFIM}

\begin{figure*}
	\centering
\includegraphics[width=\linewidth]{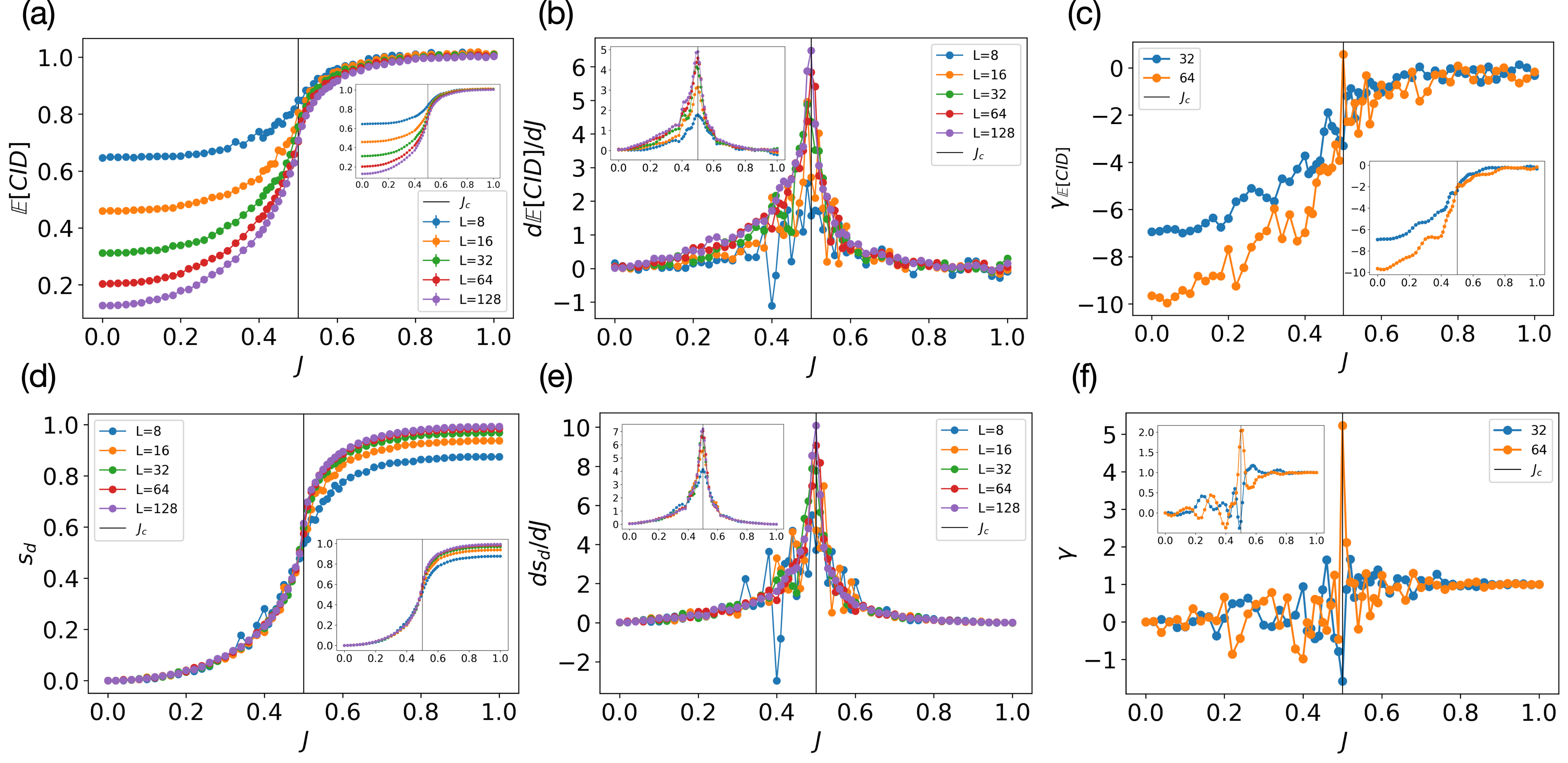}
	\caption{	(a) $\mathbb{E}[\text{CID}]$ as a function of the tuning parameter $J$ for the ground state of the $(1+1)$-D TFIM in Eq.\eqref{Eq:ham_TFIM}. (b) The derivative $d \mathbb{E}[\text{CID}]/ d J$ as a function of $J$ estimated using the finite-difference method. (c) The subleading term $\gamma_{\mathbb{E}[\text{CID}]} = 2L [\mathbb{E}[\text{CID}](2L) - \mathbb{E}[\text{CID}](L)]$ for $L = 32, 64$ as a function of $J$.  (d)-(f): Same quantities as (a)-(c) with $\mathbb{E}[\text{CID}]$ replaced by the diagonal entropy density $s_d = - \sum_{x_{\mathbf{j}}} \rho_{x_{\mathbf{j}}} \log \rho_{x_{\mathbf{j}}}/L \approx -\sum_{x_\mathbf{j} \in \text{samples}} \log(\rho_{x_\mathbf{j}})/(L N_s) $, where $N_s$ is the number of samples.
 All insets show the results obtained from smoothing the raw data by locally averaging the nearby data points (see the main text for details).
    }
\label{Fig:tfim_LZ_direct_all}
\end{figure*}

To illustrate the usefulness of CID and provide support for the above replicated field-theory analysis, we apply the dephasing channel in Eq.~\eqref{Eq:dephasing} to the ground state of the (1+1)-D TFIM in Eq.~\eqref{Eq:ham_TFIM}. The expected phase diagram of the resulting density matrix $\rho = \rho(J,r)$ is shown in Fig.~\ref{Fig:tfim_image}(a). The probability $\rho_{x_\mathbf{j}} = \langle x_\mathbf{j} |\rho(J,r) |x_\mathbf{j}\rangle = |\langle x_\mathbf{j} |\Psi(J)\rangle |^2$ is independent of $r$ and can be computed by mapping the system to free fermions. Typical measurement outcomes based on the probability distribution can then be obtained using the standard Metropolis–Hastings algorithm (see App.\ref{sec:app_numerics} for details).
Fig.~\ref{Fig:tfim_image}(b) displays five samples based on the probability distribution $\rho_{x_\mathbf{j}}$ for a total system size of $L = 128$, with $J = 0.4$ and $0.6$ (the critical point is at $J_c = 0.5$). In these images, yellow points indicate positions where $x_j = -1$, corresponding to Ising vortices in the dual picture. It is evident that Ising vortices are suppressed for $J < J_c$ and proliferate for $J > J_c$, consistent with our intuition discussed in the previous subsection.

Next, we compute $\mathbb{E}[\text{CID}]$ for these images. The results are shown in Fig.~\ref{Fig:tfim_LZ_direct_all}(a). We also compute the diagonal entropy $s_d$ directly as $s_d\approx -\sum_{x_\mathbf{j} \in \text{samples}} \log(\rho_{x_\mathbf{j}})/(L N_s)$, where $N_s$ is the number of samples. The results are shown in Fig.~\ref{Fig:tfim_LZ_direct_all}(d). Overall, for large system sizes ($L \approx 100$), $\mathbb{E}[\text{CID}]$ provides a good approximation of $s_d$. For all system sizes, $\mathbb{E}[\text{CID}] > s_d$, as expected based on exact results \cite{shannon1948mathematical}. Furthermore, the difference $\mathbb{E}[\text{CID}] - s_d$ decreases with increasing $L$, a trend similar to previous work on classical systems \cite{stefano2019quantifying,martiniani2020correlation}. We will revisit the systematic study of the difference $\mathbb{E}[\text{CID}] - s_d$ as a function of system size $L$ in Sec.~\ref{sec:complexity}.

The above field-theory analysis predicts that the singular part of $s_d$ scales as $|(J-J_c) \log(|J-J_c|)|$. Therefore, to expose the singularity and locate the critical point, one needs to take the first derivative of $s_d$. We use finite-difference differentiation to compute the first derivative of $\mathbb{E}[\text{CID}]$ with respect to $J$. As shown in Fig.~\ref{Fig:tfim_LZ_direct_all}(c), despite the noisy data, we observe clear signatures of a singularity at $J \approx J_c = 0.5$. We also compare it with $ds_d/dJ$ [Fig.~\ref{Fig:tfim_LZ_direct_all}(f)], and both exhibit similar qualitative behavior. The noise can be mitigated by locally averaging the data (see the next paragraph for the procedure), and the results for both CID and direct calculations are shown in the insets. Although we do not have sufficient resolution to fit the data to the analytical prediction, the location of the critical point and the presence of a singularity are apparent in our numerical results.

Alternatively, one can locate the transition point by exploiting the ergodicity breaking in the ferromagnetic phase, which leads to the expectation that \( S_d = \alpha L \log(2) - \gamma \), with \( \gamma = \log(2) \) in the ferromagnetic phase and \( \gamma = 0 \) in the paramagnetic phase~\cite{stephan2009shannon}. Therefore, one may use \( \gamma = 2L [s_d(2L) - s_d(L)] \) as an order parameter for the transition. We first calculate this difference based on our direct estimation of \( S_d \) (i.e., without using the Lempel-Ziv algorithm) and then discuss how well it is reproduced by \( \mathbb{E}[\text{CID}] \). The result for \( \gamma = 2L [s_d(2L) - s_d(L)] \) as a function of \( J \) is shown in Fig.~\ref{Fig:tfim_LZ_direct_all}(e). Note that the subtraction amplifies noise, so even a small amount can significantly degrade the quality of the results. However, it is still evident that \( \gamma \approx 0 \) (\( \log 2 \)) when \( J < J_c \) (\( J > J_c \)). To improve the quality of the results, we reduce noise by smoothing the raw data. Specifically, we locally average $s_d$ through $s^{(r+1)}_d(J = n \Delta J) = [s^{(r)}_d( n \Delta J - \Delta J ) + s^{(r)}_d(n \Delta J) + s^{(r)}_d(n \Delta J + \Delta J)]/3 $, where the superscript $r$ indicates the number of iteration so that the raw data  $s_d(n \Delta J) $ corresponds to $r = 0$.
The resulting \( s_d^{(3)} \) and \( \gamma^{(3)} \) are shown in the insets of Fig.~\ref{Fig:tfim_LZ_direct_all}(d) and (e), respectively, and the difference between \( J < J_c \) and \( J > J_c \) for \( \gamma^{(3)} \) becomes more pronounced. 
Therefore, the subleading term of the diagonal entropy indeed serves as an order parameter that distinguishes the symmetric and SSB phases. As an aside, the subleading correction to $S_d$ in the SSB phase is already visible in the finite-size effects for $s_d = S_d/L$ in Fig.~\ref{Fig:tfim_LZ_direct_all}(d), where $s_d \approx s^{\infty}_d - \log(2)/L$ increases as a function of $L$ at fixed $J$. Note that the diagonal density matrix has an exact $\mathbb{Z}_2$ symmetry; thus, at $J = 1$, $\rho_d = \sum_{x_\mathbf{j}\ \text{s.t.}\ \prod_j x_j = 1} |x_\mathbf{j}\rangle \langle x_\mathbf{j}|/2^{L-1}$. It follows that $S_d = (L-1) \log(2) $ at $J = 1$, implying that $s_d = S_d / L = 1 - \log 2 / L$.

However, as shown in Fig.~\ref{Fig:tfim_LZ_direct_all}(b), we are unable to extract the subleading $\log(2)$ entropy contribution using $\gamma_{\mathbb{E}[\text{CID}]} = 2L [\mathbb{E}[\text{CID}](2L) - \mathbb{E}[\text{CID}](L)]$. The inset shows the data after applying an averaging procedure similar to that used for $s_d$ described above. Although $\gamma_{\mathbb{E}[\text{CID}]}$ exhibits a sharp feature across $J = J_c$, it does not take the expected values of $0$ ($\log 2$) for $J < J_c$ ($J > J_c$), and it significantly depends on the system size for $J < J_c$. We suspect that the global constraint $\prod_j x_j = 1$ is not captured by the LZ-77 coding algorithm, which compresses data locally, leading to its failure in detecting the correct subleading term.

\section{Strong-to-Weak SSB and  standard SSB in 2+1-D Ising Paramagnet}
\label{sec:chamon}

\begin{figure*}
	\centering
\includegraphics[width=\linewidth]{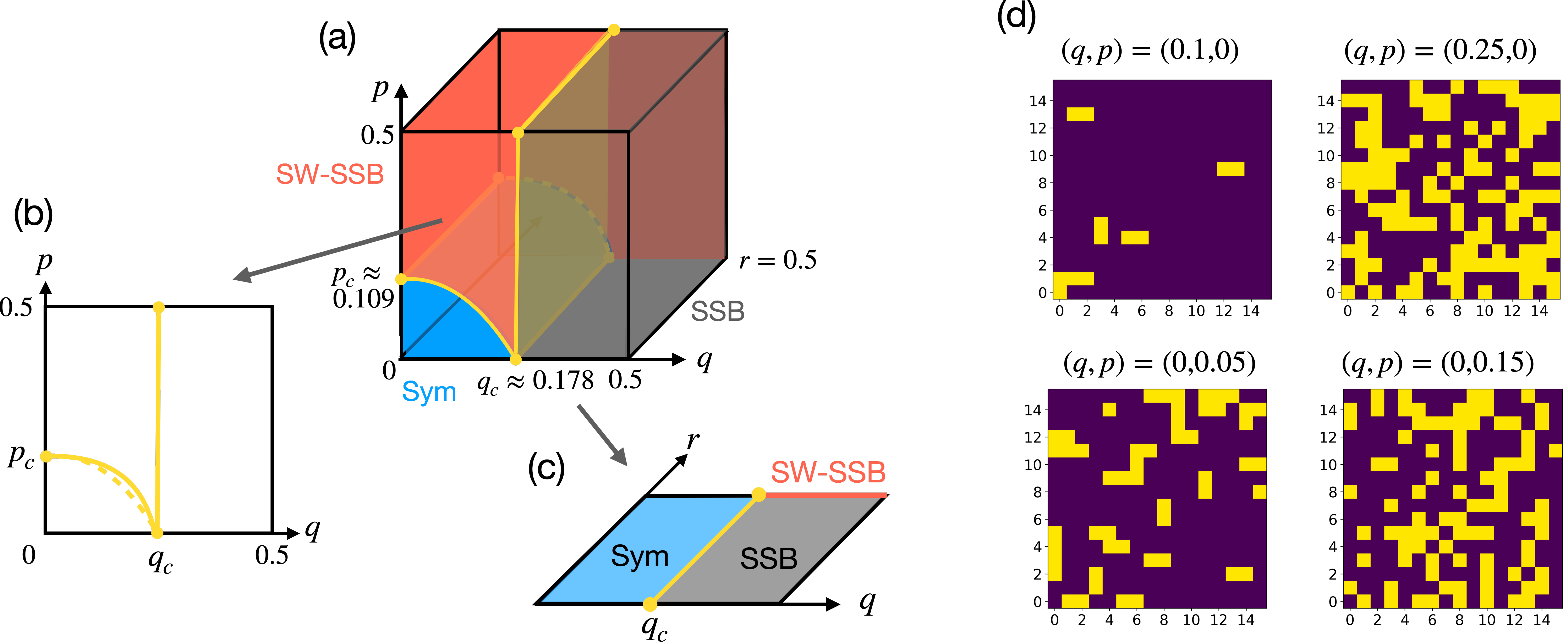}
	\caption{	
   (a) The three-dimensional phase diagram for the mixed state $\sigma(p,q,r) = \mathcal{E}_{r}[\rho(p,q)]$, where $\mathcal{E}_{r}[\cdot] = \prod_j \mathcal{E}_{j,r}[\cdot]$, $\mathcal{E}_{j,r}[\cdot] = (1-r)[\cdot] + r X_j [\cdot] X_j$ (see Eqs.~\ref{eq:cc_state},\ref{eq:zzchannel}).
 The target mixed state $\rho$ and the diagonal mixed state $\rho_d$ corresponds to the $r = 0$ and $r = 0.5$ plane, respectively.
 (b) The phase boundaries for $\rho$ (solid line) and $\rho_d$ (dashed lines). Note that $\rho$ exhibit three phases (Sym, SSB, and SW-SSB phase) while $\rho_d$ only exhibit two phases (Sym and SW-SSB phase). 
 (c) The phase diagram of $\rho(q,p = 0,r)$. Note that the SW-SSB phase only exists along the $r = 0.5$ line with $q>q_c$. 
 (d) The typical images $x_\mathbf{j}$ based on the probability distribution $\rho_{x_\mathbf{j}}(q,p)$ at $(q,p) = (0.1, 0), (0.25, 0), (0, 0.05)$, and $(0, 0.15)$. 
	}
\label{Fig:perturb_phase_image}
\end{figure*}

In this section, we study the phase diagram of a (2+1)-D paramagnet that exhibits a decoherence-induced transition from a paramagnet to an SW-SSB phase, as well as a transition from a paramagnet to a conventional ferromagnet. We will elucidate the close connection between the phase diagram of the decohered density matrix and that of the corresponding diagonal density matrix in the symmetric basis. We will then use this connection to investigate the phase diagram using $\mathbb{E}[\text{CID}]$ and the diagonal entropy $S_D$.

The Hilbert space of our model consists of qubits on the vertices of a 2D square lattice. The state of our interest is obtained as follows: we start with the pure state  

\be 
|\Psi(q)\rangle = \prod_{\langle i,j\rangle}[(1-q)I + q Z_i Z_j] |x_\mathbf{j} = 1\rangle \label{eq:cc_state}
\ee 
and then subject it to the following strongly symmetric channel on all edges: 

\be 
\mathcal{E}_{\langle i,j\rangle }(p)[\cdot] = (1-p)(\cdot) + p Z_i Z_j (\cdot) Z_i Z_j. \label{eq:zzchannel}
\ee 

The resulting density matrix $\rho(q,p) = \prod_{\langle i, j\rangle} \mathcal{E}_{\langle i,j\rangle }(p)[|\Psi(q)\rangle \langle \Psi(q)|]$ thus depends on two tuning parameters. Analogous to the discussion in Sec.~\ref{sec:tfim}, we are again interested in the relationship between the ``target mixed state'' $\rho(p,q)$ and the diagonal mixed state $\rho_d$, which is obtained by subjecting $\rho(p,q)$ to a maximal dephasing channel in the Pauli-$X$ basis. Thus, it is useful to consider a more general three-parameter mixed state $\sigma(p,q,r) = \mathcal{E}_{r}[\rho(p,q)]$, where the channel $\mathcal{E}_{r}$ is defined in Eq.~\eqref{Eq:dephasing}. The planes $r = 0$ and $r = 0.5$ correspond to $\rho(p,q)$ and $\rho_d(p,q)$, respectively.

In the following, we provide arguments that the phase diagram of the state $\sigma(p,q,r)$ schematically corresponds to Fig.~\ref{Fig:perturb_phase_image}(a). In particular, we will see that the locations of all solid yellow points along the phase boundaries are determined exactly via mapping to appropriate classical statistical mechanics models. We first consider the $r = 0$ plane, which corresponds to the target mixed state $\rho(q,p) = \sigma(q,p,0)$, and then use a combination of statistical mechanics mappings, quantum information inequalities, and perturbative arguments to analyze the three-dimensional global phase diagram.

\textbf{Phase diagram in the $\mathbf{r = 0}$ plane:} 
The conjectured phase diagram for the target mixed state $\rho(q,p)$ has been discussed in Ref.~\cite{chen2024unconventional} (more precisely, Ref.~\cite{chen2024unconventional} studied the phase diagram of the Wegner dual of $\rho(q,p)$). Along the $p = 0$ line, $\rho(q,0) = |\Psi(q)\rangle \langle \Psi(q)|$ is the ground state of a Rokhsar-Kivelson-type Hamiltonian \cite{rokhsar1988superconductivity}, and the paramagnetic-to-ferromagnetic transition (which occurs at $q_c = (1-\sqrt{\sqrt{2} - 1})/2 \approx 0.178$) is described by the $(2+0)$-D Ising universality (and not the $(3+0)$-D Ising model), a phenomenon of dimensional reduction commonly observed in ``conformal quantum critical points'' \cite{ardonne2004topological,castelnovo2008quantum}. On the other hand, the $q = 0$ line corresponds to decohering the zero-correlation-length paramagnetic state, which has already been studied in Ref.~\cite{lee2023quantum}. When $p > p_c \approx 0.109$, the system is in the SW-SSB phase characterized by the long-range order in the fidelity correlator without long-range order in the two-point correlations \cite{lessa2024strong}. From the results along the $p = 0$ and $q = 0$ lines, it is natural to expect that the system across the entire $(q,p)$-plane exhibits three phases: symmetric (Sym), SW-SSB, and SSB phases. The phase boundary between the SSB and SW-SSB phases is straight because non-maximal local decoherence cannot alter long-distance correlations of local operators; therefore, the correlations along the line $q = q_c$ must decay as a power law. The $r = 0$ plane in Fig.~\ref{Fig:perturb_phase_image}(a) presents the conjectured phase diagram. We emphasize that the straight yellow boundary at $q = q_c$ is precisely determined by the long-range order of the two-point correlation function, whereas the curved yellow line is drawn only schematically.

\textbf{Phase diagram for $\mathbf{r \neq 0}$:}
Given the above understanding of the $r = 0$ plane, we can now use the property that $\sigma(q,p,r) = \mathcal{E}_r[\rho(q,p)]$ to infer the topology of the entire phase diagram. Similar to the discussion in Sec.~\ref{sec:tfim}, a straightforward calculation shows that $\tr [\sigma(q,p,r)Z_i Z_j ]= (1-2r)^2 \tr [\rho(q,p)Z_i Z_j ]$. Therefore, $\sigma(q,p,r<0.5)$ exhibits long- (short-) range order in the two-point correlations if and only if $\rho(q,p)$ exhibits long- (short-) range order in the two-point correlations. For the special case $r = 0.5$ (which corresponds to the diagonal mixed state $\rho_d$), the channel $\mathcal{E}_{r=0.5} [\cdot]$ completely eliminates the two-point correlations. This implies that the system in the $r = 0.5$ plane will never exhibit an SSB phase. On the other hand, it can also be shown that the fidelity $F(\sigma, Z_i Z_j \sigma Z_i Z_j)$ is an upper bound of the fidelity $F(\rho, Z_i Z_j \rho Z_i Z_j)$, using the same argument as in Sec.~\ref{sec:tfim}. It follows that the transition of $\sigma(p,q, r > 0)$ out of the symmetric phase cannot occur before the transition of $\rho(p,q) = \sigma(p,q, r = 0)$ out of the symmetric phase.

We now elaborate on the relationship between the diagonal mixed state $\rho_d(q,p) = \sigma(q,p,r = 0.5)$ and the target mixed state $\rho(q,p)$. First, consider these density matrices at $q = 0$: $\rho(q = 0,p) \propto \sum_{x_\mathbf{j}\ \text{s.t.} \prod_j x_{j} =1} \mathcal{Z}_{x_\mathbf{j}}(p) |x_\mathbf{j}\rangle \langle x_\mathbf{j}|$ \cite{lee2023quantum}, where $\mathcal{Z}_{x_\mathbf{j} }(p) = \sum_{s_{\tilde{i}}} e^{\beta \sum_{\langle \tilde{i}, \tilde{j} \rangle} J_{\langle \tilde{i}, \tilde{j} \rangle} s_{\tilde{i}} s_{\tilde{j}} },\ \tanh(\beta) = 1-2p$ is the partition function of the random bond Ising model on the dual lattice $ \mathbf{{\tilde{i}}} $, with $\{ J_{\langle \tilde{i}, \tilde{j} \rangle } \}$ representing any bond configuration satisfying $\prod_{ \langle \tilde{i}, \tilde{j} \rangle  \in j} J_{\langle \tilde{i}, \tilde{j} \rangle }   = x_{j}$. Since this density matrix is already diagonal in the Pauli-$X$ basis, it follows that $\rho(0,p) = \rho_d(0,p)$, and thus the diagonal entropy is exactly equal to the von Neumann entropy $S(\rho) = -\tr(\rho \log \rho)$ of the target mixed state. Furthermore, $S(p,q) = S_d(p,q) + O(q^2)$ for small $q$. This follows from the symmetry $S(p,q) = S(p,-q)$, and similarly $S_d(p,-q) = S_d(p,q)$. This symmetry is implemented by the unitary operator $\prod_{i \in A} X_i$, where $i \in A$ denotes sites on the $A$ sublattice of the square lattice. Thus, the phase boundaries determined by the singularities of $S$ and $S_d$ have zero slope at $(p = p_c, q = 0)$ and coincide at linear order in $q$.

Remarkably, the diagonal entropy also correctly captures the transition for $\rho$ when $p = 0$, i.e., along the pure-state transition line. This can be seen by expressing the eigenvalues of $\rho_d$ analytically: $\rho_{x_\mathbf{j}} = |\langle x_\mathbf{j}| \Psi\rangle |^2 = |\langle x_\mathbf{j} = 1| \prod_j Z_j^{(1-x_j)/2} |\Psi\rangle |^2 \propto |\langle x_\mathbf{j} = 1| \prod_j Z_j^{(1-x_j)/2} e^{K \sum_{\langle i,j\rangle} Z_i Z_j} |x_\mathbf{j} = 1\rangle |^2$, where $\tanh(K) = q/(1-q)$. By inserting a complete Pauli-$Z$ basis $I = \sum_{z_\mathbf{j}} |z_\mathbf{j} \rangle \langle z_\mathbf{j}|$ between $e^{K \sum_{\langle i,j\rangle} Z_i Z_j}$ and $|x_\mathbf{j} = 1\rangle$, we obtain
\begin{align}
\label{Eq:rho2_chamon_corr}
\rho_{x_\mathbf{j}} & \propto |\sum_{z_\mathbf{j}} \prod_j z_j^{(1-x_j)/2} e^{K \sum_{\langle i,j \rangle} z_i z_j}|^2 \\   
\label{Eq:rho2_RBIM}
& \propto |\sum_{s_{\tilde{
i}}} e^{\beta \sum_{\langle \tilde{i}, \tilde{j} \rangle} J_{\langle \tilde{i}, \tilde{j} \rangle} s_{\tilde{i}} s_{\tilde{j}} }|^2 = \mathcal{Z}^2_{x_\mathbf{j}}(q).
\end{align}
Here, $\tanh(\beta) = 1-2q$, and $\{ J_{\langle \tilde{i}, \tilde{j} \rangle } \}$ represents any bond configuration satisfying $\prod_{ \langle \tilde{i}, \tilde{j} \rangle  \in j} J_{\langle \tilde{i}, \tilde{j} \rangle }  = x_{j}$.
Note that Eq.~\eqref{Eq:rho2_RBIM} follows from the standard Kramers–Wannier duality. By comparing $\rho(0,p) \propto \sum_{x_\mathbf{j}\ \text{s.t.} \prod_j x_{j} =1} \mathcal{Z}_{x_\mathbf{j}}(p) |x_\mathbf{j}\rangle \langle x_\mathbf{j}|$ with Eq.~\eqref{Eq:rho2_RBIM}, we find that $\rho_d(q,0) = \rho^2(0,p = q)/\tr[\rho^2(0,p = q)]$ is proportional to the \textit{square} of the fixed-point paramagnetic state under decoherence. Ref.~\cite{chen2023separability} shows that the density matrix $\rho^2(0,p)/\tr[\rho^2(0,p)]$ undergoes a separability transition at $p = q_c$, where it can (cannot) be expressed as a convex sum of GHZ states for $p > q_c$ ($p < q_c$). This serves as a signature of SW-SSB, leading to the conclusion that the SSB transition in the pure state $\rho(q,0) = |\Psi(q)\rangle \langle \Psi(q)|$ manifests as the SW-SSB transition in the diagonal mixed state $\rho_d(q,0)$.

Based on (i) the fact that the slopes of the phase boundaries for $\rho_d$ and $\rho$ are the same at $(q, p) = (0, p_c)$, (ii) the co-occurrence of the transition points for $\rho_d$ and $\rho$ at $q = q_c$, and (iii) the absence of a spontaneous symmetry-breaking (SSB) phase in $\rho_d$, we plot the schematic phase boundaries of $\rho$ and $\rho_d$ in Fig.~\ref{Fig:perturb_phase_image}(b) using solid and dashed lines, respectively. We emphasize that the constraints we derived does not rule out the possibility that the dashed and solid lines overlap along finite segments. Furthermore, since $\rho_d$ along the $p = 0$ line remains in the symmetric phase for $q < q_c$, it immediately follows that $\sigma(p = 0, q < q_c, r)$ is also in the symmetric phase (recall that $\sigma(q,p,r) = \mathcal{E}_r[\rho(q,p)]$). If $\sigma(p = 0, q < q_c, r)$ were in an SSB or SW-SSB phase, it would imply that $\rho_d = \sigma(p = 0, q < q_c, r = 0.5)$ is in an SW-SSB phase, which contradicts our earlier discussion. Combined with the fact that $\rho(q > q_c, 0, 0)$ is in an SW-SSB phase, this allows us to plot the exact phase diagram of $\sigma$ in the $p = 0$ plane, as shown in Fig.~\ref{Fig:perturb_phase_image}(c).

An interesting question concerns the location of the phase boundary and the corresponding universality class of the transition between the symmetric and SW-SSB phases when moving away from the $q = 0$ plane. It is plausible that this transition continues to be described by the Nishimori critical point, though we are not aware of any rigorous results. Notably, the entire $r = 0.5$ plane (corresponding to the diagonal mixed state $\rho_d$) can be efficiently simulated using hybrid tensor network and Monte Carlo methods with a bond dimension of $\chi = 4$. Along the $p = 0.5$ and $q = 0$ lines, the bond dimension can be further reduced to $\chi = 2$, which is the primary focus of this work. A comprehensive investigation of the entire $r = 0.5$ plane is left for future work.

\subsection{$p = 0$ line: pure state subjected to maximal Pauli-$X$ dephasing} \label{sec:ground_state_2d}

\begin{figure*}
	\centering	\includegraphics[width=\linewidth]{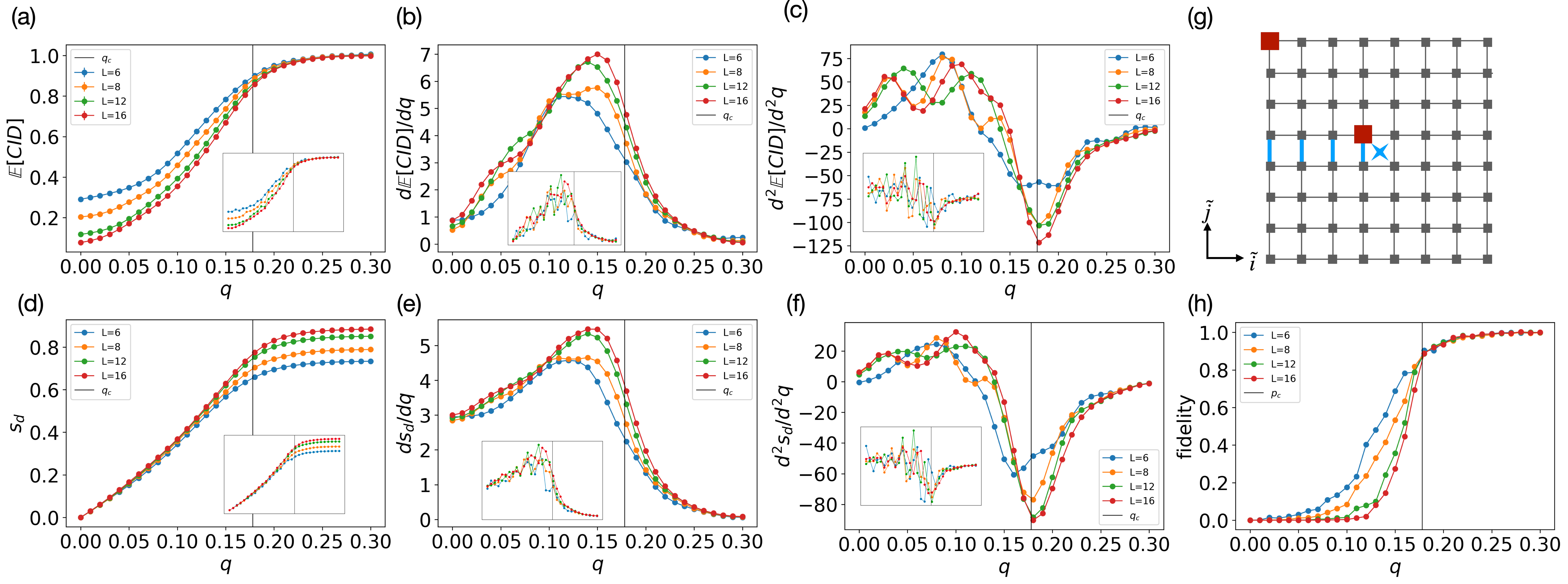}
	\caption{Numerical results for the pure state $\rho(p=0,q)$ (see Fig.~\ref{Fig:perturb_phase_image} for the phase diagram). (a)-(c) show $\mathbb{E}[\text{CID}]$, $d \mathbb{E}[\text{CID}] /dq$, and $d^2 \mathbb{E}[\text{CID}] /d^2 q$ as functions of the tuning parameter $q$, obtained by smoothing the raw data. The insets show the original data. (d)-(f) show the same quantities as (a)-(c) with $\mathbb{E}[\text{CID}]$ replaced by the diagonal entropy density $s_d$.
 (g) depicts the system on the dual lattice of size $L \times L$ with $L = 7$.  Here the Ising vortices reside on plaquettes, and there are $(L+1) \times (L+1)$ spins on the dual lattice. 
 The blue cross indicates an isolated Ising vortex at the center, which can be created by flipping the  bonds labeled in the blue color. 
The red squares indicate the locations $\tilde{i}$ and $\tilde{j}$ when computing the disorder-averaged spin-spin correlation function $[\langle s_{\tilde{i}} s_{\tilde{j}} \rangle ]$.
(h) shows the disorder-averaged free energy cost $[\langle e^{-\Delta F_l /2}\rangle]$ as a function of $q$.
 }
	\label{Fig:rho2_LZ_direct_all}
\end{figure*}

\subsubsection{Universal scaling behavior of diagonal entropy}
Before using Lempel-Ziv to numerically estimate $S_d$, let us first analytically analyze $S_d$ along the $p = 0$ line. We emphasize that, while both the $(1+1)$-D TFIM and the current $(2+0)$-D deformed paramagnetic state can be mapped to the 2D Ising model, their corresponding diagonal entropies are not equivalent. This distinction is already evident in the subtle differences between Eq.~\eqref{Eq:rhod_tfim_corr} and Eq.~\eqref{Eq:rho2_chamon_corr}. Although both probability distributions are mapped to the correlators of the 2D $\phi^4$ theory, Eq.~\eqref{Eq:rhod_tfim_corr} involves correlators solely on the 1D temporal boundary, whereas Eq.~\eqref{Eq:rho2_chamon_corr} involves correlators on the \textit{2D temporal} boundary. We will demonstrate that, due to this subtle difference, the singular part of $S_d$ scales as $(q - q_c)^2 \log(|q - q_c|)$. In other words, \textit{two derivatives} are required to observe the logarithmic divergence, which contrasts with the single derivative required in the $(1+1)$-D TFIM.

Similar to Sec.~\ref{sec:tfim_analytical}, we employ the replica trick to study $S_d$:
\begin{equation}
S_d = \lim_{n \to 1} \frac{\log\left(\sum_{x_{\mathbf{j}}} \rho^{n}_{x_{\mathbf{j}}}\right)}{1-n} = \lim_{n \to 1} \frac{F(2n) - n F(2)}{1-n},
\end{equation}
where $F(2n) \equiv \log\left(\sum_{x_{\mathbf{j}}}\mathcal{Z}^{2n}_{x_{\mathbf{j}}}\right)$ with $\mathcal{Z}^{2n}_{x_{\mathbf{j}}}$ given in Eq.~\eqref{Eq:rho2_chamon_corr}. We note that the corresponding statistical mechanical model for general half-integral $n$ has already been derived in Ref.~\cite{fan2023diagnostics}. Furthermore, it is well known that $F(2)$ is equivalent to the free energy of the 2D Ising model. However, $F(2+2\delta)$ with $\delta \ll 1$ has not been discussed and is crucial for understanding the behavior of the diagonal entropy. We now briefly derive the statistical mechanical model for $F(2n)$, following Ref.~\cite{fan2023diagnostics}. We then use effective field theory to study the behavior of $S_d$ closed to the critical point. 
Since our approach is fairly general and can be applied to the Ising model in any dimension (in other words, it is applicable to studying the diagonal entropy in the Pauli-$X$ basis of the pure state $|\Psi(q)\rangle = \prod_{\langle i,j\rangle}[(1-q)I + q Z_i Z_j] |x_\mathbf{j} = 1\rangle
$ in any dimension), we will first consider the problem in general $D$-dimensions and later restrict to $D = 2$. 

Unlike the $(1+1)$-D TFIM case, the multi-point correlations in Eq.~\eqref{Eq:rho2_chamon_corr} now exist on all sites instead of being confined to a single line. This facilitates a direct calculation in the original model without applying Kramers-Wannier transformation. In particular, using Eq.\eqref{Eq:rho2_chamon_corr} and the property that $z^{(1-x)/2}= x^{(1-z)/2}$ for $x, z = \pm 1$, one finds
\begin{equation}
Z^{2n}_{x_\mathbf{j}} = \sum_{z^{(\alpha)}_\mathbf{j}} (\prod_j x_j^{\sum_{\alpha}(1-z^{(\alpha)}_j)/2}) e^{K \sum_{\langle i,j \rangle} (\sum_{\alpha} z^{(\alpha)}_i z^{(\alpha)}_j)}.
\end{equation}
It follows that the terms surviving after summing over $x_{\mathbf{j}}$ are  
constrained to satisfy $\prod_{\alpha = 1}^{2n} z^{(\alpha)}_j = 1, \forall j$.
Therefore, one can parameterize $z^{2n}_j = \prod_{\alpha = 1}^{2n-1} z^{(\alpha)}_j$ and write
\begin{equation}
\label{Eq:chamon_replica}
\sum_{x_\mathbf{j}} Z^{2n}_{x_\mathbf{j}} = \sum_{z^{(\alpha)}_\mathbf{j}} e^{K \sum_{ \langle i,j\rangle} \big( \sum_{\alpha = 1}^{2n-1} s^{(\alpha)}_i s^{(\alpha)}_j + \prod_{\alpha = 1}^{2n-1} s^{(\alpha)}_i s^{(\alpha)}_j \big)}.
\end{equation}
Eq.\eqref{Eq:chamon_replica} implies $F(2n) = \log(\sum_{x_\mathbf{j}} Z^{2n}_{x_\mathbf{j}})$ is equivalent to  $(2n-1)$ copies of Ising model $H_{\text{Ising}}^{(\alpha)} = -K \sum_{\langle i,j \rangle } s^{(\alpha)}_i s^{(\alpha)}_j$ interacting with one another through $H_{\text{int}} = -K\sum_{\langle i,j \rangle }(\prod_{\alpha = 1}^{2n-1} s^{(\alpha)}_i s^{(\alpha)}_j) $.
The analytical understanding of $S_d$ can then be obtained by studying the effect of $H_{\text{int}}$ on the decoupled Ising model $H_0 = \sum_{\alpha} H_{\text{Ising}}^{(\alpha)}$ near the Ising critical point.
Specifically, by expanding $s^{(\alpha)}_i s^{(\alpha)}_j \sim 1 + \epsilon^{(\alpha)}(r)$, where $\epsilon^{(\alpha)}(r)$ is the energy operator of the $\alpha$-th Ising model, and retaining the most relevant and second most relevant terms, one finds that the effect of $H_{\text{int}}$ is described by the following action:
\begin{equation}
\mathcal{S}_{\text{int}} =  \int d^D r \Big( \lambda_0 \sum_{\alpha} \epsilon^{(\alpha)}(r) +\lambda\sum_{\alpha \neq \beta} \epsilon^{(\alpha)}(r)  \epsilon^{(\beta)}(r) \Big).
\end{equation}
The $\lambda_0$ term is proportional to the local energy density and only shifts the value of $K_c$ without modifying the universal behavior. The renormalization group equation for $\lambda$ has been discussed in Ref.\cite{Cardy_1996_ch8} and takes the form:
\begin{equation}
\label{Eq:beta_function}
\frac{d \lambda}{d l} = \left( \frac{2}{\nu} - D \right) \lambda + \left( 4(2n-3) + 2C^2 \right) \lambda^2 + O(\lambda^3),
\end{equation}
where $\nu$ is the correlation-length exponent and $C$ is the coefficient in the following operator product expansion (OPE):
\begin{equation}
\label{Eq:EE_ope}
\epsilon^{(\alpha)} \epsilon^{(\beta)} \sim \delta_{\alpha, \beta} + C \delta_{\alpha, \beta} \epsilon^{(\alpha)} + \cdots.
\end{equation}
The coefficient linear in $\lambda$ is determined by the scaling dimension of $E$, $\Delta_E = 2/\nu$. On the other hand, the coefficient quadratic in $\lambda$ is determined from the OPE $(\sum_{\alpha \neq \beta} \epsilon^{(\alpha)}(r)  \epsilon^{(\beta)}(r) ) (\sum_{\gamma \neq \delta} \epsilon^{(\gamma)}(r)  \epsilon^{(\delta)}(r)) \sim \left( 4(2n-3) + 2C^2 \right) \sum_{\alpha \neq \beta} \epsilon^{(\alpha)}(r)  \epsilon^{(\beta)}(r)   + \cdots$, which can be obtained using Eq.\eqref{Eq:EE_ope}.

We now restrict to the $D = 2$ situation. Since $\nu = 2$, one finds $(2/\nu) - D = 0$, and thus the $\lambda$ term is a marginal interaction. To determine whether it's marginally relevant or irrelevant, we consider the second-order term in Eq.\eqref{Eq:beta_function}.
Interestingly, the self-duality of the 2D Ising model implies $C = 0$, as Eq.~\eqref{Eq:EE_ope} should be invariant under $\epsilon^{(\alpha)} \rightarrow -\epsilon^{(\alpha)}$. Therefore, when $n = 1 + \delta$, with $\delta \ll 1$, one finds $4(2n-3) + 2C^2 \approx -4 + 8\delta < 0$, implying that the $\lambda$ term is marginally irrelevant. Furthermore, the $\lambda$ term cannot be dangerously irrelevant, as the limit $\lambda \rightarrow 0$ corresponds to the case of decoupled Ising models.
Therefore, we expect the singular part of $F(2+2\delta)$ to exhibit the same scaling form as the singular part of $F(2)$, i.e. 2D Ising model's free energy. This implies $F(2+2\delta) = c_{2+2\delta} (q-q_c)^2 \log(|q-q_c|)$, where $c_{2+2\delta}$ is a non-universal coefficient. Following essentially the same logic as Eq.~\eqref{Eq:sd_replicalimit}, one finds that the singular part of the diagonal entropy density scales as $s^{\infty,\text{sing}}_d \approx c'_2 (q-q_c)^2 |\log(|q-q_c|)|$. 
This implies that \textit{two derivatives} are required to locate the critical point, in contrast to the case of $(1+1)$-D TFIM (Sec.\ref{sec:tfim}) where only one derivative was needed.

\subsubsection{Numerical estimation}

We now numerically study $\mathbb{E}[\text{CID}]$ and $s_d$ for the mixed state obtained by measuring the pure state along the $p = 0$ line in the Pauli-$X$ basis.
To generate  samples with the Born probability distribution, we employ a tensor-network-based approach that computes $\rho_{x_\mathbf{j}}$ using the dual model in Eq.\eqref{Eq:rho2_RBIM} {(see App.\ref{sec:app_numerics} for details)}.

We note that in the dual model, the Ising vortices reside on the plaquettes, and there are $(L+1) \times (L+1)$ spins on the dual lattice.
Moreover, we don't enforce the constraint $\prod_{j} x_j = 1$, as it does not affect the entropy density.
With access to $\rho_{x_\mathbf{j}}$, we generate samples using the Metropolis-Hasting algorithm, similar to Sec.\ref{sec:tfim}. 
Fig.\ref{Fig:perturb_phase_image}(d) shows typical samples for a 17 $\times$ 17 system based on the probability distribution $\rho_{x_\mathbf{j}}(q,0)$ at $q = 0.1$ and $q = 0.25$ (recall the transition point occurs at $q_c \approx 0.178$).
\label{sec:Nishimori}
\begin{figure*}
	\centering	\includegraphics[width=\linewidth]{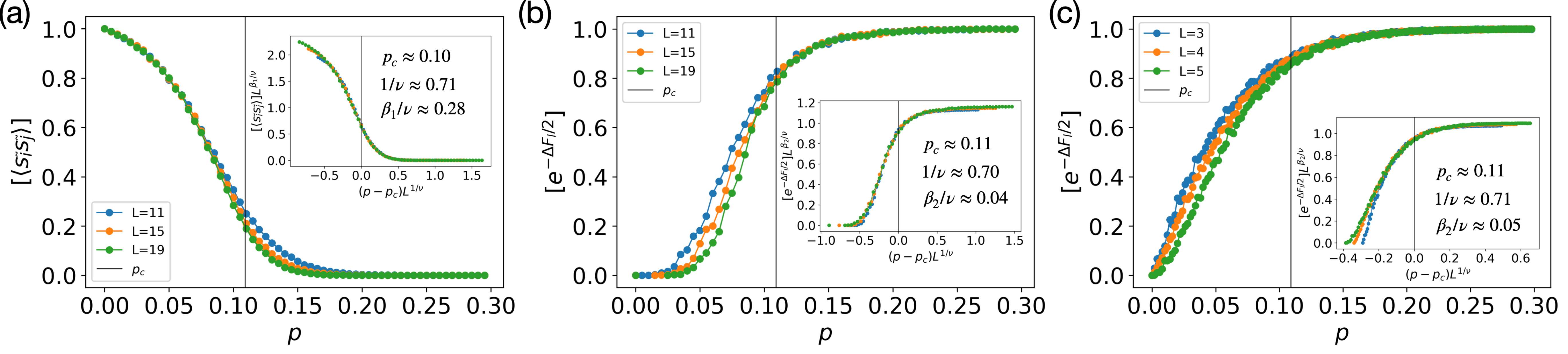}
	\caption{(a) The disorder averaged spin-spin correlation function $ [\langle s_{\tilde{i}} s_{\tilde{j}} \rangle ] $ and (b) the free energy cost $[ \langle e^{-\Delta F_l /2} \rangle ] $ as functions of the tuning parameter $p$ for the state $\rho(p,q=0)$ (see Fig.~\ref{Fig:perturb_phase_image} for the phase diagram).
The vertical line in each figure indicates the estimated critical point $p_c^{*}$, extracted from the data collapse in a window $ u = (p-p^{*}_c)L^{1/\nu} = [-0.4,0.4]$, using autoScale.py \cite{melchert2009autoscale}.
(c) A  figure similar to (b), but for smaller system sizes, which can be obtained using a direct tomographic approach.	}
	\label{Fig:rbim_collapse}
\end{figure*}

\begin{figure*}
	\centering	\includegraphics[width=\linewidth]{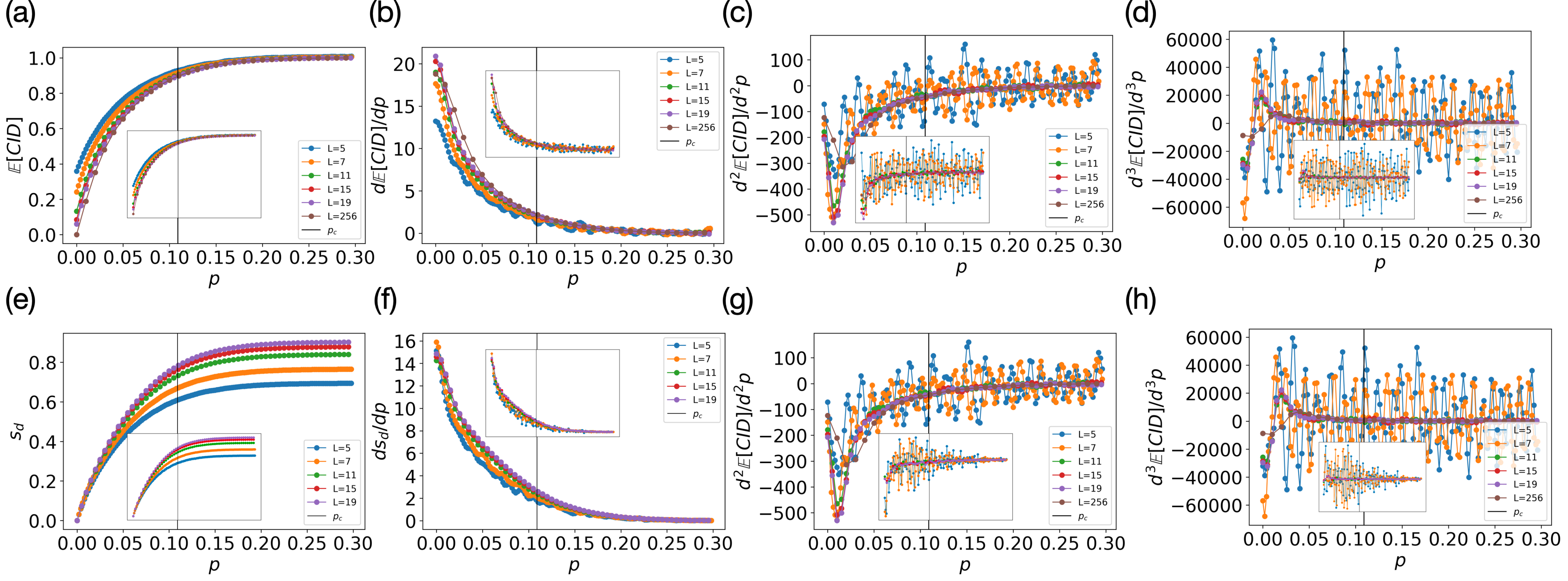}
	\caption{Diagonal entropy/CID and its derivatives for the state $\rho(p,q=0)$ (see Fig.~\ref{Fig:perturb_phase_image} for the phase diagram).	(a)-(d) show $\mathbb{E}[\text{CID}]$, $d \mathbb{E}[\text{CID}] /dp$, $d^2 \mathbb{E}[\text{CID}] /d^2 p$, and $d^3 \mathbb{E}[\text{CID}] /d^3 p$ as functions of the tuning parameter $q$ obtained by smoothing the raw data. The insets show the original data. (e)-(h) show the same quantities as (a)-(d) with $\mathbb{E}[\text{CID}]$ replaced by the diagonal entropy density $s_d$.
	}
	\label{Fig:rho_LZ_direct_all}
\end{figure*}

To confirm our previous analysis that two derivatives are required to observe the logarithmic divergence, we use finite-difference differentiation to compute $\mathbb{E}[\text{CID}]$, $d\mathbb{E}[\text{CID}]/dq$, and $d^2\mathbb{E}[\text{CID}]/dq^2$. The results are shown in Fig.~\ref{Fig:rho2_LZ_direct_all}(a), (b), and (c), respectively.
Since the raw data and its  derivatives are quite noisy, we display them in insets, while the main plots show the results after three iterations of smoothing.
We also compute the diagonal entropy $s_d$ directly using the formula $s_d \approx -\sum_{x_\mathbf{j} \in \text{samples}} \log(\rho_{x_\mathbf{j}})/(L^2 N_s)$ and take its derivatives, where $\log(\rho_{x_\mathbf{j}})$ is calculated using the aforementioned tensor-network-based approach. The results are displayed in Fig.~\ref{Fig:rho2_LZ_direct_all}(d), (e), and (f).
It is evident that the second derivative of both $\mathbb{E}[\text{CID}]$ and $s_d$ shows a peak at $q = q_c \approx 0.178$.
To further support the identification of the peak with the  $\rho_d$ transition, we compute the fidelity $F(\rho_d, Z_i Z_j \rho_d Z_i Z_j)$, which serves as the order parameter  distinguishing SW-SSB phase from the symmetric phase. 
Recall that in the dual model (Eq.\eqref{Eq:rho2_RBIM}), $\sqrt{\rho_{x_\mathbf{j}}} \propto \mathcal{Z}_{x_\mathbf{j} }(q) = \sum_{s_{\tilde{
i}}} e^{\beta \sum_{\langle \tilde{i}, \tilde{j} \rangle} J_{\langle \tilde{i}, \tilde{j} \rangle} s_{\tilde{i}} s_{\tilde{j}} } = \mathcal{Z}_{J_{\langle \tilde{i},\tilde{j} \rangle}}(q),\ \tanh(\beta) = 1-2q$, and $\{ J_{\langle \tilde{i}, \tilde{j} \rangle } \} $ can be any bond configuration satisfying $\prod_{ \langle \tilde{i}, \tilde{j} \rangle  \in j} J_{\langle \tilde{i}, \tilde{j} \rangle }   = x_{j} $.
It turns out that 
$F(\rho_d, Z_i Z_j \rho_d Z_i Z_j)$ is related to the free energy cost of inserting an isolated Ising vortex, weighted by the Renyi-2 probability: {$[ \langle e^{-\Delta F_l } \rangle ] = 
\sum_{J_{\langle \tilde{i}, \tilde{j} \rangle} } \mathcal{Z}^2_{J_{\langle \tilde{i}, \tilde{j} \rangle}} e^{-(\ln \mathcal{Z}_{J_{\langle \tilde{i}, \tilde{j} \rangle}} -\ln \mathcal{Z}_{J_{\langle \tilde{i}, \tilde{j} \rangle,l}})} /\sum_{J_{\langle \tilde{i}, \tilde{j} \rangle}} \mathcal{Z}^2_{J_{\langle \tilde{i}, \tilde{j} \rangle}} 
=  \sum_{J_{\langle \tilde{i}, \tilde{j} \rangle} } \mathcal{Z}_{J_{\langle \tilde{i}, \tilde{j} \rangle}} \mathcal{Z}_{J_{\langle \tilde{i}, \tilde{j} \rangle, l}} /\sum_{J_{\langle \tilde{i}, \tilde{j} \rangle} } \mathcal{Z}^2_{J_{\langle \tilde{i}, \tilde{j} \rangle}}  $}, where $J_{\langle \tilde{i}, \tilde{j} \rangle, l}$ denotes the configuration obtained from creating a single Ising vortex on top of the configuration $J_{\langle \tilde{i}, \tilde{j} \rangle} $ along path $l$ that runs from the location of the Ising vortex to the boundary [see Fig.\ref{Fig:rho2_LZ_direct_all}(g)]. Fig.\ref{Fig:rho2_LZ_direct_all}(h) shows the free energy cost $[\langle e^{-\Delta F_l /2}\rangle]$ as a function of perturbed strength $q$.
Clearly, $\langle e^{-\Delta F_l}\rangle$ vanishes for $q<q_c$ while it is a non-zero constant for $q>q_c$, consistent with the peak in the 2nd order derivatives of both $\mathbb{E}[\text{CID}]$ and $s_d$.

\subsection{$q = 0$ line: SW-SSB transition and Nishimori critical point}

Before attempting to use $\mathbb{E}[\text{CID}]$ and $s_d$ to study the mixed state along the $q = 0$ line, we first discuss some known results about the connection between the mixed state and the RBIM along the Nishimori line.
Recall
$\rho(0,p) \propto \sum_{x_\mathbf{j}\ \text{s.t.} \prod_j x_{j} =1} \mathcal{Z}_{x_\mathbf{j}}(p) |x_\mathbf{j}\rangle \langle x_\mathbf{j}|$ so that $S_d = S \propto \sum_{x_\mathbf{j}\ \text{s.t.} \prod_j x_{j} =1}  \mathcal{Z}_{x_\mathbf{j}} \log(\mathcal{Z}_{x_\mathbf{j}})$, which is the disorder-averaged free energy of the RBIM along the Nishimori line (in the even sector $\prod_j x_{j} =1$). Using symmetry arguments, Refs.\cite{le1988location, le1989varepsilon} showed that the singular part of the free energy density along the Nishimori line in $d = 2$ follows the scaling form $f_{\text{s}}(p,L) = L^{-d}g[ (p-p_c)L^{1/\nu}]$.
The correlation length exponent $\nu$ has been numerically extracted to be $\nu \approx 1/0.655 \approx 
1.5$ by calculating the disorder-averaged susceptibilities \cite{hasenbusch2008multicritical}.
Choosing $(p - p_c)L^{1/\nu} = 1$, one finds \( f_s(p) = (p - p_c)^{\nu d} g(1) \sim (p - p_c)^3 \). 
Therefore, unlike the $p = 0$ line, to observe a jump arising from the singular part of the free energy, one should differentiate $s_d \propto f$ with respect to $p$ three times.
We anticipate that this may present a challenge in detecting the phase transition using the diagonal entropy, as the noise amplified by the three iterative finite differences may be too large, rendering the signal of the phase transition undetectable.

To study either CID and $s_d$, we first generate the configurations according to the probability distribution $\rho_{x_\mathbf{j}} \propto \mathcal{Z}_{x_\mathbf{j}}$.
Unlike the previous examples, these configurations can be generated rather efficiently without using the Metropolis-Hastings algorithm.
In particular, one can first sample bond configurations $J_{\langle \tilde{i}, \tilde{j} \rangle} $ on the dual lattice independently through the probability distribution $P(J_{\langle \tilde{i}, \tilde{j} \rangle} =1) = 1-p $, $P(J_{\langle \tilde{i}, \tilde{j} \rangle} =-1) = p $, and then project them onto the vortex configurations $x_\mathbf{j}$ using $x_j = \prod_{ \langle \tilde{i}, \tilde{j} \rangle  \in j} J_{\langle \tilde{i}, \tilde{j} \rangle }  $.
The probability of the bond configuration $J_{\mathbf{ \langle \tilde{i}, \tilde{j} \rangle }} = \{ J_{ \langle \tilde{i}, \tilde{j} \rangle } \} $ from the above sampling is simply $P({J_{\mathbf{ \langle \tilde{i}, \tilde{j} \rangle }}}) = \prod_{\langle i,j \rangle} (1-p)^{(1+J_{\langle \tilde{i}, \tilde{j} \rangle})/2} p^{(1-J_{\langle \tilde{i}, \tilde{j} \rangle})/2} \propto \prod_{\langle i,j \rangle} [p/(1-p)]^{(1-J_{\langle \tilde{i}, \tilde{j} \rangle})/2} \propto e^{\beta \sum_{\langle \tilde{i}, \tilde{j} \rangle} J_{\langle \tilde{i}, \tilde{j} \rangle}} $, where $e^{2\beta} = p/(1-p)$.
Since different bond configurations $J_{\mathbf{\langle \tilde{i}, \tilde{j} \rangle}} $ with the same Ising vortex configuration $x_\mathbf{j}$ can be obtained via the gauge transformation $J_{{\langle \tilde{i}, \tilde{j} \rangle}} \rightarrow J_{{\langle \tilde{i}, \tilde{j} \rangle}} \sigma_{\tilde{i}} \sigma_{\tilde{j}}, \sigma_{\tilde{i}} = \pm 1$, the probability $\rho_{x_\mathbf{j}}$ can then be written as $\rho_{x_\mathbf{j}} \propto \sum_{\sigma_{\mathbf{\tilde{i}}} } e^{\beta \sum_{\langle \tilde{i}, \tilde{j} \rangle} J_{\langle \tilde{i}, \tilde{j} \rangle} \sigma_{\tilde{i}} \sigma_{\tilde{i}}}  $, which is nothing but the partition function of the RBIM.
We benchmark this approach by  studying the disorder averaged spin-spin correlation function $ [\langle s_{\tilde{i}} s_{\tilde{j}} \rangle ] = \sum_{J_{\langle \mathbf{\tilde{k}, \tilde{l} }\rangle}} (\sum_{s_{\mathbf{\tilde{
k}}}} s_{\tilde{i}} s_{\tilde{j}}  e^{\beta \sum_{\langle \tilde{k}, \tilde{l} \rangle} J_{\langle \tilde{k}, \tilde{l} \rangle} s_{\tilde{k}} s_{\tilde{l}} })/ \sum_{J_{\langle \mathbf{\tilde{k}, \tilde{l}} \rangle}} \mathcal{Z}_{J_{\langle \tilde{k}, \tilde{l} \rangle}}$,  and the free energy cost of a single Ising vortex  $[ \langle e^{-\Delta F_l /2} \rangle ] =  \sum_{J_{\langle \mathbf{\tilde{i}, \tilde{j}} \rangle} } \sqrt{\mathcal{Z}_{J_{\langle \tilde{i}, \tilde{j} \rangle}} \mathcal{Z}_{J_{\langle \tilde{i}, \tilde{j} \rangle, l}} }/\sum_{J_{\langle \mathbf{\tilde{i}, \tilde{j}} \rangle}} \mathcal{Z}_{J_{\langle \tilde{i}, \tilde{j} \rangle}}  $. These calculations also involve the tensor-network-based approach mentioned in Sec.\ref{sec:ground_state_2d}. The results are shown in Fig.\ref{Fig:rbim_collapse}(a) and Fig.\ref{Fig:rbim_collapse}(b), respectively.
Each figure's vertical line marks the estimated critical point $p_c^{}$, obtained from the data collapse (shown in the inset) within a window $u = (p - p_c^{*})L^{1/\nu} = [-0.4, 0.4]$, using autoScale.py \cite{melchert2009autoscale}. 
For both $[\langle s_{\tilde{i}} s_{\tilde{j}} \rangle ]$ and $[ \langle e^{-\Delta F_l /2} \rangle ]$, the location of the critical point and correlation length exponent $\nu$ thus obtained are in reasonable agreement with the previous results \cite{honecker2001universality,cho1997criticality,hasenbusch2008multicritical}.
Note that for the spin-spin correlation function $[\langle s_{\tilde{i}} s_{\tilde{j}} \rangle ]$, we fix $\tilde{i} = (0, L)$ and $\tilde{j} = ([L/2],[L/2])$, where $[x]$ is the floor function that outputs the greatest integer less than or equal to $x$ (see Fig.\ref{Fig:rho2_LZ_direct_all}(g) for the system of $L\times L$ Ising vortices with $L = 7$).

We now turn to our main object of interest: Lempel-Ziv entropy $\mathbb{E}[\text{CID}]$. Fig.\ref{Fig:perturb_phase_image}(d) shows a few typical images $x_\mathbf{j}$ based on the probability distribution $\rho_{x_\mathbf{j}}(0,p) \propto \mathcal{Z}_{x_\mathbf{j}}(p)$ at $p = 0.05$ and $p = 0.15$. To extract quantitative results from these images, we compute $\mathbb{E}[\text{CID}]$ and its corresponding finite-difference derivatives.
The results are shown in Figs.\ref{Fig:rho_LZ_direct_all}(a)-(d), where the main figures show the results after smoothing the data iteratively three times while the insets show the raw data. We do not see a sharp peak close to the critical point in the third derivative of CID.  
Three possible reasons could explain the absence of a peak in the third derivative. The first possibility is that the noise from the finite-difference differentiation, clearly visible in Figs.\ref{Fig:rho_LZ_direct_all}(c),(d),  overwhelms the signatures of a critical point. The second possibility is that even the third derivative of the free energy density does not show any divergence along the Nishimori line. The third possibility is that the Lempel-Ziv algorithm fails to estimate the entropy accurately. To test the third possibility, we next compute $s_d$ and its corresponding derivatives [see Fig.\ref{Fig:rho_LZ_direct_all}(e)-(h)].
We find that the results are similar to the ones for CID, and we are unable to locate the critical point from the third derivatives of $s_d$ either. In Sec.\ref{sec:complexity} we study the difference $|s_d - \mathbb{E}[\text{CID}]|$ in a systematic manner and find that it goes to zero in the thermodynamic limit. Therefore, CID is an excellent approximation for the diagonal entropy density, and the aforementioned third possibility is ruled out.
\begin{figure*}
	\centering	\includegraphics[width=\linewidth]{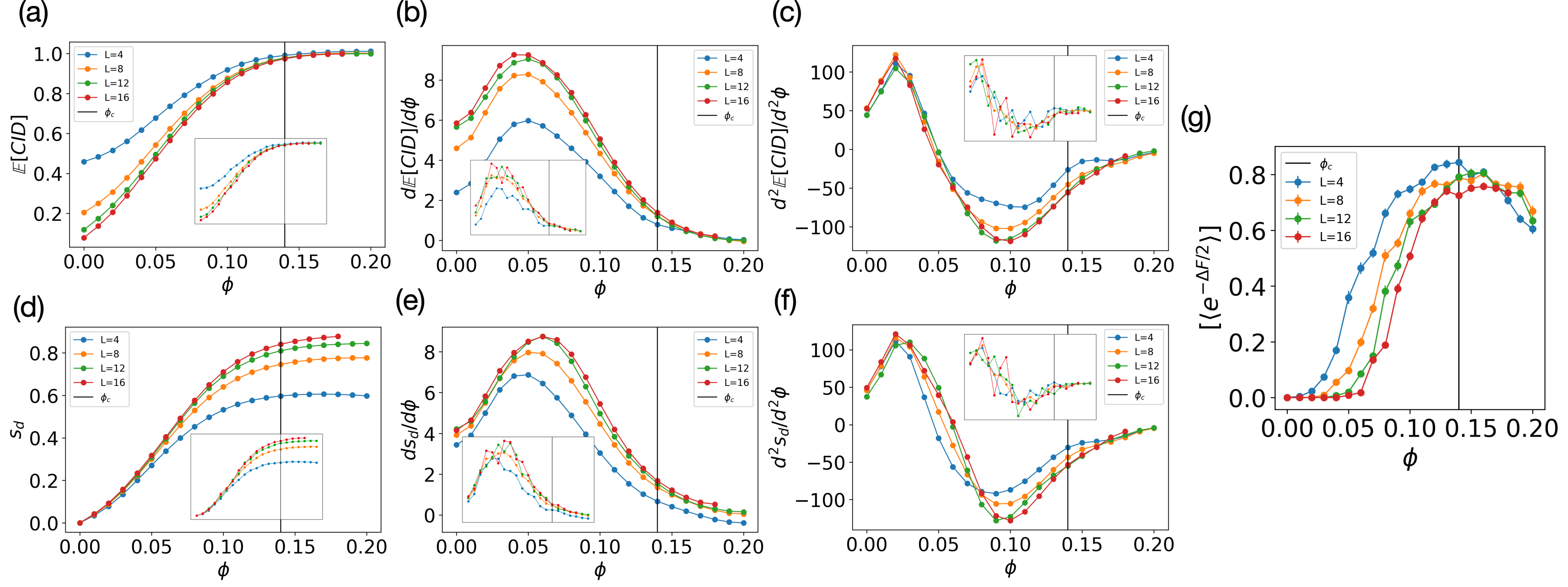}
	\caption{Numerical results for the state $\rho(\phi)$ discussed in Sec.\ref{sec:coherent}. (a)-(c) show $\mathbb{E}[\text{CID}]$, $d \mathbb{E}[\text{CID}] /d \phi$, and $d^2 \mathbb{E}[\text{CID}] /d^2 \phi$, as functions of the tuning parameter $\phi$ obtained by smoothing the raw data. The insets show the original data. (d)-(f) show same quantities as (a)-(d) with $\mathbb{E}[\text{CID}]$ replaced by the diagonal entropy density $s_d$.
 (g) shows the disorder-averaged free energy cost $[\langle e^{-\Delta F_l /2}\rangle]$ as a function of $\phi$.
	}
\label{Fig:coherent_LZ_direct_smooth}
\end{figure*}

Nonetheless, one difference we observe between $s_d$ and $\mathbb{E}[\text{CID}]$  is that fluctuations in the derivatives of $s_d$ seem stronger when $p<p_c$, especially for small system sizes, which may be due to the singular behavior in the \textit{subleading} terms of the free energy (which we know are singular). Besides, as mentioned earlier, the probability distribution $\rho_{x_\mathbf{j}} \propto \mathcal{Z}_{x_\mathbf{j}}$ can be generated efficiently without directly computing $\mathcal{Z}_{x_\mathbf{j}}$. 
This allows us to study CID for very large system sizes.
In Fig.\ref{Fig:rho_LZ_direct_all}
(a)-(d), we also compute $\mathbb{E}[\text{CID}]$ and its derivatives for $L = 256$, and the results are similar to those for smaller system sizes.
We conclude that the absence of a peak in the third derivative of either $s_d$ and $\mathbb{E}[\text{CID}]$ is not due to finite-size effects but may result from extremely large noise caused by finite-difference differentiation or the absence of divergence in the third derivative of the free energy density.

The above discussion indicates that without prior knowledge of the corresponding statistical mechanics model, a scalable approach to locate the SW-SSB transition (or its dual --- the decoding transition in a topological error correcting code) is probably rather challenging. Nonetheless, we know that at least the sub-volume terms in the diagonal entropy are sensitive to this transition~\cite{dennis2002} (as shown in Figs.\ref{Fig:rbim_collapse}(a),(b)), even though they are not accessible via the Lempel-Ziv algorithm. To gauge the difficulty of using a direct tomographic approach to locate the transition, we estimate the probability distribution of measurement outcomes through $\rho_{x_\mathbf{j}} = N_{x_\mathbf
j}/N_s$, where $N_s$ is the number of samples and $N_{x_\mathbf{j}}$ is the number of times each outcome $x_\mathbf{j}$ appears.
The free energy cost associated with a domain wall, $[ \langle e^{-\Delta F_l /2} \rangle ]  = \sum_{x_\mathbf{j}}\sqrt{\rho_{x_\mathbf{j}} \rho_{x_{\mathbf{j},l}}}$, can then be obtained through direct computation.
Using this approach, we study the free energy cost for $L = 3,4,5$, and find that for $L = 4$, around $2^{22} \approx 4.2\times 10^6$ samples are required for convergence.
Fig.\ref{Fig:rbim_collapse}(c) shows $[ \langle e^{-\Delta F_l /2} \rangle ]$ as a function of $p$ for $L= 3,4$, and $5$.
The vertical line indicates the estimated critical point $p_c^{*}$, obtained from the data collapse (shown in the inset)  using autoScale.py \cite{melchert2009autoscale}.

\section{Paramagnetic state under local coherent error} \label{sec:coherent}

In this section, we briefly consider another example of a decoherence-induced transition whose universality class is expected to be different from that of the examples discussed earlier \cite{venn2023coherent}.
The mixed state \(\rho(\phi)\) as a function of the tuning parameter \(\phi\) is obtained by subjecting the fixed-point paramagnet \(|x_\mathbf{j} = 1\rangle \langle x_\mathbf{j} = 1|\) to the strongly symmetric finite-depth channel \(\mathcal{E}_d[U(\phi) (\cdot) U^\dagger(\phi)]\), where \(U(\phi) = e^{i\phi \sum_{\langle i,j \rangle } Z_i Z_j}\).
One can interpret \(U(\phi)\) as a unitary operator that creates coherent errors in the fixed-point paramagnet.
Such an error can be completely reversed by applying the inverse unitary operator \(U^\dagger(\phi)\).
However, if one is unaware of the noise and directly measures the system in the Pauli-$X$ basis, the observer will obtain the mixed state \(\rho(\phi)\).
The gauged version of this model was studied in Ref.\cite{venn2023coherent}, where it was  mapped to the (2+0)-D RBIM with complex couplings along the Nishimori line.
Ref. \cite{venn2023coherent} also numerically demonstrated that a phase transition occurs at \(\phi = \phi_c \approx 0.14\pi\).
Their results, when translated to the ungauged model of our interest, imply the existence of two phases that can be distinguished by the behavior of the fidelity $F(\rho, Z_i Z_j \rho Z_i Z_j)$.
When $\phi < \phi_c$, the fidelity decays exponentially as a function of $|i-j|$.
On the other hand, $F(\rho, Z_i Z_j \rho Z_i Z_j)$ decays \textit{polynomially} as $|i-j|$ when $\phi >\phi_c$.
We note that this differs from the SW-SSB state mentioned in Secs.~\ref{sec:tfim} and \ref{sec:chamon} where $\lim_{|i-j| \rightarrow \infty } F(\rho, Z_i Z_j \rho Z_i Z_j)$ saturates to a finite constant. Therefore, the mixed state $\rho(\phi)$ in the regime $\phi >\phi_c$ can be regarded as possessing an algebraic/quasi-long-range SW-SSB order.

Since $\rho(\phi)$ is diagonal in the Pauli-$X$ basis for all $\phi$, its diagonal entropy in the Pauli-$X$ basis equals its von Neumann entropy.
The spectrum can then be expressed as $\rho_{x_\mathbf{j}} = |\langle x_{\mathbf{j}} |U(\phi)|x_\mathbf{j} = 1\rangle |^2$.
Inserting a complete Pauli-$Z$ basis $I = \sum_{z_\mathbf{j}} |z_\mathbf{j} \rangle \langle z_\mathbf{j}|$ between $U(\phi)$ and $|x_\mathbf{j}\rangle$, one finds
\begin{align}
    \rho_{x_\mathbf{j}} & = |\sum_{z_\mathbf{j}}  \prod_j z_{\mathbf{j}}^{(1-x_j)/2} e^{i \phi \sum_{\langle i,j\rangle } z_i z_j}|^2 \\
    \label{Eq:complex_RBIM}
    & \propto |\sum_{s_{\tilde{
i}}} e^{\beta \sum_{\langle \tilde{i}, \tilde{j} \rangle} J_{\langle \tilde{i}, \tilde{j} \rangle} s_{\tilde{i}} s_{\tilde{j}} }|^2 = |\mathcal{Z}_{x_\mathbf{j}}(\phi)|^2,
\end{align}
where $e^{2 \beta} = i \tan (\phi)$ and $\{ J_{\langle \tilde{i}, \tilde{j} \rangle } \} $ can be any bond configuration satisfying $\prod_{ \langle \tilde{i}, \tilde{j} \rangle  \in j} J_{\langle \tilde{i}, \tilde{j} \rangle }  = x_{j} $.
Similar to Eq.\eqref{Eq:rho2_RBIM}, Eq.\eqref{Eq:complex_RBIM} can be derived from the standard Kramers-Wannier duality.

Figs.~\ref{Fig:coherent_LZ_direct_smooth}(a), (b), and (c) display $\mathbb{E}[\text{CID}]$, $d\mathbb{E}[\text{CID}]/d\phi$, and $d^2 \mathbb{E}[\text{CID}]/d^2 \phi$ respectively (after locally averaging the data as previously described). 
The same quantities estimated by the direct evaluation $s_d \approx -  \sum_{x_\mathbf{j} \in \text{samples}} \log( \rho_{x_\mathbf{j}})/(L^2 N_s)$ are shown in Fig.\ref{Fig:coherent_LZ_direct_smooth}(d), (e), and (f). Similar to Sec.\ref{sec:ground_state_2d}, the generation of samples as well as the calculation of $s_d$ are done using a tensor-network-based approach.

In both calculations, we find a peak in the 2nd derivative of $s_d$ at $\phi \approx 0.09$.
We also compute the free energy cost of a single Ising vortex $[ \langle e^{-\Delta F_l /2} \rangle ] =  \sum_{J_{\langle \mathbf{\tilde{i}, \tilde{j}} \rangle} } \sqrt{\mathcal{Z}_{J_{\langle \tilde{i}, \tilde{j} \rangle}} \mathcal{Z}_{J_{\langle \tilde{i}, \tilde{j} \rangle, l}} }/\sum_{J_{\langle \mathbf{\tilde{i}, \tilde{j}} \rangle}} \mathcal{Z}_{J_{\langle \tilde{i}, \tilde{j} \rangle}}  $, which is the same quantity studied in Sec.\ref{sec:chamon}.
We find that $[ \langle e^{-\Delta F_l /2} \rangle ]$  becomes close to zero close to $\phi \approx 0.09$, consistent with the peak estimated from the second derivative of $s_d$.
The location of the critical point $\approx 0.09$ we found is lower than the one estimated in Ref.\cite{venn2023coherent} ($\phi_c \approx 0.14$).
We do not know the origin of this discrepancy. One possibility is the limited system sizes accessible to us. The main challenge in simulating larger systems is faithfully generating images with the correct Born probability distribution; estimating entropy using Lempel-Ziv is straightforward once such images are available. Nonetheless, similar to the previous examples, the entropy density estimated using Lempel-Ziv approaches the one calculated directly, as shown more systematically in the next section.

\section{{Complexity of estimating diagonal entropy}} 
\label{sec:complexity}

\begin{figure}
	\centering
\includegraphics[width=0.9\linewidth]{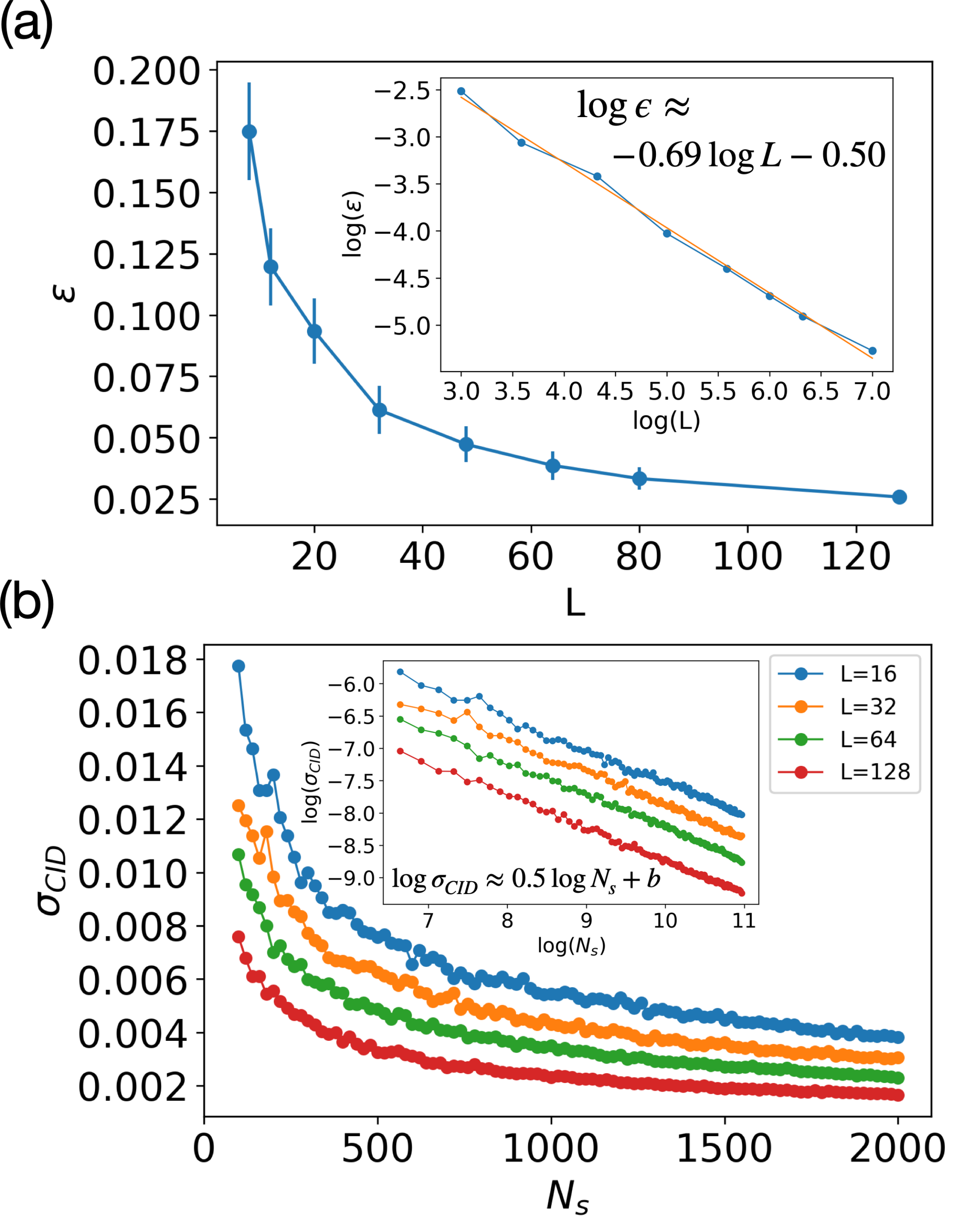}
	\caption{
(a) $\epsilon = |\mathbb{E}[\text{CID}](L,N_s) - s_d|$ as a function of the system size $L$ with $N_s = 50 + 5L$.
 (b) Standard deviation for $\mathbb{E}[\text{CID}](L,N_s)$ due to finite number of samples, $\sigma_{\text{CID}}$, as a function of the number of sample $N_s$ for different system sizes $L$. Both plots are for 1+1-D TFIM at $J = 0.6$.   Both plots are for 1+1-D TFIM at $J = 0.6$.
 }
\label{Fig:error_tfim_rigorous}
\end{figure}

\begin{figure}
	\centering
\includegraphics[width=1.0\linewidth]{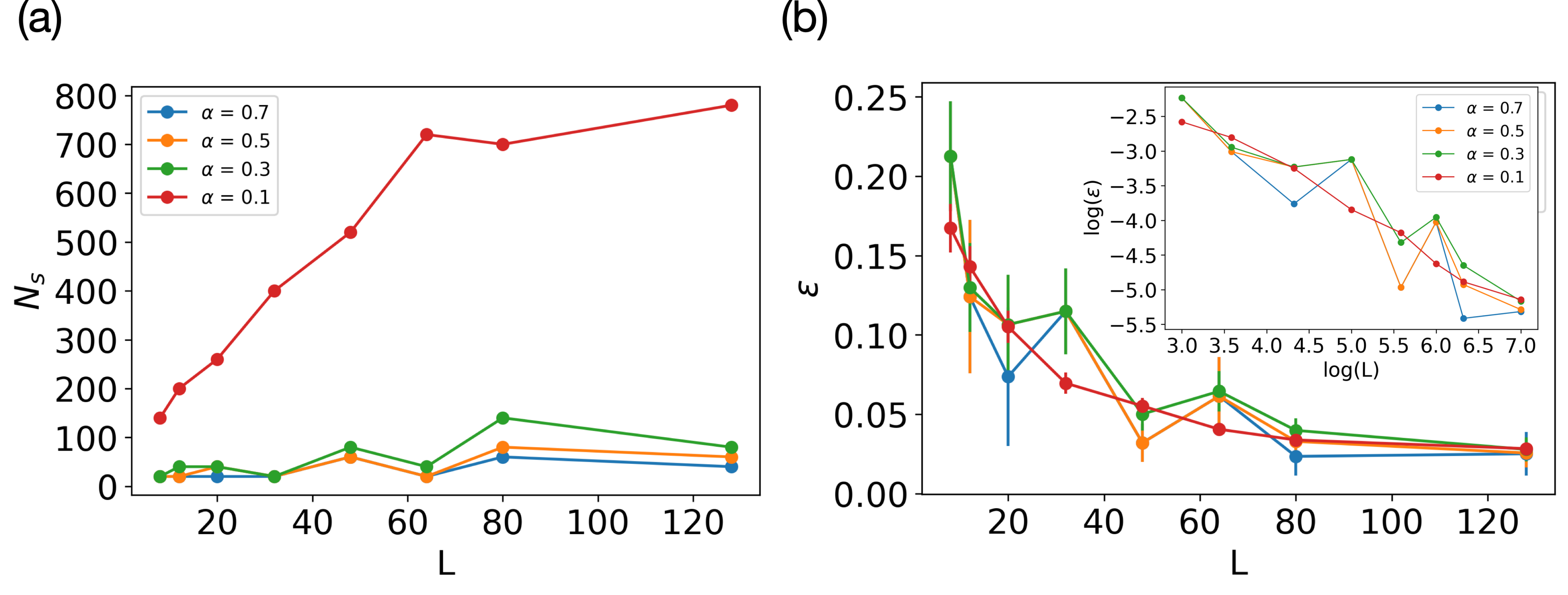}
	\caption{ (a) $N_s^{\alpha}$, the minimum number of samples such that $\sigma_{\text{CID}}(L, N_s^{\alpha}) \leq \alpha \cdot \epsilon(L, N_s^{\alpha})$, as a function of $L$ for different values of $\alpha$.
(b) $\epsilon^{\alpha}(L) = |\mathbb{E}[\text{CID}](L, N^{\alpha}_s(L)) - s_d(L)| $ as a function of the system size $L$. Both plots are for 1+1-D TFIM at $J = 0.6$.
 }
\label{Fig:error_tfim_heuristic}
\end{figure}

\begin{figure}
	\centering
\includegraphics[width=\linewidth]{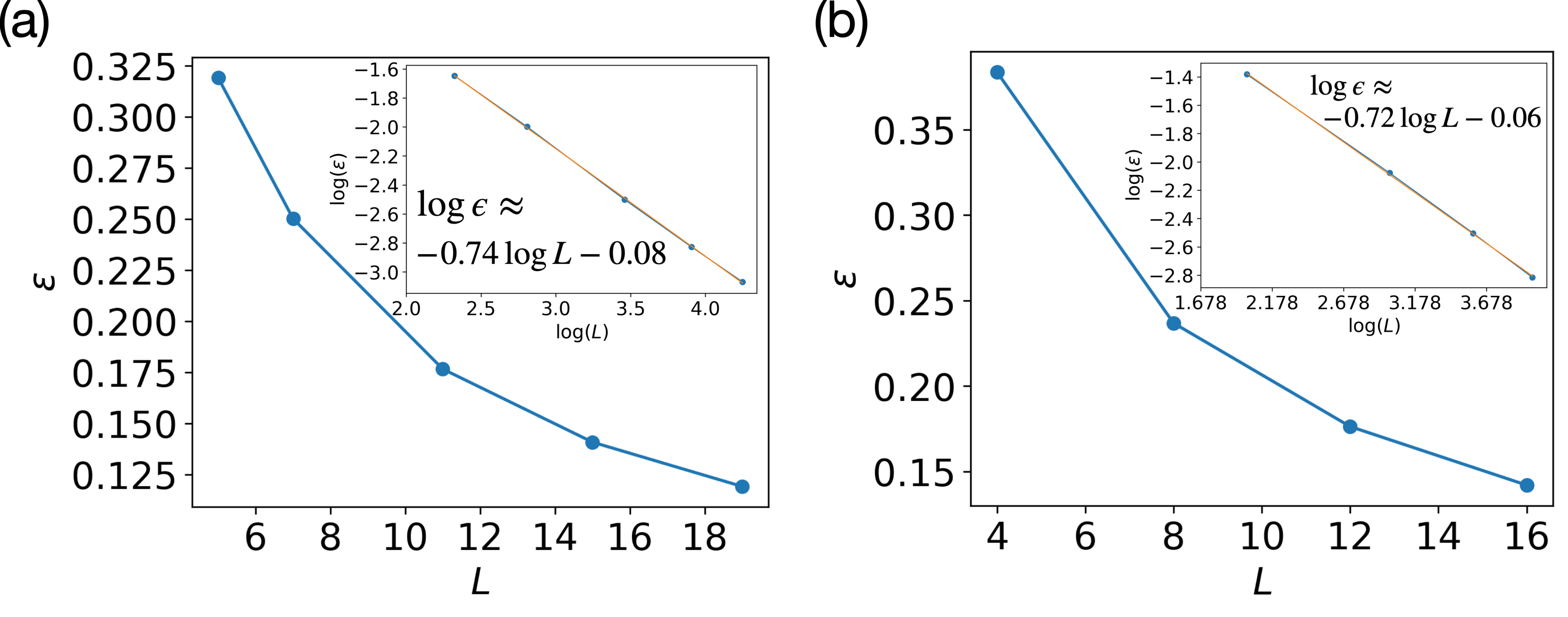}
	\caption{ 
Plots analogous to Fig.~\ref{Fig:error_tfim_rigorous}(a) for the real and the complex RBIM along the Nishimori line with $p = 0.1$ and $\phi = 0.1 \pi$, respectively. Here we fix $N_s = 2000$.
 }
\label{Fig:error_RBIM_real_complex}
\end{figure}

We now analyze the complexity of using Lempel-Ziv compression scheme to estimate the diagonal entropy density as a function of the system size $L$ and the number of samples $N_s$. We  primarily focus on the 1+1-D example from Sec.\ref{sec:tfim} with $J = 0.6$ (on a classical computer, generating samples with the correct probability is the limiting factor, and among the examples we studied, it is easiest to generate the samples in the 1+1-D example). We aim to understand how the difference $\epsilon(L, N_s) = |\mathbb{E}[\text{CID}](L, N_s) - s_d(L)| $ behaves, where $\mathbb{E}[\text{CID}](L, N_s)$ denotes the value of CID for a system of linear size $L$, averaged over $N_s$ samples. We will numerically substantiate the following claim: there exists a function $N_s(L)$ which scales at most polynomially with $L$ such that $\epsilon(L, N_s(L))$ decays to zero as $1/L^{a}$ where $a$ is some positive number.

Fig.\ref{Fig:error_tfim_rigorous}(a) shows $\epsilon(L, N_s = a L + b)$, where $a, b$ are some numbers. We find that $\epsilon(L, N_s = a L + b) \approx 1/L^{0.7}$. This is strongly suggestive that one only needs at most polynomially many samples so that  $\epsilon(L, N_s)$ decays as 1/poly($L$). A natural question is whether one can further optimize $N_s(L)$, as suggested by studies on classical many-body systems \cite{levine_rutgers_talk}. To explore the possibility of a function $N_s(L)$ that grows even slower than poly($L$), we first note that there are two sources of deviations between $\mathbb{E}[\text{CID}](L, N_s)$ and $s_d$. The first source is finite $N_s$ at a fixed $L$, and the second source is finite $L$ (in the following discussion, we ignore the Monte Carlo standard deviation associated with $s_d$, since it's much smaller than either of these two sources of deviations). Let's first discuss the behavior of the finite sampling error. Fig.\ref{Fig:error_tfim_rigorous}(b) shows the standard deviation $\sigma_{\text{CID}}(L, N_s)$ as a function of $N_s$ for different $L$. We find that for all system sizes, $\sigma_{\text{CID}}(L, N_s) \approx \sigma_0(L)/\sqrt{N_s}$, which is the standard behavior for a sampling error. The function $\sigma_0(L)$ is a monotonically decreasing function of $L$. When $N_s$ is large enough, the main source of deviation is finite $L$. This motivates the following question: how many samples are needed at a fixed $L$, such that $\sigma_{\text{CID}}(L, N_s) \leq \alpha \epsilon(L, N_s)$, where $\alpha$ is some positive number. We denote the function that saturates this inequality as $N^{\alpha}_s(L)$. When the number of samples is bigger than $N^{\alpha}_s(L)$, the limiting factor is finite $L$.
Fig.\ref{Fig:error_tfim_heuristic}(a) shows $N^{\alpha}_s(L)$ for various $\alpha$ as a function of $L$, and Fig. shows  $\epsilon^{\alpha}(L) = |\mathbb{E}[\text{CID}](L, N^{\alpha}_s(L)) - s_d(L)| $. Although the plots thus obtained are more jagged than the ones for $N_s \sim  L $,  they are suggestive that $\epsilon^{\alpha}(L) = |\mathbb{E}[\text{CID}](L, N^{\alpha}_s(L)) - s_d(L)| $ decays to zero with a similar slope, while the asymptotic behavior of $N^{\alpha}_s(L) $ seems to grow slower than poly($L$) when $\alpha \gtrsim 0.3$. These behaviors clearly illustrate the advantage of \(\mathbb{E}[\text{CID}]\) over a tomographic approach to estimate entropy:  in tomography, the number of samples required to achieve the same error  increases \textit{exponentially} as the system size increases.

The aforementioned scaling behavior also underlines why extracting \textit{subleading} terms in the diagonal entropy is not feasible using the image compression method. For example, if $S_d = s^{\infty}_d L + \log(2) + O(1/L)$, then our numerical results imply that the estimated diagonal entropy equals $L \mathbb{E}[\text{CID}] = s^{\infty}_d L + O(L^{\alpha})$ where $\alpha \approx 0.3$. It is not obvious that the exponent $\alpha$ is universal, and one might be able to improve upon it by using alternative variants of the compression scheme, as briefly discussed in Sec.\ref{sec:discuss}.

In addition to the 1+1-D TFIM, we also studied the scaling behavior of $\epsilon(L, N_s)$ for the following two states (a) the state $\rho(p,q=0)$ discussed in Sec.~\ref{sec:Nishimori}, whose diagonal entropy corresponds to the disorder-averaged free energy of the 2D RBIM along the Nishimori line with real couplings. (b) the state $\rho(\phi)$ discussed in Sec.\ref{sec:coherent}, whose diagonal entropy corresponds to the free energy of the 2D RBIM along the Nishimori line with complex couplings (in the latter case, we have access to images for system sizes only up to $L \approx 16$). For simplicity, we choose $N_s = 2000$, so that $\sigma_{\text{CID}}(L, N_s) \ll \alpha \epsilon(L, N_s)$, and the limiting factor is the finite system size (as an aside, the plot for $\epsilon(L, N_s = 2000)$ in the 1+1-D TFIM case is essentially identical to the one shown in Fig.\ref{Fig:error_tfim_rigorous}(a) for $\epsilon(L, N_s \sim L)$). We find that for both the real and the complex Nishimori cases, $\epsilon(L, N_s = 2000)$ decays polynomially with $L$ at the system sizes accessible to us, see Figs.~\ref{Fig:error_RBIM_real_complex}(a), (b).

The above results also have implications for estimating the derivatives of $s_d$. Let us consider estimating $ds_d/dJ$ using a finite-difference scheme where $J$ is a tuning parameter. Approximating $ds_d/dJ$ at $J = J_0$ as

\be 
\frac{ds_d}{dJ}\Big|_{J=J_0} \approx \frac{s_d(J+\Delta J)-s_d(J-\Delta J)}{2\Delta J},
\ee 
there are two sources of errors: (a) Error due to inaccurate estimation of $s_d$ --- this scales as $\epsilon(L)/\Delta J$, where we have assumed that the number of samples $N_s$ is large enough so that the limiting factor is finite system size. (b) Truncation error due to non-zero $\Delta J$ --- this scales as $|s^{(3)}| (\Delta J)^2$ where $s^{(3)} = d^3s_d/dJ^3|_{J=J_0}$. The total error then scales as $\epsilon(L)/\Delta J + |s^{(3)}| (\Delta J)^2$. This analysis implies that the optimal value of $\Delta(J) \sim (\epsilon(L)/|s^{(3)}|)^{1/3}$. Choosing this optimal value, the error itself scales as $\epsilon(L)^{2/3} |s^{(3)}|^{1/3}$. Thus, as long as $s^{(3)}$ does not diverge with the system size $L$, derivative estimates can be made arbitrarily precise by considering larger systems. However, in the vicinity of a critical point, $s^{(3)}$ typically does diverge, potentially limiting the accuracy of $ds_d/dJ$. A similar analysis can be carried out for higher derivatives. A systematic numerical study using this optimal approach is left for future work.

\section{Discussion} \label{sec:discuss}
In this paper we explored Lempel-Ziv's lossless compression algorithm as a scalable scheme to estimate the Shannon entropy density of the probability distribution corresponding to the outcomes of projective measurements on a quantum state (``diagonal entropy''). We verified the validity of this scheme for several problems, some inspired by recent ideas in the phases of open quantum systems, such as strong-to-weak symmetry breaking. We also developed a renormalization group and replica-based approach to diagonal entropy for certain problems where we applied our scheme. Partly motivated from our exploration of the diagonal entropy, we also studied the phase diagram of a 2+1-D quantum paramagnet subjected to decoherence, and which exhibits a rich phase diagram consisting of strong-to-weak symmetry breaking as well as standard paramagnet to ferromagnet transition (Fig.\ref{Fig:perturb_phase_image}).

A basic question is: when does the compression algorithm lead to the asymptotically correct value of the diagonal entropy density $s_d$? Results in Refs.\cite{Wyner1989asymptotic,Nobel1992recurrence,ornstein2002entropy,wyner1991fixed,wyner2002sliding,ornstein1990universal} show that the LZ compression scheme correctly estimates the entropy density for a stationary, ergodic source. In colloquial terms, stationarity implies that the samples are statistically translationally invariant, while ergodicity implies that the sample average of any function equals the space average in a fixed sample. Notably, these results do not assume that the source is Markovian. They also provide intuition to construct examples where LZ scheme fails to estimate the entropy density. For instance, let's consider a mixed state of  $N$ qubits that violates ergodicity: $\rho = \frac{1}{2}\left(|x_A\rangle \langle x_A| + |x_B\rangle \langle x_B|\right)$, where $x_A$ and $x_B$ are two specific bit-strings in the Pauli-$X$ basis that are generated as follows: on a given site $i$, $x_A(i) = 1$ with probability $p$, and $-1$ with probability $1-p$. Similarly,  $x_B(i) = 1$ with probability $p$, and $-1$ with probability $1-p$. A projective measurement on $\rho$ in the Pauli-$X$ basis will always result in one of these two images ($x_A$ or $x_B$) and the Lempel-Ziv compression on either of these images will result in a CID that equals $-p \log(p) - (1-p) \log(1-p)$. However, the diagonal entropy of this state is $\log(2)$, and therefore, the compression scheme fails to obtain the correct diagonal entropy. 

There is a slightly different, physics-based intuition that provides guidance on when LZ compression scheme works: in all our examples in Secs. III-VI, a measurement outcome \( x_{\mathbf{j}} = (x_1, \dots, x_N) \) is closely tied with its Born probability \( \rho_{x_{\mathbf{j}}} \). Specifically, \( \rho_{x_{\mathbf{j}}} \) is proportional to the multi-point correlator with respect to a local action. If the measurement outcome itself is uncorrelated with the Born probability, then we expect that the image compression will fail to provide an estimate for the corresponding Shannon entropy.
 In contrast, in the counterexample provided in the last paragraph, the Born probability of the images $x_A$ and $x_B$ $(= 1/2)$ is completely unrelated to the structure of bits encoded in the image. We conjecture that if the Born probability distribution, $\langle x_\mathbf{j}|\rho|x_\mathbf{j}\rangle$, corresponding to the measurement outcomes $x_\mathbf{j}$ can be related to the correlators of a local Hamiltonian (see Eq.~\eqref{Eq:boundary_continuum}), then the computable information density (CID) obtained via Lempel-Ziv compression scheme will be a faithful approximation of the Shannon entropy density. Note that the set of Gibbs states of local Hamiltonians is a proper subset of such states --- for example, the density matrix corresponding to decohered toric code/paramagnet is \textit{not} a Gibbs state of a local Hamiltonian, but nonetheless, the corresponding probability distribution can be related to the correlator of a local theory (RBIM along the Nishimori line), and correspondingly, the compression scheme works in this case (see Fig.\ref{Fig:error_RBIM_real_complex}). It is notable that in the example considered in Sec.\ref{sec:coherent}, the action, though local, is not even real. It would be worthwhile to seek counterexamples to this conjecture to sharpen it further.

For testing our scheme, we picked examples where we were able to calculate the diagonal entropy by other means, either due to exact solvability of the model (e.g. 1+1-D TFIM) or by using tensor network methods (e.g. 2+1-D paramagnet subjected to local decoherence). This is primarily because our goal was to test the algorithm, and also because we do not have access to a quantum machine that can generate images of a many-body system with Born probability. Relatedly, for 2+1-D systems, we were limited by system sizes for which images can be generated with the correct Born distribution using classical methods. Much more interesting are examples where the images corresponding to the measurement outcomes can be generated on quantum emulators such as cold atomic systems (or for that matter, on a quantum computer), but \textit{cannot} be generated on a classical computer, e.g., due to the presence of quantum Monte Carlo sign problem \cite{troyer2005computational,pan2022sign} or due to ergodicity issues even in the absence of sign problem \cite{hastings2013obstructions,hastings2021power,gilyen2021sub}. However, in the absence of another scalable algorithm for estimating $s_d$ in these systems, it might be difficult to test whether an image compression algorithm’s result truly converges to the correct $s_d$ in the thermodynamic limit.

It is interesting to speculate about a potential relation between the singularity of the diagonal entropy and the ability to perform error correction using measurement outcomes. 
For example, as discussed in Ref.\cite{lavasani2024stability}, one can perform error correction in the 1+1-D TFIM using measurement outcomes of operators $Z_i Z_{i+1}$ (the stabilizers of the repetition code), right upto the critical point (see also Ref.\cite{sang2024approximate} for related ideas). Could it be that the non-singular nature of the diagonal entropy associated with Pauli-$X$ measurements for $J < J_c = 1/2$ is an indication that such a recovery should be possible? (note that from the low-energy perspective the operators $X_i$ and $Z_i Z_{i+1}$ are indistinguishable as they are both Ising symmetric).

Another aspect worth closer investigation is the nature of singularity of the free energy in the RBIM along the Nishimori line. Although it is well established that the subleading terms in the free energy (such as domain wall free energy) are singular across the Nishimori critical point, it is not clear that the free energy density (i.e. the coefficient of the volume-law term in the free energy) itself is singular. A striking and unusual feature of this problem is that the expectation value of any local operator or correlation functions of local operators are non-singular across the critical point. Although we were able to accurately estimate $s_d$, which showed no singular behavior in the vicinity of the transition, we found estimating its derivatives with respect to the tuning parameter rather challenging. It might be interesting to calculate the derivatives of $s_d$ using tensor networks assisted by automatic differentiation methods \cite{liao2019differentiable}.

Another notable aspect is the replica calculation of the diagonal entropy. In the examples we studied, the couplings between various replicas were irrelevant in the replica limit, and therefore, it is reasonable to assume that the replica limit gives the correct answer for the diagonal entropy. However, this need not be the case for other problems. For example, in the 3+1-D version of the decohered paramagnet (Eq.\ref{eq:cc_state}) subjected to maximal dephasing, the coupling between replicas is relevant (Eq.\ref{Eq:beta_function} with $D = 3$, and $\nu \approx 0.62$). It is not obvious if this implies that the diagonal entropy is determined by a new, non-Ising fixed point, or whether it signals a potential issue with the replica limit.

Finally, our implementation of the Lempel-Ziv algorithm employed a raster scan, similar to Ref.\cite{stefano2019quantifying}. This method is best suited for 1+1-D systems. Nonetheless, it was adequate for our purposes, as demonstrated by the monotonic, power-law decay of \(\epsilon(L) = |\mathbb{E}[\text{CID}](L) - s_d(L)|\) across all studied systems, including those in 2+1-D. As mentioned in Ref.\cite{levine_rutgers_talk}, in higher dimensions a ``pattern-matching'' scheme to implement Lempel-Ziv may work better than a raster scan. It will be interesting to implement such a scheme to estimate diagonal entropy of quantum systems in two or higher dimensions.

\acknowledgments We thank Ania Jayich, David Weld and John McGreevy for  discussions, and Sagar Vijay and Yi-Zhuang You for feedback. 


\begin{thebibliography}{118}%
\makeatletter
\providecommand \@ifxundefined [1]{%
 \@ifx{#1\undefined}
}%
\providecommand \@ifnum [1]{%
 \ifnum #1\expandafter \@firstoftwo
 \else \expandafter \@secondoftwo
 \fi
}%
\providecommand \@ifx [1]{%
 \ifx #1\expandafter \@firstoftwo
 \else \expandafter \@secondoftwo
 \fi
}%
\providecommand \natexlab [1]{#1}%
\providecommand \enquote  [1]{``#1''}%
\providecommand \bibnamefont  [1]{#1}%
\providecommand \bibfnamefont [1]{#1}%
\providecommand \citenamefont [1]{#1}%
\providecommand \href@noop [0]{\@secondoftwo}%
\providecommand \href [0]{\begingroup \@sanitize@url \@href}%
\providecommand \@href[1]{\@@startlink{#1}\@@href}%
\providecommand \@@href[1]{\endgroup#1\@@endlink}%
\providecommand \@sanitize@url [0]{\catcode `\\12\catcode `\$12\catcode
  `\&12\catcode `\#12\catcode `\^12\catcode `\_12\catcode `\%12\relax}%
\providecommand \@@startlink[1]{}%
\providecommand \@@endlink[0]{}%
\providecommand \url  [0]{\begingroup\@sanitize@url \@url }%
\providecommand \@url [1]{\endgroup\@href {#1}{\urlprefix }}%
\providecommand \urlprefix  [0]{URL }%
\providecommand \Eprint [0]{\href }%
\providecommand \doibase [0]{https://doi.org/}%
\providecommand \selectlanguage [0]{\@gobble}%
\providecommand \bibinfo  [0]{\@secondoftwo}%
\providecommand \bibfield  [0]{\@secondoftwo}%
\providecommand \translation [1]{[#1]}%
\providecommand \BibitemOpen [0]{}%
\providecommand \bibitemStop [0]{}%
\providecommand \bibitemNoStop [0]{.\EOS\space}%
\providecommand \EOS [0]{\spacefactor3000\relax}%
\providecommand \BibitemShut  [1]{\csname bibitem#1\endcsname}%
\let\auto@bib@innerbib\@empty
\bibitem [{\citenamefont {Boltzmann}(1877)}]{boltzmann1877beziehung}%
  \BibitemOpen
  \bibfield  {author} {\bibinfo {author} {\bibfnamefont {L.}~\bibnamefont
  {Boltzmann}},\ }\bibfield  {title} {\bibinfo {title} {{\"U}ber die beziehung
  zwischen dem zweiten hauptsatze des mechanischen w{\"a}rmetheorie und der
  wahrscheinlichkeitsrechnung, respective den s{\"a}tzen {\"u}ber das
  w{\"a}rmegleichgewicht},\ }\href@noop {} {\bibfield  {journal} {\bibinfo
  {journal} {Kk Hof-und Staatsdruckerei}\ } (\bibinfo {year}
  {1877})}\BibitemShut {NoStop}%
\bibitem [{\citenamefont {Gibbs}(1902)}]{gibbs1902elementary}%
  \BibitemOpen
  \bibfield  {author} {\bibinfo {author} {\bibfnamefont {J.~W.}\ \bibnamefont
  {Gibbs}},\ }\bibfield  {title} {\bibinfo {title} {Elementary principles in
  statistical mechanics: developed with especial reference to the rational
  foundations of thermodynamics},\ }\href
  {https://www.cambridge.org/core/books/elementary-principles-in-statistical-mechanics/E7AB3B94A215CF4E3A193C092F29856A}
  {\bibfield  {journal} {\bibinfo  {journal} {C. Scribner's sons}\ } (\bibinfo
  {year} {1902})}\BibitemShut {NoStop}%
\bibitem [{\citenamefont {Neumann}(1927)}]{Neumann1927}%
  \BibitemOpen
  \bibfield  {author} {\bibinfo {author} {\bibfnamefont {J.~v.}\ \bibnamefont
  {Neumann}},\ }\bibfield  {title} {{\selectlanguage {german}\bibinfo {title}
  {Thermodynamik quantenmechanischer gesamtheiten}},\ }\href
  {http://eudml.org/doc/59231} {\bibfield  {journal} {\bibinfo  {journal}
  {Nachrichten von der Gesellschaft der Wissenschaften zu Göttingen,
  Mathematisch-Physikalische Klasse}\ }\textbf {\bibinfo {volume} {1927}},\
  \bibinfo {pages} {273} (\bibinfo {year} {1927})}\BibitemShut {NoStop}%
\bibitem [{\citenamefont {Holzhey}\ \emph {et~al.}(1994)\citenamefont
  {Holzhey}, \citenamefont {Larsen},\ and\ \citenamefont
  {Wilczek}}]{Holzhey94}%
  \BibitemOpen
  \bibfield  {author} {\bibinfo {author} {\bibfnamefont {C.}~\bibnamefont
  {Holzhey}}, \bibinfo {author} {\bibfnamefont {F.}~\bibnamefont {Larsen}},\
  and\ \bibinfo {author} {\bibfnamefont {F.}~\bibnamefont {Wilczek}},\
  }\bibfield  {title} {\bibinfo {title} {Geometric and renormalized entropy in
  conformal field theory},\ }\href
  {https://doi.org/http://dx.doi.org/10.1016/0550-3213(94)90402-2} {\bibfield
  {journal} {\bibinfo  {journal} {Nuclear Physics B}\ }\textbf {\bibinfo
  {volume} {424}},\ \bibinfo {pages} {443 } (\bibinfo {year}
  {1994})}\BibitemShut {NoStop}%
\bibitem [{\citenamefont {Calabrese}\ and\ \citenamefont
  {Cardy}(2004)}]{Calabrese04}%
  \BibitemOpen
  \bibfield  {author} {\bibinfo {author} {\bibfnamefont {P.}~\bibnamefont
  {Calabrese}}\ and\ \bibinfo {author} {\bibfnamefont {J.}~\bibnamefont
  {Cardy}},\ }\bibfield  {title} {\bibinfo {title} {Entanglement entropy and
  quantum field theory},\ }\href
  {http://stacks.iop.org/1742-5468/2004/i=06/a=P06002} {\bibfield  {journal}
  {\bibinfo  {journal} {Journal of Statistical Mechanics: Theory and
  Experiment}\ }\textbf {\bibinfo {volume} {2004}},\ \bibinfo {pages} {P06002}
  (\bibinfo {year} {2004})}\BibitemShut {NoStop}%
\bibitem [{\citenamefont {Shannon}(1948)}]{shannon1948mathematical}%
  \BibitemOpen
  \bibfield  {author} {\bibinfo {author} {\bibfnamefont {C.~E.}\ \bibnamefont
  {Shannon}},\ }\bibfield  {title} {\bibinfo {title} {A mathematical theory of
  communication},\ }\href@noop {} {\bibfield  {journal} {\bibinfo  {journal}
  {The Bell system technical journal}\ }\textbf {\bibinfo {volume} {27}},\
  \bibinfo {pages} {379} (\bibinfo {year} {1948})}\BibitemShut {NoStop}%
\bibitem [{\citenamefont {Ziv}\ and\ \citenamefont
  {Lempel}(1977)}]{ziv1977universal}%
  \BibitemOpen
  \bibfield  {author} {\bibinfo {author} {\bibfnamefont {J.}~\bibnamefont
  {Ziv}}\ and\ \bibinfo {author} {\bibfnamefont {A.}~\bibnamefont {Lempel}},\
  }\bibfield  {title} {\bibinfo {title} {A universal algorithm for sequential
  data compression},\ }\href {https://doi.org/10.1109/TIT.1977.1055714}
  {\bibfield  {journal} {\bibinfo  {journal} {IEEE Transactions on information
  theory}\ }\textbf {\bibinfo {volume} {23}},\ \bibinfo {pages} {337} (\bibinfo
  {year} {1977})}\BibitemShut {NoStop}%
\bibitem [{\citenamefont {Schlosser}\ \emph {et~al.}(2001)\citenamefont
  {Schlosser}, \citenamefont {Reymond}, \citenamefont {Protsenko},\ and\
  \citenamefont {Grangier}}]{schlosser2001sub}%
  \BibitemOpen
  \bibfield  {author} {\bibinfo {author} {\bibfnamefont {N.}~\bibnamefont
  {Schlosser}}, \bibinfo {author} {\bibfnamefont {G.}~\bibnamefont {Reymond}},
  \bibinfo {author} {\bibfnamefont {I.}~\bibnamefont {Protsenko}},\ and\
  \bibinfo {author} {\bibfnamefont {P.}~\bibnamefont {Grangier}},\ }\bibfield
  {title} {\bibinfo {title} {Sub-poissonian loading of single atoms in a
  microscopic dipole trap},\ }\href {https://doi.org/10.1038/35082512}
  {\bibfield  {journal} {\bibinfo  {journal} {Nature}\ }\textbf {\bibinfo
  {volume} {411}},\ \bibinfo {pages} {1024} (\bibinfo {year}
  {2001})}\BibitemShut {NoStop}%
\bibitem [{\citenamefont {Bakr}\ \emph {et~al.}(2009)\citenamefont {Bakr},
  \citenamefont {Gillen}, \citenamefont {Peng}, \citenamefont {F{\"o}lling},\
  and\ \citenamefont {Greiner}}]{bakr2009quantum}%
  \BibitemOpen
  \bibfield  {author} {\bibinfo {author} {\bibfnamefont {W.~S.}\ \bibnamefont
  {Bakr}}, \bibinfo {author} {\bibfnamefont {J.~I.}\ \bibnamefont {Gillen}},
  \bibinfo {author} {\bibfnamefont {A.}~\bibnamefont {Peng}}, \bibinfo {author}
  {\bibfnamefont {S.}~\bibnamefont {F{\"o}lling}},\ and\ \bibinfo {author}
  {\bibfnamefont {M.}~\bibnamefont {Greiner}},\ }\bibfield  {title} {\bibinfo
  {title} {A quantum gas microscope for detecting single atoms in a
  hubbard-regime optical lattice},\ }\href
  {https://doi.org/10.1038/nature08482} {\bibfield  {journal} {\bibinfo
  {journal} {Nature}\ }\textbf {\bibinfo {volume} {462}},\ \bibinfo {pages}
  {74} (\bibinfo {year} {2009})}\BibitemShut {NoStop}%
\bibitem [{\citenamefont {Sherson}\ \emph {et~al.}(2010)\citenamefont
  {Sherson}, \citenamefont {Weitenberg}, \citenamefont {Endres}, \citenamefont
  {Cheneau}, \citenamefont {Bloch},\ and\ \citenamefont
  {Kuhr}}]{sherson2010single}%
  \BibitemOpen
  \bibfield  {author} {\bibinfo {author} {\bibfnamefont {J.~F.}\ \bibnamefont
  {Sherson}}, \bibinfo {author} {\bibfnamefont {C.}~\bibnamefont {Weitenberg}},
  \bibinfo {author} {\bibfnamefont {M.}~\bibnamefont {Endres}}, \bibinfo
  {author} {\bibfnamefont {M.}~\bibnamefont {Cheneau}}, \bibinfo {author}
  {\bibfnamefont {I.}~\bibnamefont {Bloch}},\ and\ \bibinfo {author}
  {\bibfnamefont {S.}~\bibnamefont {Kuhr}},\ }\bibfield  {title} {\bibinfo
  {title} {Single-atom-resolved fluorescence imaging of an atomic mott
  insulator},\ }\href {https://doi.org/10.1038/nature09378} {\bibfield
  {journal} {\bibinfo  {journal} {Nature}\ }\textbf {\bibinfo {volume} {467}},\
  \bibinfo {pages} {68} (\bibinfo {year} {2010})}\BibitemShut {NoStop}%
\bibitem [{\citenamefont {Cheuk}\ \emph {et~al.}(2015)\citenamefont {Cheuk},
  \citenamefont {Nichols}, \citenamefont {Okan}, \citenamefont {Gersdorf},
  \citenamefont {Ramasesh}, \citenamefont {Bakr}, \citenamefont {Lompe},\ and\
  \citenamefont {Zwierlein}}]{cheuk2015quantum}%
  \BibitemOpen
  \bibfield  {author} {\bibinfo {author} {\bibfnamefont {L.~W.}\ \bibnamefont
  {Cheuk}}, \bibinfo {author} {\bibfnamefont {M.~A.}\ \bibnamefont {Nichols}},
  \bibinfo {author} {\bibfnamefont {M.}~\bibnamefont {Okan}}, \bibinfo {author}
  {\bibfnamefont {T.}~\bibnamefont {Gersdorf}}, \bibinfo {author}
  {\bibfnamefont {V.~V.}\ \bibnamefont {Ramasesh}}, \bibinfo {author}
  {\bibfnamefont {W.~S.}\ \bibnamefont {Bakr}}, \bibinfo {author}
  {\bibfnamefont {T.}~\bibnamefont {Lompe}},\ and\ \bibinfo {author}
  {\bibfnamefont {M.~W.}\ \bibnamefont {Zwierlein}},\ }\bibfield  {title}
  {\bibinfo {title} {Quantum-gas microscope for fermionic atoms},\ }\href
  {https://doi.org/10.1103/PhysRevLett.114.193001} {\bibfield  {journal}
  {\bibinfo  {journal} {Phys. Rev. Lett.}\ }\textbf {\bibinfo {volume} {114}},\
  \bibinfo {pages} {193001} (\bibinfo {year} {2015})}\BibitemShut {NoStop}%
\bibitem [{\citenamefont {Haller}\ \emph {et~al.}(2015)\citenamefont {Haller},
  \citenamefont {Hudson}, \citenamefont {Kelly}, \citenamefont {Cotta},
  \citenamefont {Peaudecerf}, \citenamefont {Bruce},\ and\ \citenamefont
  {Kuhr}}]{haller2015single}%
  \BibitemOpen
  \bibfield  {author} {\bibinfo {author} {\bibfnamefont {E.}~\bibnamefont
  {Haller}}, \bibinfo {author} {\bibfnamefont {J.}~\bibnamefont {Hudson}},
  \bibinfo {author} {\bibfnamefont {A.}~\bibnamefont {Kelly}}, \bibinfo
  {author} {\bibfnamefont {D.~A.}\ \bibnamefont {Cotta}}, \bibinfo {author}
  {\bibfnamefont {B.}~\bibnamefont {Peaudecerf}}, \bibinfo {author}
  {\bibfnamefont {G.~D.}\ \bibnamefont {Bruce}},\ and\ \bibinfo {author}
  {\bibfnamefont {S.}~\bibnamefont {Kuhr}},\ }\bibfield  {title} {\bibinfo
  {title} {Single-atom imaging of fermions in a quantum-gas microscope},\
  }\href@noop {} {\bibfield  {journal} {\bibinfo  {journal} {Nature Physics}\
  }\textbf {\bibinfo {volume} {11}},\ \bibinfo {pages} {738} (\bibinfo {year}
  {2015})}\BibitemShut {NoStop}%
\bibitem [{\citenamefont {Parsons}\ \emph {et~al.}(2015)\citenamefont
  {Parsons}, \citenamefont {Huber}, \citenamefont {Mazurenko}, \citenamefont
  {Chiu}, \citenamefont {Setiawan}, \citenamefont {Wooley-Brown}, \citenamefont
  {Blatt},\ and\ \citenamefont {Greiner}}]{parsons2015site}%
  \BibitemOpen
  \bibfield  {author} {\bibinfo {author} {\bibfnamefont {M.~F.}\ \bibnamefont
  {Parsons}}, \bibinfo {author} {\bibfnamefont {F.}~\bibnamefont {Huber}},
  \bibinfo {author} {\bibfnamefont {A.}~\bibnamefont {Mazurenko}}, \bibinfo
  {author} {\bibfnamefont {C.~S.}\ \bibnamefont {Chiu}}, \bibinfo {author}
  {\bibfnamefont {W.}~\bibnamefont {Setiawan}}, \bibinfo {author}
  {\bibfnamefont {K.}~\bibnamefont {Wooley-Brown}}, \bibinfo {author}
  {\bibfnamefont {S.}~\bibnamefont {Blatt}},\ and\ \bibinfo {author}
  {\bibfnamefont {M.}~\bibnamefont {Greiner}},\ }\bibfield  {title} {\bibinfo
  {title} {Site-resolved imaging of fermionic li 6 in an optical lattice},\
  }\href@noop {} {\bibfield  {journal} {\bibinfo  {journal} {Physical review
  letters}\ }\textbf {\bibinfo {volume} {114}},\ \bibinfo {pages} {213002}
  (\bibinfo {year} {2015})}\BibitemShut {NoStop}%
\bibitem [{\citenamefont {Edge}\ \emph {et~al.}(2015)\citenamefont {Edge},
  \citenamefont {Anderson}, \citenamefont {Jervis}, \citenamefont {McKay},
  \citenamefont {Day}, \citenamefont {Trotzky},\ and\ \citenamefont
  {Thywissen}}]{edge2015imaging}%
  \BibitemOpen
  \bibfield  {author} {\bibinfo {author} {\bibfnamefont {G.~J.}\ \bibnamefont
  {Edge}}, \bibinfo {author} {\bibfnamefont {R.}~\bibnamefont {Anderson}},
  \bibinfo {author} {\bibfnamefont {D.}~\bibnamefont {Jervis}}, \bibinfo
  {author} {\bibfnamefont {D.~C.}\ \bibnamefont {McKay}}, \bibinfo {author}
  {\bibfnamefont {R.}~\bibnamefont {Day}}, \bibinfo {author} {\bibfnamefont
  {S.}~\bibnamefont {Trotzky}},\ and\ \bibinfo {author} {\bibfnamefont {J.~H.}\
  \bibnamefont {Thywissen}},\ }\bibfield  {title} {\bibinfo {title} {Imaging
  and addressing of individual fermionic atoms in an optical lattice},\
  }\href@noop {} {\bibfield  {journal} {\bibinfo  {journal} {Physical Review
  A}\ }\textbf {\bibinfo {volume} {92}},\ \bibinfo {pages} {063406} (\bibinfo
  {year} {2015})}\BibitemShut {NoStop}%
\bibitem [{\citenamefont {Gross}\ and\ \citenamefont
  {Bakr}(2021)}]{gross2021quantum}%
  \BibitemOpen
  \bibfield  {author} {\bibinfo {author} {\bibfnamefont {C.}~\bibnamefont
  {Gross}}\ and\ \bibinfo {author} {\bibfnamefont {W.~S.}\ \bibnamefont
  {Bakr}},\ }\bibfield  {title} {\bibinfo {title} {Quantum gas microscopy for
  single atom and spin detection},\ }\href@noop {} {\bibfield  {journal}
  {\bibinfo  {journal} {Nature Physics}\ }\textbf {\bibinfo {volume} {17}},\
  \bibinfo {pages} {1316} (\bibinfo {year} {2021})}\BibitemShut {NoStop}%
\bibitem [{\citenamefont {Browaeys}\ and\ \citenamefont
  {Lahaye}(2020)}]{browaeys2020many}%
  \BibitemOpen
  \bibfield  {author} {\bibinfo {author} {\bibfnamefont {A.}~\bibnamefont
  {Browaeys}}\ and\ \bibinfo {author} {\bibfnamefont {T.}~\bibnamefont
  {Lahaye}},\ }\bibfield  {title} {\bibinfo {title} {Many-body physics with
  individually controlled rydberg atoms},\ }\href@noop {} {\bibfield  {journal}
  {\bibinfo  {journal} {Nature Physics}\ }\textbf {\bibinfo {volume} {16}},\
  \bibinfo {pages} {132} (\bibinfo {year} {2020})}\BibitemShut {NoStop}%
\bibitem [{\citenamefont {Kjaergaard}\ \emph {et~al.}(2020)\citenamefont
  {Kjaergaard}, \citenamefont {Schwartz}, \citenamefont {Braum{\"u}ller},
  \citenamefont {Krantz}, \citenamefont {Wang}, \citenamefont {Gustavsson},\
  and\ \citenamefont {Oliver}}]{kjaergaard2020superconducting}%
  \BibitemOpen
  \bibfield  {author} {\bibinfo {author} {\bibfnamefont {M.}~\bibnamefont
  {Kjaergaard}}, \bibinfo {author} {\bibfnamefont {M.~E.}\ \bibnamefont
  {Schwartz}}, \bibinfo {author} {\bibfnamefont {J.}~\bibnamefont
  {Braum{\"u}ller}}, \bibinfo {author} {\bibfnamefont {P.}~\bibnamefont
  {Krantz}}, \bibinfo {author} {\bibfnamefont {J.~I.-J.}\ \bibnamefont {Wang}},
  \bibinfo {author} {\bibfnamefont {S.}~\bibnamefont {Gustavsson}},\ and\
  \bibinfo {author} {\bibfnamefont {W.~D.}\ \bibnamefont {Oliver}},\ }\bibfield
   {title} {\bibinfo {title} {Superconducting qubits: Current state of play},\
  }\href
  {https://www.annualreviews.org/doi/abs/10.1146/annurev-conmatphys-031119-050605}
  {\bibfield  {journal} {\bibinfo  {journal} {Annual Review of Condensed Matter
  Physics}\ }\textbf {\bibinfo {volume} {11}},\ \bibinfo {pages} {369}
  (\bibinfo {year} {2020})}\BibitemShut {NoStop}%
\bibitem [{\citenamefont {Blatt}\ and\ \citenamefont
  {Roos}(2012)}]{blatt2012quantum}%
  \BibitemOpen
  \bibfield  {author} {\bibinfo {author} {\bibfnamefont {R.}~\bibnamefont
  {Blatt}}\ and\ \bibinfo {author} {\bibfnamefont {C.~F.}\ \bibnamefont
  {Roos}},\ }\bibfield  {title} {\bibinfo {title} {Quantum simulations with
  trapped ions},\ }\href@noop {} {\bibfield  {journal} {\bibinfo  {journal}
  {Nature Physics}\ }\textbf {\bibinfo {volume} {8}},\ \bibinfo {pages} {277}
  (\bibinfo {year} {2012})}\BibitemShut {NoStop}%
\bibitem [{\citenamefont {St{\'e}phan}\ \emph {et~al.}(2009)\citenamefont
  {St{\'e}phan}, \citenamefont {Furukawa}, \citenamefont {Misguich},\ and\
  \citenamefont {Pasquier}}]{stephan2009shannon}%
  \BibitemOpen
  \bibfield  {author} {\bibinfo {author} {\bibfnamefont {J.-M.}\ \bibnamefont
  {St{\'e}phan}}, \bibinfo {author} {\bibfnamefont {S.}~\bibnamefont
  {Furukawa}}, \bibinfo {author} {\bibfnamefont {G.}~\bibnamefont {Misguich}},\
  and\ \bibinfo {author} {\bibfnamefont {V.}~\bibnamefont {Pasquier}},\
  }\bibfield  {title} {\bibinfo {title} {Shannon and entanglement entropies of
  one-and two-dimensional critical wave functions},\ }\href@noop {} {\bibfield
  {journal} {\bibinfo  {journal} {Physical Review B—Condensed Matter and
  Materials Physics}\ }\textbf {\bibinfo {volume} {80}},\ \bibinfo {pages}
  {184421} (\bibinfo {year} {2009})}\BibitemShut {NoStop}%
\bibitem [{\citenamefont {Zaletel}\ \emph {et~al.}(2011)\citenamefont
  {Zaletel}, \citenamefont {Bardarson},\ and\ \citenamefont
  {Moore}}]{zaletel2011logarithmic}%
  \BibitemOpen
  \bibfield  {author} {\bibinfo {author} {\bibfnamefont {M.~P.}\ \bibnamefont
  {Zaletel}}, \bibinfo {author} {\bibfnamefont {J.~H.}\ \bibnamefont
  {Bardarson}},\ and\ \bibinfo {author} {\bibfnamefont {J.~E.}\ \bibnamefont
  {Moore}},\ }\bibfield  {title} {\bibinfo {title} {Logarithmic terms in
  entanglement entropies of 2d quantum critical points and shannon entropies of
  spin chains},\ }\href@noop {} {\bibfield  {journal} {\bibinfo  {journal}
  {Physical review letters}\ }\textbf {\bibinfo {volume} {107}},\ \bibinfo
  {pages} {020402} (\bibinfo {year} {2011})}\BibitemShut {NoStop}%
\bibitem [{\citenamefont {Alcaraz}\ and\ \citenamefont
  {Rajabpour}(2013)}]{alcaraz2013universal}%
  \BibitemOpen
  \bibfield  {author} {\bibinfo {author} {\bibfnamefont {F.~C.}\ \bibnamefont
  {Alcaraz}}\ and\ \bibinfo {author} {\bibfnamefont {M.~A.}\ \bibnamefont
  {Rajabpour}},\ }\bibfield  {title} {\bibinfo {title} {Universal behavior of
  the shannon mutual information of critical quantum chains},\ }\href@noop {}
  {\bibfield  {journal} {\bibinfo  {journal} {Physical Review Letters}\
  }\textbf {\bibinfo {volume} {111}},\ \bibinfo {pages} {017201} (\bibinfo
  {year} {2013})}\BibitemShut {NoStop}%
\bibitem [{\citenamefont {Luitz}\ \emph
  {et~al.}(2014{\natexlab{a}})\citenamefont {Luitz}, \citenamefont
  {Laflorencie},\ and\ \citenamefont {Alet}}]{luitz2014participation}%
  \BibitemOpen
  \bibfield  {author} {\bibinfo {author} {\bibfnamefont {D.~J.}\ \bibnamefont
  {Luitz}}, \bibinfo {author} {\bibfnamefont {N.}~\bibnamefont {Laflorencie}},\
  and\ \bibinfo {author} {\bibfnamefont {F.}~\bibnamefont {Alet}},\ }\bibfield
  {title} {\bibinfo {title} {Participation spectroscopy and entanglement
  hamiltonian of quantum spin models},\ }\href@noop {} {\bibfield  {journal}
  {\bibinfo  {journal} {Journal of Statistical Mechanics: Theory and
  Experiment}\ }\textbf {\bibinfo {volume} {2014}},\ \bibinfo {pages} {P08007}
  (\bibinfo {year} {2014}{\natexlab{a}})}\BibitemShut {NoStop}%
\bibitem [{\citenamefont {Sierant}\ and\ \citenamefont
  {Turkeshi}(2022)}]{PhysRevLett.128.130605}%
  \BibitemOpen
  \bibfield  {author} {\bibinfo {author} {\bibfnamefont {P.}~\bibnamefont
  {Sierant}}\ and\ \bibinfo {author} {\bibfnamefont {X.}~\bibnamefont
  {Turkeshi}},\ }\bibfield  {title} {\bibinfo {title} {Universal behavior
  beyond multifractality of wave functions at measurement-induced phase
  transitions},\ }\href {https://doi.org/10.1103/PhysRevLett.128.130605}
  {\bibfield  {journal} {\bibinfo  {journal} {Phys. Rev. Lett.}\ }\textbf
  {\bibinfo {volume} {128}},\ \bibinfo {pages} {130605} (\bibinfo {year}
  {2022})}\BibitemShut {NoStop}%
\bibitem [{\citenamefont {Sierant}\ \emph {et~al.}(2022)\citenamefont
  {Sierant}, \citenamefont {Schir\`o}, \citenamefont {Lewenstein},\ and\
  \citenamefont {Turkeshi}}]{PhysRevB.106.214316}%
  \BibitemOpen
  \bibfield  {author} {\bibinfo {author} {\bibfnamefont {P.}~\bibnamefont
  {Sierant}}, \bibinfo {author} {\bibfnamefont {M.}~\bibnamefont {Schir\`o}},
  \bibinfo {author} {\bibfnamefont {M.}~\bibnamefont {Lewenstein}},\ and\
  \bibinfo {author} {\bibfnamefont {X.}~\bibnamefont {Turkeshi}},\ }\bibfield
  {title} {\bibinfo {title} {Measurement-induced phase transitions in
  $(d+1)$-dimensional stabilizer circuits},\ }\href
  {https://doi.org/10.1103/PhysRevB.106.214316} {\bibfield  {journal} {\bibinfo
   {journal} {Phys. Rev. B}\ }\textbf {\bibinfo {volume} {106}},\ \bibinfo
  {pages} {214316} (\bibinfo {year} {2022})}\BibitemShut {NoStop}%
\bibitem [{\citenamefont {Liu}\ \emph {et~al.}(2025)\citenamefont {Liu},
  \citenamefont {Sierant}, \citenamefont {Stornati}, \citenamefont
  {Lewenstein},\ and\ \citenamefont {Płodzień}}]{Liu_2025}%
  \BibitemOpen
  \bibfield  {author} {\bibinfo {author} {\bibfnamefont {Y.}~\bibnamefont
  {Liu}}, \bibinfo {author} {\bibfnamefont {P.}~\bibnamefont {Sierant}},
  \bibinfo {author} {\bibfnamefont {P.}~\bibnamefont {Stornati}}, \bibinfo
  {author} {\bibfnamefont {M.}~\bibnamefont {Lewenstein}},\ and\ \bibinfo
  {author} {\bibfnamefont {M.}~\bibnamefont {Płodzień}},\ }\bibfield  {title}
  {\bibinfo {title} {Quantum algorithms for inverse participation ratio
  estimation in multiqubit and multiqudit systems},\ }\bibfield  {journal}
  {\bibinfo  {journal} {Physical Review A}\ }\textbf {\bibinfo {volume}
  {111}},\ \href {https://doi.org/10.1103/physreva.111.052614}
  {10.1103/physreva.111.052614} (\bibinfo {year} {2025})\BibitemShut {NoStop}%
\bibitem [{\citenamefont {Turkeshi}\ and\ \citenamefont
  {Sierant}(2024{\natexlab{a}})}]{PhysRevLett.132.140401}%
  \BibitemOpen
  \bibfield  {author} {\bibinfo {author} {\bibfnamefont {X.}~\bibnamefont
  {Turkeshi}}\ and\ \bibinfo {author} {\bibfnamefont {P.}~\bibnamefont
  {Sierant}},\ }\bibfield  {title} {\bibinfo {title} {Error-resilience phase
  transitions in encoding-decoding quantum circuits},\ }\href
  {https://doi.org/10.1103/PhysRevLett.132.140401} {\bibfield  {journal}
  {\bibinfo  {journal} {Phys. Rev. Lett.}\ }\textbf {\bibinfo {volume} {132}},\
  \bibinfo {pages} {140401} (\bibinfo {year} {2024}{\natexlab{a}})}\BibitemShut
  {NoStop}%
\bibitem [{\citenamefont {Turkeshi}\ and\ \citenamefont
  {Sierant}(2024{\natexlab{b}})}]{Turkeshi_2024}%
  \BibitemOpen
  \bibfield  {author} {\bibinfo {author} {\bibfnamefont {X.}~\bibnamefont
  {Turkeshi}}\ and\ \bibinfo {author} {\bibfnamefont {P.}~\bibnamefont
  {Sierant}},\ }\bibfield  {title} {\bibinfo {title} {Hilbert space
  delocalization under random unitary circuits},\ }\href
  {https://doi.org/10.3390/e26060471} {\bibfield  {journal} {\bibinfo
  {journal} {Entropy}\ }\textbf {\bibinfo {volume} {26}},\ \bibinfo {pages}
  {471} (\bibinfo {year} {2024}{\natexlab{b}})}\BibitemShut {NoStop}%
\bibitem [{\citenamefont {Tirrito}\ \emph {et~al.}(2024)\citenamefont
  {Tirrito}, \citenamefont {Turkeshi},\ and\ \citenamefont
  {Sierant}}]{tirrito2024anticoncentrationmagicspreadingergodic}%
  \BibitemOpen
  \bibfield  {author} {\bibinfo {author} {\bibfnamefont {E.}~\bibnamefont
  {Tirrito}}, \bibinfo {author} {\bibfnamefont {X.}~\bibnamefont {Turkeshi}},\
  and\ \bibinfo {author} {\bibfnamefont {P.}~\bibnamefont {Sierant}},\ }\href
  {https://arxiv.org/abs/2412.10229} {\bibinfo {title} {Anticoncentration and
  magic spreading under ergodic quantum dynamics}} (\bibinfo {year} {2024}),\
  \Eprint {https://arxiv.org/abs/2412.10229} {arXiv:2412.10229 [quant-ph]}
  \BibitemShut {NoStop}%
\bibitem [{\citenamefont {Sierant}\ \emph {et~al.}(2025)\citenamefont
  {Sierant}, \citenamefont {Lewenstein}, \citenamefont {Scardicchio},
  \citenamefont {Vidmar},\ and\ \citenamefont {Zakrzewski}}]{sierant2025many}%
  \BibitemOpen
  \bibfield  {author} {\bibinfo {author} {\bibfnamefont {P.}~\bibnamefont
  {Sierant}}, \bibinfo {author} {\bibfnamefont {M.}~\bibnamefont {Lewenstein}},
  \bibinfo {author} {\bibfnamefont {A.}~\bibnamefont {Scardicchio}}, \bibinfo
  {author} {\bibfnamefont {L.}~\bibnamefont {Vidmar}},\ and\ \bibinfo {author}
  {\bibfnamefont {J.}~\bibnamefont {Zakrzewski}},\ }\bibfield  {title}
  {\bibinfo {title} {Many-body localization in the age of classical
  computing},\ }\href
  {https://iopscience.iop.org/article/10.1088/1361-6633/ad9756} {\bibfield
  {journal} {\bibinfo  {journal} {Reports on Progress in Physics}\ }\textbf
  {\bibinfo {volume} {88}},\ \bibinfo {pages} {026502} (\bibinfo {year}
  {2025})}\BibitemShut {NoStop}%
\bibitem [{\citenamefont {Wang}\ \emph {et~al.}(2025)\citenamefont {Wang},
  \citenamefont {Vasseur}, \citenamefont {Trebst}, \citenamefont {Ludwig},\
  and\ \citenamefont
  {Zhu}}]{wang2025decoherenceinducedselfdualcriticalitytopological}%
  \BibitemOpen
  \bibfield  {author} {\bibinfo {author} {\bibfnamefont {Q.}~\bibnamefont
  {Wang}}, \bibinfo {author} {\bibfnamefont {R.}~\bibnamefont {Vasseur}},
  \bibinfo {author} {\bibfnamefont {S.}~\bibnamefont {Trebst}}, \bibinfo
  {author} {\bibfnamefont {A.~W.~W.}\ \bibnamefont {Ludwig}},\ and\ \bibinfo
  {author} {\bibfnamefont {G.-Y.}\ \bibnamefont {Zhu}},\ }\href
  {https://arxiv.org/abs/2502.14034} {\bibinfo {title} {Decoherence-induced
  self-dual criticality in topological states of matter}} (\bibinfo {year}
  {2025}),\ \Eprint {https://arxiv.org/abs/2502.14034} {arXiv:2502.14034
  [quant-ph]} \BibitemShut {NoStop}%
\bibitem [{\citenamefont {Fisher}\ and\ \citenamefont
  {Ferdinand}(1967)}]{fisher1967interfacial}%
  \BibitemOpen
  \bibfield  {author} {\bibinfo {author} {\bibfnamefont {M.~E.}\ \bibnamefont
  {Fisher}}\ and\ \bibinfo {author} {\bibfnamefont {A.~E.}\ \bibnamefont
  {Ferdinand}},\ }\bibfield  {title} {\bibinfo {title} {Interfacial, boundary,
  and size effects at critical points},\ }\href
  {https://doi.org/10.1103/PhysRevLett.19.169} {\bibfield  {journal} {\bibinfo
  {journal} {Phys. Rev. Lett.}\ }\textbf {\bibinfo {volume} {19}},\ \bibinfo
  {pages} {169} (\bibinfo {year} {1967})}\BibitemShut {NoStop}%
\bibitem [{\citenamefont {Dennis}\ \emph {et~al.}(2002)\citenamefont {Dennis},
  \citenamefont {Kitaev}, \citenamefont {Landahl},\ and\ \citenamefont
  {Preskill}}]{dennis2002}%
  \BibitemOpen
  \bibfield  {author} {\bibinfo {author} {\bibfnamefont {E.}~\bibnamefont
  {Dennis}}, \bibinfo {author} {\bibfnamefont {A.}~\bibnamefont {Kitaev}},
  \bibinfo {author} {\bibfnamefont {A.}~\bibnamefont {Landahl}},\ and\ \bibinfo
  {author} {\bibfnamefont {J.}~\bibnamefont {Preskill}},\ }\bibfield  {title}
  {\bibinfo {title} {Topological quantum memory},\ }\href
  {https://doi.org/10.1063/1.1499754} {\bibfield  {journal} {\bibinfo
  {journal} {Journal of Mathematical Physics}\ }\textbf {\bibinfo {volume}
  {43}},\ \bibinfo {pages} {4452} (\bibinfo {year} {2002})}\BibitemShut
  {NoStop}%
\bibitem [{\citenamefont {Wang}\ \emph {et~al.}(2003)\citenamefont {Wang},
  \citenamefont {Harrington},\ and\ \citenamefont
  {Preskill}}]{wang2003confinement}%
  \BibitemOpen
  \bibfield  {author} {\bibinfo {author} {\bibfnamefont {C.}~\bibnamefont
  {Wang}}, \bibinfo {author} {\bibfnamefont {J.}~\bibnamefont {Harrington}},\
  and\ \bibinfo {author} {\bibfnamefont {J.}~\bibnamefont {Preskill}},\
  }\bibfield  {title} {\bibinfo {title} {Confinement-higgs transition in a
  disordered gauge theory and the accuracy threshold for quantum memory},\
  }\href {https://doi.org/10.1016%2Fs0003-4916%2802%2900019-2} {\bibfield
  {journal} {\bibinfo  {journal} {Annals of Physics}\ }\textbf {\bibinfo
  {volume} {303}},\ \bibinfo {pages} {31} (\bibinfo {year} {2003})}\BibitemShut
  {NoStop}%
\bibitem [{\citenamefont {Lee}\ \emph {et~al.}(2023)\citenamefont {Lee},
  \citenamefont {Jian},\ and\ \citenamefont {Xu}}]{lee2023quantum}%
  \BibitemOpen
  \bibfield  {author} {\bibinfo {author} {\bibfnamefont {J.~Y.}\ \bibnamefont
  {Lee}}, \bibinfo {author} {\bibfnamefont {C.-M.}\ \bibnamefont {Jian}},\ and\
  \bibinfo {author} {\bibfnamefont {C.}~\bibnamefont {Xu}},\ }\bibfield
  {title} {\bibinfo {title} {Quantum criticality under decoherence or weak
  measurement},\ }\href {https://doi.org/10.1103/PRXQuantum.4.030317}
  {\bibfield  {journal} {\bibinfo  {journal} {PRX Quantum}\ }\textbf {\bibinfo
  {volume} {4}},\ \bibinfo {pages} {030317} (\bibinfo {year}
  {2023})}\BibitemShut {NoStop}%
\bibitem [{\citenamefont {Fan}\ \emph {et~al.}(2024)\citenamefont {Fan},
  \citenamefont {Bao}, \citenamefont {Altman},\ and\ \citenamefont
  {Vishwanath}}]{fan2023diagnostics}%
  \BibitemOpen
  \bibfield  {author} {\bibinfo {author} {\bibfnamefont {R.}~\bibnamefont
  {Fan}}, \bibinfo {author} {\bibfnamefont {Y.}~\bibnamefont {Bao}}, \bibinfo
  {author} {\bibfnamefont {E.}~\bibnamefont {Altman}},\ and\ \bibinfo {author}
  {\bibfnamefont {A.}~\bibnamefont {Vishwanath}},\ }\bibfield  {title}
  {\bibinfo {title} {Diagnostics of mixed-state topological order and breakdown
  of quantum memory},\ }\href {https://doi.org/10.1103/PRXQuantum.5.020343}
  {\bibfield  {journal} {\bibinfo  {journal} {PRX Quantum}\ }\textbf {\bibinfo
  {volume} {5}},\ \bibinfo {pages} {020343} (\bibinfo {year}
  {2024})}\BibitemShut {NoStop}%
\bibitem [{\citenamefont {Bao}\ \emph {et~al.}(2023)\citenamefont {Bao},
  \citenamefont {Fan}, \citenamefont {Vishwanath},\ and\ \citenamefont
  {Altman}}]{bao2023mixed}%
  \BibitemOpen
  \bibfield  {author} {\bibinfo {author} {\bibfnamefont {Y.}~\bibnamefont
  {Bao}}, \bibinfo {author} {\bibfnamefont {R.}~\bibnamefont {Fan}}, \bibinfo
  {author} {\bibfnamefont {A.}~\bibnamefont {Vishwanath}},\ and\ \bibinfo
  {author} {\bibfnamefont {E.}~\bibnamefont {Altman}},\ }\bibfield  {title}
  {\bibinfo {title} {Mixed-state topological order and the errorfield double
  formulation of decoherence-induced transitions},\ }\href
  {https://arxiv.org/abs/2301.05687} {\bibfield  {journal} {\bibinfo  {journal}
  {arXiv preprint arXiv:2301.05687}\ } (\bibinfo {year} {2023})}\BibitemShut
  {NoStop}%
\bibitem [{\citenamefont {Chen}\ and\ \citenamefont
  {Grover}(2024{\natexlab{a}})}]{chen2023separability}%
  \BibitemOpen
  \bibfield  {author} {\bibinfo {author} {\bibfnamefont {Y.-H.}\ \bibnamefont
  {Chen}}\ and\ \bibinfo {author} {\bibfnamefont {T.}~\bibnamefont {Grover}},\
  }\bibfield  {title} {\bibinfo {title} {Separability transitions in
  topological states induced by local decoherence},\ }\href
  {https://doi.org/10.1103/PhysRevLett.132.170602} {\bibfield  {journal}
  {\bibinfo  {journal} {Phys. Rev. Lett.}\ }\textbf {\bibinfo {volume} {132}},\
  \bibinfo {pages} {170602} (\bibinfo {year} {2024}{\natexlab{a}})}\BibitemShut
  {NoStop}%
\bibitem [{\citenamefont {Sang}\ \emph
  {et~al.}(2024{\natexlab{a}})\citenamefont {Sang}, \citenamefont {Zou},\ and\
  \citenamefont {Hsieh}}]{sang2023mixed}%
  \BibitemOpen
  \bibfield  {author} {\bibinfo {author} {\bibfnamefont {S.}~\bibnamefont
  {Sang}}, \bibinfo {author} {\bibfnamefont {Y.}~\bibnamefont {Zou}},\ and\
  \bibinfo {author} {\bibfnamefont {T.~H.}\ \bibnamefont {Hsieh}},\ }\bibfield
  {title} {\bibinfo {title} {Mixed-state quantum phases: Renormalization and
  quantum error correction},\ }\href
  {https://doi.org/10.1103/PhysRevX.14.031044} {\bibfield  {journal} {\bibinfo
  {journal} {Phys. Rev. X}\ }\textbf {\bibinfo {volume} {14}},\ \bibinfo
  {pages} {031044} (\bibinfo {year} {2024}{\natexlab{a}})}\BibitemShut
  {NoStop}%
\bibitem [{\citenamefont {Ma}\ \emph {et~al.}(2023)\citenamefont {Ma},
  \citenamefont {Zhang}, \citenamefont {Bi}, \citenamefont {Cheng},\ and\
  \citenamefont {Wang}}]{ma2023topological}%
  \BibitemOpen
  \bibfield  {author} {\bibinfo {author} {\bibfnamefont {R.}~\bibnamefont
  {Ma}}, \bibinfo {author} {\bibfnamefont {J.-H.}\ \bibnamefont {Zhang}},
  \bibinfo {author} {\bibfnamefont {Z.}~\bibnamefont {Bi}}, \bibinfo {author}
  {\bibfnamefont {M.}~\bibnamefont {Cheng}},\ and\ \bibinfo {author}
  {\bibfnamefont {C.}~\bibnamefont {Wang}},\ }\bibfield  {title} {\bibinfo
  {title} {Topological phases with average symmetries: the decohered, the
  disordered, and the intrinsic},\ }\href {https://arxiv.org/abs/2305.16399}
  {\bibfield  {journal} {\bibinfo  {journal} {arXiv preprint arXiv:2305.16399}\
  } (\bibinfo {year} {2023})}\BibitemShut {NoStop}%
\bibitem [{\citenamefont {Lessa}\ \emph {et~al.}(2024)\citenamefont {Lessa},
  \citenamefont {Ma}, \citenamefont {Zhang}, \citenamefont {Bi}, \citenamefont
  {Cheng},\ and\ \citenamefont {Wang}}]{lessa2024strong}%
  \BibitemOpen
  \bibfield  {author} {\bibinfo {author} {\bibfnamefont {L.~A.}\ \bibnamefont
  {Lessa}}, \bibinfo {author} {\bibfnamefont {R.}~\bibnamefont {Ma}}, \bibinfo
  {author} {\bibfnamefont {J.-H.}\ \bibnamefont {Zhang}}, \bibinfo {author}
  {\bibfnamefont {Z.}~\bibnamefont {Bi}}, \bibinfo {author} {\bibfnamefont
  {M.}~\bibnamefont {Cheng}},\ and\ \bibinfo {author} {\bibfnamefont
  {C.}~\bibnamefont {Wang}},\ }\bibfield  {title} {\bibinfo {title}
  {Strong-to-weak spontaneous symmetry breaking in mixed quantum states},\
  }\href {https://arxiv.org/abs/2405.03639} {\bibfield  {journal} {\bibinfo
  {journal} {arXiv preprint arXiv:2405.03639}\ } (\bibinfo {year}
  {2024})}\BibitemShut {NoStop}%
\bibitem [{\citenamefont {Sala}\ \emph {et~al.}(2024)\citenamefont {Sala},
  \citenamefont {Gopalakrishnan}, \citenamefont {Oshikawa},\ and\ \citenamefont
  {You}}]{sala2024spontaneous}%
  \BibitemOpen
  \bibfield  {author} {\bibinfo {author} {\bibfnamefont {P.}~\bibnamefont
  {Sala}}, \bibinfo {author} {\bibfnamefont {S.}~\bibnamefont
  {Gopalakrishnan}}, \bibinfo {author} {\bibfnamefont {M.}~\bibnamefont
  {Oshikawa}},\ and\ \bibinfo {author} {\bibfnamefont {Y.}~\bibnamefont
  {You}},\ }\bibfield  {title} {\bibinfo {title} {Spontaneous strong symmetry
  breaking in open systems: Purification perspective},\ }\href
  {https://doi.org/10.1103/PhysRevB.110.155150} {\bibfield  {journal} {\bibinfo
   {journal} {Phys. Rev. B}\ }\textbf {\bibinfo {volume} {110}},\ \bibinfo
  {pages} {155150} (\bibinfo {year} {2024})}\BibitemShut {NoStop}%
\bibitem [{\citenamefont {Gu}\ \emph {et~al.}(2024)\citenamefont {Gu},
  \citenamefont {Wang},\ and\ \citenamefont
  {Wang}}]{gu2024spontaneoussymmetrybreakingopen}%
  \BibitemOpen
  \bibfield  {author} {\bibinfo {author} {\bibfnamefont {D.}~\bibnamefont
  {Gu}}, \bibinfo {author} {\bibfnamefont {Z.}~\bibnamefont {Wang}},\ and\
  \bibinfo {author} {\bibfnamefont {Z.}~\bibnamefont {Wang}},\ }\bibfield
  {title} {\bibinfo {title} {Spontaneous symmetry breaking in open quantum
  systems: strong, weak, and strong-to-weak},\ }\href
  {https://arxiv.org/abs/2406.19381} {\bibfield  {journal} {\bibinfo  {journal}
  {arXiv preprint arXiv:2412.18397}\ } (\bibinfo {year} {2024})}\BibitemShut
  {NoStop}%
\bibitem [{\citenamefont {Huang}\ \emph {et~al.}(2024)\citenamefont {Huang},
  \citenamefont {Qi}, \citenamefont {Zhang},\ and\ \citenamefont
  {Lucas}}]{huang2024hydrodynamicseffectivefieldtheory}%
  \BibitemOpen
  \bibfield  {author} {\bibinfo {author} {\bibfnamefont {X.}~\bibnamefont
  {Huang}}, \bibinfo {author} {\bibfnamefont {M.}~\bibnamefont {Qi}}, \bibinfo
  {author} {\bibfnamefont {J.-H.}\ \bibnamefont {Zhang}},\ and\ \bibinfo
  {author} {\bibfnamefont {A.}~\bibnamefont {Lucas}},\ }\bibfield  {title}
  {\bibinfo {title} {Hydrodynamics as the effective field theory of
  strong-to-weak spontaneous symmetry breaking},\ }\href
  {https://arxiv.org/abs/2407.08760} {\bibfield  {journal} {\bibinfo  {journal}
  {arXiv preprint 2407.08760}\ } (\bibinfo {year} {2024})}\BibitemShut
  {NoStop}%
\bibitem [{\citenamefont {Kuno}\ \emph {et~al.}(2024)\citenamefont {Kuno},
  \citenamefont {Orito},\ and\ \citenamefont {Ichinose}}]{kuno2024strong}%
  \BibitemOpen
  \bibfield  {author} {\bibinfo {author} {\bibfnamefont {Y.}~\bibnamefont
  {Kuno}}, \bibinfo {author} {\bibfnamefont {T.}~\bibnamefont {Orito}},\ and\
  \bibinfo {author} {\bibfnamefont {I.}~\bibnamefont {Ichinose}},\ }\bibfield
  {title} {\bibinfo {title} {Strong-to-weak symmetry breaking states in
  stochastic dephasing stabilizer circuits},\ }\href
  {https://doi.org/10.1103/PhysRevB.110.094106} {\bibfield  {journal} {\bibinfo
   {journal} {Phys. Rev. B}\ }\textbf {\bibinfo {volume} {110}},\ \bibinfo
  {pages} {094106} (\bibinfo {year} {2024})}\BibitemShut {NoStop}%
\bibitem [{\citenamefont {Zhang}\ \emph
  {et~al.}(2024{\natexlab{a}})\citenamefont {Zhang}, \citenamefont {Xu},
  \citenamefont {Zhang}, \citenamefont {Xu}, \citenamefont {Bi},\ and\
  \citenamefont {Luo}}]{zhang2024strongtoweakspontaneousbreaking1form}%
  \BibitemOpen
  \bibfield  {author} {\bibinfo {author} {\bibfnamefont {C.}~\bibnamefont
  {Zhang}}, \bibinfo {author} {\bibfnamefont {Y.}~\bibnamefont {Xu}}, \bibinfo
  {author} {\bibfnamefont {J.-H.}\ \bibnamefont {Zhang}}, \bibinfo {author}
  {\bibfnamefont {C.}~\bibnamefont {Xu}}, \bibinfo {author} {\bibfnamefont
  {Z.}~\bibnamefont {Bi}},\ and\ \bibinfo {author} {\bibfnamefont {Z.-X.}\
  \bibnamefont {Luo}},\ }\bibfield  {title} {\bibinfo {title} {Strong-to-weak
  spontaneous breaking of 1-form symmetry and intrinsically mixed topological
  order},\ }\href {https://arxiv.org/abs/2409.17530} {\bibfield  {journal}
  {\bibinfo  {journal} {arXiv preprint arXiv:2409.17530}\ } (\bibinfo {year}
  {2024}{\natexlab{a}})}\BibitemShut {NoStop}%
\bibitem [{\citenamefont {Zhang}\ \emph
  {et~al.}(2024{\natexlab{b}})\citenamefont {Zhang}, \citenamefont {Xu},\ and\
  \citenamefont
  {Xu}}]{zhang2024fluctuationdissipationtheoreminformationgeometry}%
  \BibitemOpen
  \bibfield  {author} {\bibinfo {author} {\bibfnamefont {J.-H.}\ \bibnamefont
  {Zhang}}, \bibinfo {author} {\bibfnamefont {C.}~\bibnamefont {Xu}},\ and\
  \bibinfo {author} {\bibfnamefont {Y.}~\bibnamefont {Xu}},\ }\bibfield
  {title} {\bibinfo {title} {Fluctuation-dissipation theorem and information
  geometry in open quantum systems},\ }\href {https://arxiv.org/abs/2409.18944}
  {\bibfield  {journal} {\bibinfo  {journal} {arXiv preprint arXiv:2409.18944}\
  } (\bibinfo {year} {2024}{\natexlab{b}})}\BibitemShut {NoStop}%
\bibitem [{\citenamefont {Liu}\ \emph {et~al.}(2024)\citenamefont {Liu},
  \citenamefont {Chen}, \citenamefont {Zhang}, \citenamefont {Zhou},\ and\
  \citenamefont {Zhang}}]{liu2024diagnosingstrongtoweaksymmetrybreaking}%
  \BibitemOpen
  \bibfield  {author} {\bibinfo {author} {\bibfnamefont {Z.}~\bibnamefont
  {Liu}}, \bibinfo {author} {\bibfnamefont {L.}~\bibnamefont {Chen}}, \bibinfo
  {author} {\bibfnamefont {Y.}~\bibnamefont {Zhang}}, \bibinfo {author}
  {\bibfnamefont {S.}~\bibnamefont {Zhou}},\ and\ \bibinfo {author}
  {\bibfnamefont {P.}~\bibnamefont {Zhang}},\ }\bibfield  {title} {\bibinfo
  {title} {Diagnosing strong-to-weak symmetry breaking via wightman
  correlators},\ }\href {https://arxiv.org/abs/2410.09327} {\bibfield
  {journal} {\bibinfo  {journal} {arXiv preprint arXiv:2410.09327}\ } (\bibinfo
  {year} {2024})}\BibitemShut {NoStop}%
\bibitem [{\citenamefont {Shah}\ \emph {et~al.}(2024)\citenamefont {Shah},
  \citenamefont {Fechisin}, \citenamefont {Wang}, \citenamefont {Iosue},
  \citenamefont {Watson}, \citenamefont {Wang}, \citenamefont {Ware},
  \citenamefont {Gorshkov},\ and\ \citenamefont
  {Lin}}]{shah2024instabilitysteadystatemixedstatesymmetryprotected}%
  \BibitemOpen
  \bibfield  {author} {\bibinfo {author} {\bibfnamefont {J.}~\bibnamefont
  {Shah}}, \bibinfo {author} {\bibfnamefont {C.}~\bibnamefont {Fechisin}},
  \bibinfo {author} {\bibfnamefont {Y.-X.}\ \bibnamefont {Wang}}, \bibinfo
  {author} {\bibfnamefont {J.~T.}\ \bibnamefont {Iosue}}, \bibinfo {author}
  {\bibfnamefont {J.~D.}\ \bibnamefont {Watson}}, \bibinfo {author}
  {\bibfnamefont {Y.-Q.}\ \bibnamefont {Wang}}, \bibinfo {author}
  {\bibfnamefont {B.}~\bibnamefont {Ware}}, \bibinfo {author} {\bibfnamefont
  {A.~V.}\ \bibnamefont {Gorshkov}},\ and\ \bibinfo {author} {\bibfnamefont
  {C.-J.}\ \bibnamefont {Lin}},\ }\bibfield  {title} {\bibinfo {title}
  {Instability of steady-state mixed-state symmetry-protected topological order
  to strong-to-weak spontaneous symmetry breaking},\ }\href
  {https://arxiv.org/abs/2410.12900} {\bibfield  {journal} {\bibinfo  {journal}
  {arXiv preprint arXiv:2410.12900}\ } (\bibinfo {year} {2024})}\BibitemShut
  {NoStop}%
\bibitem [{\citenamefont {Guo}\ and\ \citenamefont
  {Yang}(2024)}]{guo2024strongtoweakspontaneoussymmetrybreaking}%
  \BibitemOpen
  \bibfield  {author} {\bibinfo {author} {\bibfnamefont {Y.}~\bibnamefont
  {Guo}}\ and\ \bibinfo {author} {\bibfnamefont {S.}~\bibnamefont {Yang}},\
  }\bibfield  {title} {\bibinfo {title} {Strong-to-weak spontaneous symmetry
  breaking meets average symmetry-protected topological order},\ }\href
  {https://arxiv.org/abs/2410.13734} {\bibfield  {journal} {\bibinfo  {journal}
  {arXiv preprint arXiv:2410.13734}\ } (\bibinfo {year} {2024})}\BibitemShut
  {NoStop}%
\bibitem [{\citenamefont {Kim}\ \emph {et~al.}(2024)\citenamefont {Kim},
  \citenamefont {Altman},\ and\ \citenamefont
  {Lee}}]{kim2024errorthresholdsykcodes}%
  \BibitemOpen
  \bibfield  {author} {\bibinfo {author} {\bibfnamefont {J.}~\bibnamefont
  {Kim}}, \bibinfo {author} {\bibfnamefont {E.}~\bibnamefont {Altman}},\ and\
  \bibinfo {author} {\bibfnamefont {J.~Y.}\ \bibnamefont {Lee}},\ }\bibfield
  {title} {\bibinfo {title} {Error threshold of syk codes from strong-to-weak
  parity symmetry breaking},\ }\href {https://arxiv.org/abs/2410.24225}
  {\bibfield  {journal} {\bibinfo  {journal} {arXiv preprint arXiv:2410.24225}\
  } (\bibinfo {year} {2024})}\BibitemShut {NoStop}%
\bibitem [{\citenamefont {Ando}\ \emph {et~al.}(2024)\citenamefont {Ando},
  \citenamefont {Ryu},\ and\ \citenamefont
  {Watanabe}}]{ando2024gaugetheorymixedstate}%
  \BibitemOpen
  \bibfield  {author} {\bibinfo {author} {\bibfnamefont {T.}~\bibnamefont
  {Ando}}, \bibinfo {author} {\bibfnamefont {S.}~\bibnamefont {Ryu}},\ and\
  \bibinfo {author} {\bibfnamefont {M.}~\bibnamefont {Watanabe}},\ }\bibfield
  {title} {\bibinfo {title} {Gauge theory and mixed state criticality},\ }\href
  {https://arxiv.org/abs/2411.04360} {\bibfield  {journal} {\bibinfo  {journal}
  {arXiv preprint arXiv:2411.04360}\ } (\bibinfo {year} {2024})}\BibitemShut
  {NoStop}%
\bibitem [{\citenamefont {Chen}\ \emph {et~al.}(2024)\citenamefont {Chen},
  \citenamefont {Sun},\ and\ \citenamefont
  {Zhang}}]{chen2024strongtoweaksymmetrybreakingentanglement}%
  \BibitemOpen
  \bibfield  {author} {\bibinfo {author} {\bibfnamefont {L.}~\bibnamefont
  {Chen}}, \bibinfo {author} {\bibfnamefont {N.}~\bibnamefont {Sun}},\ and\
  \bibinfo {author} {\bibfnamefont {P.}~\bibnamefont {Zhang}},\ }\bibfield
  {title} {\bibinfo {title} {Strong-to-weak symmetry breaking and entanglement
  transitions},\ }\href {https://arxiv.org/abs/2411.05364} {\bibfield
  {journal} {\bibinfo  {journal} {arXiv preprint arXiv:2411.05364}\ } (\bibinfo
  {year} {2024})}\BibitemShut {NoStop}%
\bibitem [{\citenamefont {Orito}\ \emph {et~al.}(2025)\citenamefont {Orito},
  \citenamefont {Kuno},\ and\ \citenamefont
  {Ichinose}}]{orito2025strongweaksymmetriesspontaneous}%
  \BibitemOpen
  \bibfield  {author} {\bibinfo {author} {\bibfnamefont {T.}~\bibnamefont
  {Orito}}, \bibinfo {author} {\bibfnamefont {Y.}~\bibnamefont {Kuno}},\ and\
  \bibinfo {author} {\bibfnamefont {I.}~\bibnamefont {Ichinose}},\ }\bibfield
  {title} {\bibinfo {title} {Strong and weak symmetries and their spontaneous
  symmetry breaking in mixed states emerging from the quantum ising model under
  multiple decoherence},\ }\href {https://arxiv.org/abs/2412.12738} {\bibfield
  {journal} {\bibinfo  {journal} {arXiv preprint arXiv:2412.12738}\ } (\bibinfo
  {year} {2025})}\BibitemShut {NoStop}%
\bibitem [{\citenamefont {Sun}\ \emph {et~al.}(2025)\citenamefont {Sun},
  \citenamefont {Zhang},\ and\ \citenamefont
  {Feng}}]{sun2025schemedetectstrongtoweaksymmetry}%
  \BibitemOpen
  \bibfield  {author} {\bibinfo {author} {\bibfnamefont {N.}~\bibnamefont
  {Sun}}, \bibinfo {author} {\bibfnamefont {P.}~\bibnamefont {Zhang}},\ and\
  \bibinfo {author} {\bibfnamefont {L.}~\bibnamefont {Feng}},\ }\bibfield
  {title} {\bibinfo {title} {Scheme to detect the strong-to-weak symmetry
  breaking via randomized measurements},\ }\href
  {https://arxiv.org/abs/2412.18397} {\bibfield  {journal} {\bibinfo  {journal}
  {arXiv preprint arXiv:2412.18397}\ } (\bibinfo {year} {2025})}\BibitemShut
  {NoStop}%
\bibitem [{\citenamefont {Chen}\ \emph {et~al.}(2025)\citenamefont {Chen},
  \citenamefont {Zhu}, \citenamefont {Verresen}, \citenamefont {Seif},
  \citenamefont {B{\"a}umer}, \citenamefont {Layden}, \citenamefont
  {Tantivasadakarn}, \citenamefont {Zhu}, \citenamefont {Sheldon},
  \citenamefont {Vishwanath} \emph {et~al.}}]{chen2025nishimori}%
  \BibitemOpen
  \bibfield  {author} {\bibinfo {author} {\bibfnamefont {E.~H.}\ \bibnamefont
  {Chen}}, \bibinfo {author} {\bibfnamefont {G.-Y.}\ \bibnamefont {Zhu}},
  \bibinfo {author} {\bibfnamefont {R.}~\bibnamefont {Verresen}}, \bibinfo
  {author} {\bibfnamefont {A.}~\bibnamefont {Seif}}, \bibinfo {author}
  {\bibfnamefont {E.}~\bibnamefont {B{\"a}umer}}, \bibinfo {author}
  {\bibfnamefont {D.}~\bibnamefont {Layden}}, \bibinfo {author} {\bibfnamefont
  {N.}~\bibnamefont {Tantivasadakarn}}, \bibinfo {author} {\bibfnamefont
  {G.}~\bibnamefont {Zhu}}, \bibinfo {author} {\bibfnamefont {S.}~\bibnamefont
  {Sheldon}}, \bibinfo {author} {\bibfnamefont {A.}~\bibnamefont {Vishwanath}},
  \emph {et~al.},\ }\bibfield  {title} {\bibinfo {title} {Nishimori transition
  across the error threshold for constant-depth quantum circuits},\ }\href
  {https://www.nature.com/articles/s41567-024-02696-6} {\bibfield  {journal}
  {\bibinfo  {journal} {Nature Physics}\ }\textbf {\bibinfo {volume} {21}},\
  \bibinfo {pages} {161} (\bibinfo {year} {2025})}\BibitemShut {NoStop}%
\bibitem [{\citenamefont {da~Silva}\ \emph {et~al.}(2011)\citenamefont
  {da~Silva}, \citenamefont {Landon-Cardinal},\ and\ \citenamefont
  {Poulin}}]{da2011practical}%
  \BibitemOpen
  \bibfield  {author} {\bibinfo {author} {\bibfnamefont {M.~P.}\ \bibnamefont
  {da~Silva}}, \bibinfo {author} {\bibfnamefont {O.}~\bibnamefont
  {Landon-Cardinal}},\ and\ \bibinfo {author} {\bibfnamefont {D.}~\bibnamefont
  {Poulin}},\ }\bibfield  {title} {\bibinfo {title} {Practical characterization
  of quantum devices without tomography},\ }\href
  {https://doi.org/10.1103/PhysRevLett.107.210404} {\bibfield  {journal}
  {\bibinfo  {journal} {Phys. Rev. Lett.}\ }\textbf {\bibinfo {volume} {107}},\
  \bibinfo {pages} {210404} (\bibinfo {year} {2011})}\BibitemShut {NoStop}%
\bibitem [{\citenamefont {van Enk}\ and\ \citenamefont
  {Beenakker}(2012)}]{van2012measuring}%
  \BibitemOpen
  \bibfield  {author} {\bibinfo {author} {\bibfnamefont {S.~J.}\ \bibnamefont
  {van Enk}}\ and\ \bibinfo {author} {\bibfnamefont {C.~W.~J.}\ \bibnamefont
  {Beenakker}},\ }\bibfield  {title} {\bibinfo {title} {Measuring
  $\mathrm{Tr}{\ensuremath{\rho}}^{n}$ on single copies of $\ensuremath{\rho}$
  using random measurements},\ }\href
  {https://doi.org/10.1103/PhysRevLett.108.110503} {\bibfield  {journal}
  {\bibinfo  {journal} {Phys. Rev. Lett.}\ }\textbf {\bibinfo {volume} {108}},\
  \bibinfo {pages} {110503} (\bibinfo {year} {2012})}\BibitemShut {NoStop}%
\bibitem [{\citenamefont {Aaronson}(2018)}]{aaronson2018shadow}%
  \BibitemOpen
  \bibfield  {author} {\bibinfo {author} {\bibfnamefont {S.}~\bibnamefont
  {Aaronson}},\ }\bibfield  {title} {\bibinfo {title} {Shadow tomography of
  quantum states},\ }\href@noop {} {\bibfield  {journal} {\bibinfo  {journal}
  {Proceedings of the 50th annual ACM SIGACT symposium on theory of computing}\
  ,\ \bibinfo {pages} {325}} (\bibinfo {year} {2018})}\BibitemShut {NoStop}%
\bibitem [{\citenamefont {Elben}\ \emph {et~al.}(2018)\citenamefont {Elben},
  \citenamefont {Vermersch}, \citenamefont {Dalmonte}, \citenamefont {Cirac},\
  and\ \citenamefont {Zoller}}]{elben2018renyi}%
  \BibitemOpen
  \bibfield  {author} {\bibinfo {author} {\bibfnamefont {A.}~\bibnamefont
  {Elben}}, \bibinfo {author} {\bibfnamefont {B.}~\bibnamefont {Vermersch}},
  \bibinfo {author} {\bibfnamefont {M.}~\bibnamefont {Dalmonte}}, \bibinfo
  {author} {\bibfnamefont {J.~I.}\ \bibnamefont {Cirac}},\ and\ \bibinfo
  {author} {\bibfnamefont {P.}~\bibnamefont {Zoller}},\ }\bibfield  {title}
  {\bibinfo {title} {R{\'e}nyi entropies from random quenches in atomic hubbard
  and spin models},\ }\href@noop {} {\bibfield  {journal} {\bibinfo  {journal}
  {Physical review letters}\ }\textbf {\bibinfo {volume} {120}},\ \bibinfo
  {pages} {050406} (\bibinfo {year} {2018})}\BibitemShut {NoStop}%
\bibitem [{\citenamefont {Huang}\ \emph {et~al.}(2020)\citenamefont {Huang},
  \citenamefont {Kueng},\ and\ \citenamefont {Preskill}}]{huang2020predicting}%
  \BibitemOpen
  \bibfield  {author} {\bibinfo {author} {\bibfnamefont {H.-Y.}\ \bibnamefont
  {Huang}}, \bibinfo {author} {\bibfnamefont {R.}~\bibnamefont {Kueng}},\ and\
  \bibinfo {author} {\bibfnamefont {J.}~\bibnamefont {Preskill}},\ }\bibfield
  {title} {\bibinfo {title} {Predicting many properties of a quantum system
  from very few measurements},\ }\href
  {https://www.nature.com/articles/s41567-020-0932-7} {\bibfield  {journal}
  {\bibinfo  {journal} {Nature Physics}\ }\textbf {\bibinfo {volume} {16}},\
  \bibinfo {pages} {1050} (\bibinfo {year} {2020})}\BibitemShut {NoStop}%
\bibitem [{\citenamefont {Elben}\ \emph {et~al.}(2023)\citenamefont {Elben},
  \citenamefont {Flammia}, \citenamefont {Huang}, \citenamefont {Kueng},
  \citenamefont {Preskill}, \citenamefont {Vermersch},\ and\ \citenamefont
  {Zoller}}]{elben2023randomized}%
  \BibitemOpen
  \bibfield  {author} {\bibinfo {author} {\bibfnamefont {A.}~\bibnamefont
  {Elben}}, \bibinfo {author} {\bibfnamefont {S.~T.}\ \bibnamefont {Flammia}},
  \bibinfo {author} {\bibfnamefont {H.-Y.}\ \bibnamefont {Huang}}, \bibinfo
  {author} {\bibfnamefont {R.}~\bibnamefont {Kueng}}, \bibinfo {author}
  {\bibfnamefont {J.}~\bibnamefont {Preskill}}, \bibinfo {author}
  {\bibfnamefont {B.}~\bibnamefont {Vermersch}},\ and\ \bibinfo {author}
  {\bibfnamefont {P.}~\bibnamefont {Zoller}},\ }\bibfield  {title} {\bibinfo
  {title} {The randomized measurement toolbox},\ }\href
  {https://www.nature.com/articles/s42254-022-00535-2#citeas} {\bibfield
  {journal} {\bibinfo  {journal} {Nature Reviews Physics}\ }\textbf {\bibinfo
  {volume} {5}},\ \bibinfo {pages} {9} (\bibinfo {year} {2023})}\BibitemShut
  {NoStop}%
\bibitem [{\citenamefont {Hastings}\ \emph {et~al.}(2010)\citenamefont
  {Hastings}, \citenamefont {Gonz\'alez}, \citenamefont {Kallin},\ and\
  \citenamefont {Melko}}]{Hastings10}%
  \BibitemOpen
  \bibfield  {author} {\bibinfo {author} {\bibfnamefont {M.~B.}\ \bibnamefont
  {Hastings}}, \bibinfo {author} {\bibfnamefont {I.}~\bibnamefont
  {Gonz\'alez}}, \bibinfo {author} {\bibfnamefont {A.~B.}\ \bibnamefont
  {Kallin}},\ and\ \bibinfo {author} {\bibfnamefont {R.~G.}\ \bibnamefont
  {Melko}},\ }\bibfield  {title} {\bibinfo {title} {Measuring renyi
  entanglement entropy in quantum monte~carlo simulations},\ }\href
  {https://doi.org/10.1103/PhysRevLett.104.157201} {\bibfield  {journal}
  {\bibinfo  {journal} {Phys. Rev. Lett.}\ }\textbf {\bibinfo {volume} {104}},\
  \bibinfo {pages} {157201} (\bibinfo {year} {2010})}\BibitemShut {NoStop}%
\bibitem [{\citenamefont {Isakov}\ \emph {et~al.}(2011)\citenamefont {Isakov},
  \citenamefont {Hastings},\ and\ \citenamefont {Melko}}]{Melko_Hubbard}%
  \BibitemOpen
  \bibfield  {author} {\bibinfo {author} {\bibfnamefont {S.~V.}\ \bibnamefont
  {Isakov}}, \bibinfo {author} {\bibfnamefont {M.~B.}\ \bibnamefont
  {Hastings}},\ and\ \bibinfo {author} {\bibfnamefont {R.~G.}\ \bibnamefont
  {Melko}},\ }\bibfield  {title} {\bibinfo {title} {Topological entanglement
  entropy of a bose--hubbard spin liquid},\ }\href
  {https://doi.org/10.1038/nphys2036} {\bibfield  {journal} {\bibinfo
  {journal} {Nature Physics}\ }\textbf {\bibinfo {volume} {7}},\ \bibinfo
  {pages} {772} (\bibinfo {year} {2011})}\BibitemShut {NoStop}%
\bibitem [{\citenamefont {Zhang}\ \emph {et~al.}(2011)\citenamefont {Zhang},
  \citenamefont {Grover},\ and\ \citenamefont {Vishwanath}}]{zhang_criticalee}%
  \BibitemOpen
  \bibfield  {author} {\bibinfo {author} {\bibfnamefont {Y.}~\bibnamefont
  {Zhang}}, \bibinfo {author} {\bibfnamefont {T.}~\bibnamefont {Grover}},\ and\
  \bibinfo {author} {\bibfnamefont {A.}~\bibnamefont {Vishwanath}},\ }\bibfield
   {title} {\bibinfo {title} {Entanglement entropy of critical spin liquids},\
  }\href {https://doi.org/10.1103/PhysRevLett.107.067202} {\bibfield  {journal}
  {\bibinfo  {journal} {Phys. Rev. Lett.}\ }\textbf {\bibinfo {volume} {107}},\
  \bibinfo {pages} {067202} (\bibinfo {year} {2011})}\BibitemShut {NoStop}%
\bibitem [{\citenamefont {Jiang}\ \emph {et~al.}(2012)\citenamefont {Jiang},
  \citenamefont {Wang},\ and\ \citenamefont {Balents}}]{Jiang12}%
  \BibitemOpen
  \bibfield  {author} {\bibinfo {author} {\bibfnamefont {H.-C.}\ \bibnamefont
  {Jiang}}, \bibinfo {author} {\bibfnamefont {Z.}~\bibnamefont {Wang}},\ and\
  \bibinfo {author} {\bibfnamefont {L.}~\bibnamefont {Balents}},\ }\bibfield
  {title} {\bibinfo {title} {Identifying topological order by entanglement
  entropy},\ }\href {https://doi.org/10.1038/nphys2465} {\bibfield  {journal}
  {\bibinfo  {journal} {Nature Phys.}\ }\textbf {\bibinfo {volume} {8}},\
  \bibinfo {pages} {902} (\bibinfo {year} {2012})}\BibitemShut {NoStop}%
\bibitem [{\citenamefont {Nielsen}\ \emph {et~al.}(2012)\citenamefont
  {Nielsen}, \citenamefont {Cirac},\ and\ \citenamefont
  {Sierra}}]{nielsen2012laughlin}%
  \BibitemOpen
  \bibfield  {author} {\bibinfo {author} {\bibfnamefont {A.~E.}\ \bibnamefont
  {Nielsen}}, \bibinfo {author} {\bibfnamefont {J.~I.}\ \bibnamefont {Cirac}},\
  and\ \bibinfo {author} {\bibfnamefont {G.}~\bibnamefont {Sierra}},\
  }\bibfield  {title} {\bibinfo {title} {Laughlin spin-liquid states on
  lattices obtained from conformal field theory},\ }\href@noop {} {\bibfield
  {journal} {\bibinfo  {journal} {Physical review letters}\ }\textbf {\bibinfo
  {volume} {108}},\ \bibinfo {pages} {257206} (\bibinfo {year}
  {2012})}\BibitemShut {NoStop}%
\bibitem [{\citenamefont {Grover}(2013)}]{Grover13}%
  \BibitemOpen
  \bibfield  {author} {\bibinfo {author} {\bibfnamefont {T.}~\bibnamefont
  {Grover}},\ }\bibfield  {title} {\bibinfo {title} {Entanglement of
  interacting fermions in quantum monte~carlo calculations},\ }\href
  {https://doi.org/10.1103/PhysRevLett.111.130402} {\bibfield  {journal}
  {\bibinfo  {journal} {Phys. Rev. Lett.}\ }\textbf {\bibinfo {volume} {111}},\
  \bibinfo {pages} {130402} (\bibinfo {year} {2013})}\BibitemShut {NoStop}%
\bibitem [{\citenamefont {Assaad}\ \emph {et~al.}(2014)\citenamefont {Assaad},
  \citenamefont {Lang},\ and\ \citenamefont {Parisen~Toldin}}]{Assaad13a}%
  \BibitemOpen
  \bibfield  {author} {\bibinfo {author} {\bibfnamefont {F.~F.}\ \bibnamefont
  {Assaad}}, \bibinfo {author} {\bibfnamefont {T.~C.}\ \bibnamefont {Lang}},\
  and\ \bibinfo {author} {\bibfnamefont {F.}~\bibnamefont {Parisen~Toldin}},\
  }\bibfield  {title} {\bibinfo {title} {Entanglement spectra of interacting
  fermions in quantum monte carlo simulations},\ }\href
  {https://doi.org/10.1103/PhysRevB.89.125121} {\bibfield  {journal} {\bibinfo
  {journal} {Phys. Rev. B}\ }\textbf {\bibinfo {volume} {89}},\ \bibinfo
  {pages} {125121} (\bibinfo {year} {2014})}\BibitemShut {NoStop}%
\bibitem [{\citenamefont {Drut}\ and\ \citenamefont
  {Porter}(2016)}]{drut2016entanglement}%
  \BibitemOpen
  \bibfield  {author} {\bibinfo {author} {\bibfnamefont {J.~E.}\ \bibnamefont
  {Drut}}\ and\ \bibinfo {author} {\bibfnamefont {W.~J.}\ \bibnamefont
  {Porter}},\ }\bibfield  {title} {\bibinfo {title} {Entanglement, noise, and
  the cumulant expansion},\ }\href {https://doi.org/10.1103/PhysRevE.93.043301}
  {\bibfield  {journal} {\bibinfo  {journal} {Phys. Rev. E}\ }\textbf {\bibinfo
  {volume} {93}},\ \bibinfo {pages} {043301} (\bibinfo {year}
  {2016})}\BibitemShut {NoStop}%
\bibitem [{\citenamefont {D'Emidio}(2020)}]{d2020entanglement}%
  \BibitemOpen
  \bibfield  {author} {\bibinfo {author} {\bibfnamefont {J.}~\bibnamefont
  {D'Emidio}},\ }\bibfield  {title} {\bibinfo {title} {Entanglement entropy
  from nonequilibrium work},\ }\href
  {https://doi.org/10.1103/PhysRevLett.124.110602} {\bibfield  {journal}
  {\bibinfo  {journal} {Phys. Rev. Lett.}\ }\textbf {\bibinfo {volume} {124}},\
  \bibinfo {pages} {110602} (\bibinfo {year} {2020})}\BibitemShut {NoStop}%
\bibitem [{\citenamefont {Zhao}\ \emph {et~al.}(2022)\citenamefont {Zhao},
  \citenamefont {Chen}, \citenamefont {Wang}, \citenamefont {Yan},
  \citenamefont {Cheng},\ and\ \citenamefont {Meng}}]{zhao2022measuring}%
  \BibitemOpen
  \bibfield  {author} {\bibinfo {author} {\bibfnamefont {J.}~\bibnamefont
  {Zhao}}, \bibinfo {author} {\bibfnamefont {B.-B.}\ \bibnamefont {Chen}},
  \bibinfo {author} {\bibfnamefont {Y.-C.}\ \bibnamefont {Wang}}, \bibinfo
  {author} {\bibfnamefont {Z.}~\bibnamefont {Yan}}, \bibinfo {author}
  {\bibfnamefont {M.}~\bibnamefont {Cheng}},\ and\ \bibinfo {author}
  {\bibfnamefont {Z.~Y.}\ \bibnamefont {Meng}},\ }\bibfield  {title} {\bibinfo
  {title} {Measuring r{\'e}nyi entanglement entropy with high efficiency and
  precision in quantum monte carlo simulations},\ }\href@noop {} {\bibfield
  {journal} {\bibinfo  {journal} {npj Quantum Materials}\ }\textbf {\bibinfo
  {volume} {7}},\ \bibinfo {pages} {69} (\bibinfo {year} {2022})}\BibitemShut
  {NoStop}%
\bibitem [{\citenamefont {Zhou}\ \emph {et~al.}(2024)\citenamefont {Zhou},
  \citenamefont {Meng}, \citenamefont {Qi},\ and\ \citenamefont
  {Da~Liao}}]{zhou2024incremental}%
  \BibitemOpen
  \bibfield  {author} {\bibinfo {author} {\bibfnamefont {X.}~\bibnamefont
  {Zhou}}, \bibinfo {author} {\bibfnamefont {Z.~Y.}\ \bibnamefont {Meng}},
  \bibinfo {author} {\bibfnamefont {Y.}~\bibnamefont {Qi}},\ and\ \bibinfo
  {author} {\bibfnamefont {Y.}~\bibnamefont {Da~Liao}},\ }\bibfield  {title}
  {\bibinfo {title} {Incremental swap operator for entanglement entropy:
  Application for exponential observables in quantum monte carlo simulation},\
  }\href@noop {} {\bibfield  {journal} {\bibinfo  {journal} {Physical Review
  B}\ }\textbf {\bibinfo {volume} {109}},\ \bibinfo {pages} {165106} (\bibinfo
  {year} {2024})}\BibitemShut {NoStop}%
\bibitem [{\citenamefont {Mendes-Santos}\ \emph {et~al.}(2020)\citenamefont
  {Mendes-Santos}, \citenamefont {Giudici}, \citenamefont {Fazio},\ and\
  \citenamefont {Dalmonte}}]{mendes2020measuring}%
  \BibitemOpen
  \bibfield  {author} {\bibinfo {author} {\bibfnamefont {T.}~\bibnamefont
  {Mendes-Santos}}, \bibinfo {author} {\bibfnamefont {G.}~\bibnamefont
  {Giudici}}, \bibinfo {author} {\bibfnamefont {R.}~\bibnamefont {Fazio}},\
  and\ \bibinfo {author} {\bibfnamefont {M.}~\bibnamefont {Dalmonte}},\
  }\bibfield  {title} {\bibinfo {title} {Measuring von neumann entanglement
  entropies without wave functions},\ }\href@noop {} {\bibfield  {journal}
  {\bibinfo  {journal} {New Journal of Physics}\ }\textbf {\bibinfo {volume}
  {22}},\ \bibinfo {pages} {013044} (\bibinfo {year} {2020})}\BibitemShut
  {NoStop}%
\bibitem [{\citenamefont {Martiniani}\ \emph {et~al.}(2019)\citenamefont
  {Martiniani}, \citenamefont {Chaikin},\ and\ \citenamefont
  {Levine}}]{stefano2019quantifying}%
  \BibitemOpen
  \bibfield  {author} {\bibinfo {author} {\bibfnamefont {S.}~\bibnamefont
  {Martiniani}}, \bibinfo {author} {\bibfnamefont {P.~M.}\ \bibnamefont
  {Chaikin}},\ and\ \bibinfo {author} {\bibfnamefont {D.}~\bibnamefont
  {Levine}},\ }\bibfield  {title} {\bibinfo {title} {Quantifying hidden order
  out of equilibrium},\ }\href {https://doi.org/10.1103/PhysRevX.9.011031}
  {\bibfield  {journal} {\bibinfo  {journal} {Phys. Rev. X}\ }\textbf {\bibinfo
  {volume} {9}},\ \bibinfo {pages} {011031} (\bibinfo {year}
  {2019})}\BibitemShut {NoStop}%
\bibitem [{\citenamefont {Martiniani}\ \emph {et~al.}(2020)\citenamefont
  {Martiniani}, \citenamefont {Lemberg}, \citenamefont {Chaikin},\ and\
  \citenamefont {Levine}}]{martiniani2020correlation}%
  \BibitemOpen
  \bibfield  {author} {\bibinfo {author} {\bibfnamefont {S.}~\bibnamefont
  {Martiniani}}, \bibinfo {author} {\bibfnamefont {Y.}~\bibnamefont {Lemberg}},
  \bibinfo {author} {\bibfnamefont {P.~M.}\ \bibnamefont {Chaikin}},\ and\
  \bibinfo {author} {\bibfnamefont {D.}~\bibnamefont {Levine}},\ }\bibfield
  {title} {\bibinfo {title} {Correlation lengths in the language of computable
  information},\ }\href {https://doi.org/10.1103/PhysRevLett.125.170601}
  {\bibfield  {journal} {\bibinfo  {journal} {Physical review letters}\
  }\textbf {\bibinfo {volume} {125}},\ \bibinfo {pages} {170601} (\bibinfo
  {year} {2020})}\BibitemShut {NoStop}%
\bibitem [{\citenamefont {Sheinwald}\ \emph {et~al.}(1990)\citenamefont
  {Sheinwald}, \citenamefont {Lempel},\ and\ \citenamefont
  {Ziv}}]{sheinwald1990two}%
  \BibitemOpen
  \bibfield  {author} {\bibinfo {author} {\bibfnamefont {D.}~\bibnamefont
  {Sheinwald}}, \bibinfo {author} {\bibfnamefont {A.}~\bibnamefont {Lempel}},\
  and\ \bibinfo {author} {\bibfnamefont {J.}~\bibnamefont {Ziv}},\ }\bibfield
  {title} {\bibinfo {title} {Two-dimensional encoding by finite-state
  encoders},\ }\href {https://doi.org/10.1109/26.48892} {\bibfield  {journal}
  {\bibinfo  {journal} {IEEE Transactions on Communications}\ }\textbf
  {\bibinfo {volume} {38}},\ \bibinfo {pages} {341} (\bibinfo {year}
  {1990})}\BibitemShut {NoStop}%
\bibitem [{\citenamefont {Melchert}\ and\ \citenamefont
  {Hartmann}(2015)}]{melchert2015analysis}%
  \BibitemOpen
  \bibfield  {author} {\bibinfo {author} {\bibfnamefont {O.}~\bibnamefont
  {Melchert}}\ and\ \bibinfo {author} {\bibfnamefont {A.~K.}\ \bibnamefont
  {Hartmann}},\ }\bibfield  {title} {\bibinfo {title} {Analysis of the phase
  transition in the two-dimensional ising ferromagnet using a lempel-ziv
  string-parsing scheme and black-box data-compression utilities},\ }\href
  {https://doi.org/10.1103/PhysRevE.91.023306} {\bibfield  {journal} {\bibinfo
  {journal} {Phys. Rev. E}\ }\textbf {\bibinfo {volume} {91}},\ \bibinfo
  {pages} {023306} (\bibinfo {year} {2015})}\BibitemShut {NoStop}%
\bibitem [{\citenamefont {Castelnovo}\ and\ \citenamefont
  {Chamon}(2008)}]{castelnovo2008quantum}%
  \BibitemOpen
  \bibfield  {author} {\bibinfo {author} {\bibfnamefont {C.}~\bibnamefont
  {Castelnovo}}\ and\ \bibinfo {author} {\bibfnamefont {C.}~\bibnamefont
  {Chamon}},\ }\bibfield  {title} {\bibinfo {title} {Quantum topological phase
  transition at the microscopic level},\ }\href
  {https://doi.org/10.1103%2Fphysrevb.77.054433} {\bibfield  {journal}
  {\bibinfo  {journal} {Physical Review B}\ }\textbf {\bibinfo {volume} {77}},\
  \bibinfo {pages} {054433} (\bibinfo {year} {2008})}\BibitemShut {NoStop}%
\bibitem [{\citenamefont {Torlai}\ \emph {et~al.}(2018)\citenamefont {Torlai},
  \citenamefont {Mazzola}, \citenamefont {Carrasquilla}, \citenamefont
  {Troyer}, \citenamefont {Melko},\ and\ \citenamefont
  {Carleo}}]{torlai2018neural}%
  \BibitemOpen
  \bibfield  {author} {\bibinfo {author} {\bibfnamefont {G.}~\bibnamefont
  {Torlai}}, \bibinfo {author} {\bibfnamefont {G.}~\bibnamefont {Mazzola}},
  \bibinfo {author} {\bibfnamefont {J.}~\bibnamefont {Carrasquilla}}, \bibinfo
  {author} {\bibfnamefont {M.}~\bibnamefont {Troyer}}, \bibinfo {author}
  {\bibfnamefont {R.}~\bibnamefont {Melko}},\ and\ \bibinfo {author}
  {\bibfnamefont {G.}~\bibnamefont {Carleo}},\ }\bibfield  {title} {\bibinfo
  {title} {Neural-network quantum state tomography},\ }\href@noop {} {\bibfield
   {journal} {\bibinfo  {journal} {Nature physics}\ }\textbf {\bibinfo {volume}
  {14}},\ \bibinfo {pages} {447} (\bibinfo {year} {2018})}\BibitemShut
  {NoStop}%
\bibitem [{\citenamefont {Torlai}\ \emph {et~al.}(2019)\citenamefont {Torlai},
  \citenamefont {Timar}, \citenamefont {Van~Nieuwenburg}, \citenamefont
  {Levine}, \citenamefont {Omran}, \citenamefont {Keesling}, \citenamefont
  {Bernien}, \citenamefont {Greiner}, \citenamefont {Vuleti{\'c}},
  \citenamefont {Lukin} \emph {et~al.}}]{torlai2019integrating}%
  \BibitemOpen
  \bibfield  {author} {\bibinfo {author} {\bibfnamefont {G.}~\bibnamefont
  {Torlai}}, \bibinfo {author} {\bibfnamefont {B.}~\bibnamefont {Timar}},
  \bibinfo {author} {\bibfnamefont {E.~P.}\ \bibnamefont {Van~Nieuwenburg}},
  \bibinfo {author} {\bibfnamefont {H.}~\bibnamefont {Levine}}, \bibinfo
  {author} {\bibfnamefont {A.}~\bibnamefont {Omran}}, \bibinfo {author}
  {\bibfnamefont {A.}~\bibnamefont {Keesling}}, \bibinfo {author}
  {\bibfnamefont {H.}~\bibnamefont {Bernien}}, \bibinfo {author} {\bibfnamefont
  {M.}~\bibnamefont {Greiner}}, \bibinfo {author} {\bibfnamefont
  {V.}~\bibnamefont {Vuleti{\'c}}}, \bibinfo {author} {\bibfnamefont {M.~D.}\
  \bibnamefont {Lukin}}, \emph {et~al.},\ }\bibfield  {title} {\bibinfo {title}
  {Integrating neural networks with a quantum simulator for state
  reconstruction},\ }\href@noop {} {\bibfield  {journal} {\bibinfo  {journal}
  {Physical review letters}\ }\textbf {\bibinfo {volume} {123}},\ \bibinfo
  {pages} {230504} (\bibinfo {year} {2019})}\BibitemShut {NoStop}%
\bibitem [{\citenamefont {Luitz}\ \emph
  {et~al.}(2014{\natexlab{b}})\citenamefont {Luitz}, \citenamefont {Alet},\
  and\ \citenamefont {Laflorencie}}]{luitz2014universal}%
  \BibitemOpen
  \bibfield  {author} {\bibinfo {author} {\bibfnamefont {D.~J.}\ \bibnamefont
  {Luitz}}, \bibinfo {author} {\bibfnamefont {F.}~\bibnamefont {Alet}},\ and\
  \bibinfo {author} {\bibfnamefont {N.}~\bibnamefont {Laflorencie}},\
  }\bibfield  {title} {\bibinfo {title} {Universal behavior beyond
  multifractality in quantum many-body systems},\ }\href@noop {} {\bibfield
  {journal} {\bibinfo  {journal} {Physical Review Letters}\ }\textbf {\bibinfo
  {volume} {112}},\ \bibinfo {pages} {057203} (\bibinfo {year}
  {2014}{\natexlab{b}})}\BibitemShut {NoStop}%
\bibitem [{\citenamefont {de~Groot}\ \emph {et~al.}(2022)\citenamefont
  {de~Groot}, \citenamefont {Turz~illo},\ and\ \citenamefont
  {Schuch}}]{de2022symmetry}%
  \BibitemOpen
  \bibfield  {author} {\bibinfo {author} {\bibfnamefont {C.}~\bibnamefont
  {de~Groot}}, \bibinfo {author} {\bibfnamefont {A.}~\bibnamefont
  {Turz~illo}},\ and\ \bibinfo {author} {\bibfnamefont {N.}~\bibnamefont
  {Schuch}},\ }\bibfield  {title} {\bibinfo {title} {Symmetry {P}rotected
  {T}opological {O}rder in {O}pen {Q}uantum {S}ystems},\ }\href
  {https://doi.org/10.22331/q-2022-11-10-856} {\bibfield  {journal} {\bibinfo
  {journal} {{Quantum}}\ }\textbf {\bibinfo {volume} {6}},\ \bibinfo {pages}
  {856} (\bibinfo {year} {2022})}\BibitemShut {NoStop}%
\bibitem [{\citenamefont {Ma}\ and\ \citenamefont
  {Wang}(2023)}]{ma2022average}%
  \BibitemOpen
  \bibfield  {author} {\bibinfo {author} {\bibfnamefont {R.}~\bibnamefont
  {Ma}}\ and\ \bibinfo {author} {\bibfnamefont {C.}~\bibnamefont {Wang}},\
  }\bibfield  {title} {\bibinfo {title} {Average symmetry-protected topological
  phases},\ }\href {https://doi.org/10.1103/PhysRevX.13.031016} {\bibfield
  {journal} {\bibinfo  {journal} {Phys. Rev. X}\ }\textbf {\bibinfo {volume}
  {13}},\ \bibinfo {pages} {031016} (\bibinfo {year} {2023})}\BibitemShut
  {NoStop}%
\bibitem [{\citenamefont {Lee}\ \emph {et~al.}(2025)\citenamefont {Lee},
  \citenamefont {You},\ and\ \citenamefont {Xu}}]{lee2022symmetry}%
  \BibitemOpen
  \bibfield  {author} {\bibinfo {author} {\bibfnamefont {J.~Y.}\ \bibnamefont
  {Lee}}, \bibinfo {author} {\bibfnamefont {Y.-Z.}\ \bibnamefont {You}},\ and\
  \bibinfo {author} {\bibfnamefont {C.}~\bibnamefont {Xu}},\ }\bibfield
  {title} {\bibinfo {title} {Symmetry protected topological phases under
  decoherence},\ }\href {https://doi.org/10.22331/q-2025-01-23-1607} {\bibfield
   {journal} {\bibinfo  {journal} {Quantum}\ }\textbf {\bibinfo {volume} {9}},\
  \bibinfo {pages} {1607} (\bibinfo {year} {2025})}\BibitemShut {NoStop}%
\bibitem [{\citenamefont {You}\ \emph {et~al.}(2014)\citenamefont {You},
  \citenamefont {Bi}, \citenamefont {Rasmussen}, \citenamefont {Slagle},\ and\
  \citenamefont {Xu}}]{you2014wave}%
  \BibitemOpen
  \bibfield  {author} {\bibinfo {author} {\bibfnamefont {Y.-Z.}\ \bibnamefont
  {You}}, \bibinfo {author} {\bibfnamefont {Z.}~\bibnamefont {Bi}}, \bibinfo
  {author} {\bibfnamefont {A.}~\bibnamefont {Rasmussen}}, \bibinfo {author}
  {\bibfnamefont {K.}~\bibnamefont {Slagle}},\ and\ \bibinfo {author}
  {\bibfnamefont {C.}~\bibnamefont {Xu}},\ }\bibfield  {title} {\bibinfo
  {title} {Wave function and strange correlator of short-range entangled
  states},\ }\href {https://doi.org/10.1103/PhysRevLett.112.247202} {\bibfield
  {journal} {\bibinfo  {journal} {Phys. Rev. Lett.}\ }\textbf {\bibinfo
  {volume} {112}},\ \bibinfo {pages} {247202} (\bibinfo {year}
  {2014})}\BibitemShut {NoStop}%
\bibitem [{\citenamefont {Zhang}\ \emph {et~al.}(2022)\citenamefont {Zhang},
  \citenamefont {Qi},\ and\ \citenamefont {Bi}}]{zhang2022strange}%
  \BibitemOpen
  \bibfield  {author} {\bibinfo {author} {\bibfnamefont {J.-H.}\ \bibnamefont
  {Zhang}}, \bibinfo {author} {\bibfnamefont {Y.}~\bibnamefont {Qi}},\ and\
  \bibinfo {author} {\bibfnamefont {Z.}~\bibnamefont {Bi}},\ }\bibfield
  {title} {\bibinfo {title} {Strange correlation function for average
  symmetry-protected topological phases},\ }\href
  {https://arxiv.org/abs/2210.17485} {\bibfield  {journal} {\bibinfo  {journal}
  {arXiv preprint arXiv:2210.17485}\ } (\bibinfo {year} {2022})}\BibitemShut
  {NoStop}%
\bibitem [{\citenamefont {Chen}\ and\ \citenamefont
  {Grover}(2024{\natexlab{b}})}]{chen2023symmetry}%
  \BibitemOpen
  \bibfield  {author} {\bibinfo {author} {\bibfnamefont {Y.-H.}\ \bibnamefont
  {Chen}}\ and\ \bibinfo {author} {\bibfnamefont {T.}~\bibnamefont {Grover}},\
  }\bibfield  {title} {\bibinfo {title} {Symmetry-enforced many-body
  separability transitions},\ }\href
  {https://doi.org/10.1103/PRXQuantum.5.030310} {\bibfield  {journal} {\bibinfo
   {journal} {PRX Quantum}\ }\textbf {\bibinfo {volume} {5}},\ \bibinfo {pages}
  {030310} (\bibinfo {year} {2024}{\natexlab{b}})}\BibitemShut {NoStop}%
\bibitem [{\citenamefont {Wyner}\ and\ \citenamefont
  {Ziv}(1989)}]{Wyner1989asymptotic}%
  \BibitemOpen
  \bibfield  {author} {\bibinfo {author} {\bibfnamefont {A.}~\bibnamefont
  {Wyner}}\ and\ \bibinfo {author} {\bibfnamefont {J.}~\bibnamefont {Ziv}},\
  }\bibfield  {title} {\bibinfo {title} {Some asymptotic properties of the
  entropy of a stationary ergodic data source with applications to data
  compression},\ }\href {https://doi.org/10.1109/18.45281} {\bibfield
  {journal} {\bibinfo  {journal} {IEEE Transactions on Information Theory}\
  }\textbf {\bibinfo {volume} {35}},\ \bibinfo {pages} {1250} (\bibinfo {year}
  {1989})}\BibitemShut {NoStop}%
\bibitem [{\citenamefont {Nobel}\ and\ \citenamefont
  {Wyner}(1992)}]{Nobel1992recurrence}%
  \BibitemOpen
  \bibfield  {author} {\bibinfo {author} {\bibfnamefont {A.}~\bibnamefont
  {Nobel}}\ and\ \bibinfo {author} {\bibfnamefont {A.}~\bibnamefont {Wyner}},\
  }\bibfield  {title} {\bibinfo {title} {A recurrence theorem for dependent
  processes with applications to data compression},\ }\href
  {https://doi.org/10.1109/18.149506} {\bibfield  {journal} {\bibinfo
  {journal} {IEEE Transactions on Information Theory}\ }\textbf {\bibinfo
  {volume} {38}},\ \bibinfo {pages} {1561} (\bibinfo {year}
  {1992})}\BibitemShut {NoStop}%
\bibitem [{\citenamefont {Ornstein}\ and\ \citenamefont
  {Weiss}(2002)}]{ornstein2002entropy}%
  \BibitemOpen
  \bibfield  {author} {\bibinfo {author} {\bibfnamefont {D.~S.}\ \bibnamefont
  {Ornstein}}\ and\ \bibinfo {author} {\bibfnamefont {B.}~\bibnamefont
  {Weiss}},\ }\bibfield  {title} {\bibinfo {title} {Entropy and data
  compression schemes},\ }\href@noop {} {\bibfield  {journal} {\bibinfo
  {journal} {IEEE Transactions on information theory}\ }\textbf {\bibinfo
  {volume} {39}},\ \bibinfo {pages} {78} (\bibinfo {year} {2002})}\BibitemShut
  {NoStop}%
\bibitem [{\citenamefont {Wyner}\ and\ \citenamefont
  {Ziv}(1991)}]{wyner1991fixed}%
  \BibitemOpen
  \bibfield  {author} {\bibinfo {author} {\bibfnamefont {A.~D.}\ \bibnamefont
  {Wyner}}\ and\ \bibinfo {author} {\bibfnamefont {J.}~\bibnamefont {Ziv}},\
  }\bibfield  {title} {\bibinfo {title} {Fixed data base version of the
  lempel-ziv data compression algorithm},\ }\href@noop {} {\bibfield  {journal}
  {\bibinfo  {journal} {IEEE transactions on information theory}\ }\textbf
  {\bibinfo {volume} {37}},\ \bibinfo {pages} {878} (\bibinfo {year}
  {1991})}\BibitemShut {NoStop}%
\bibitem [{\citenamefont {Wyner}\ and\ \citenamefont
  {Ziv}(2002)}]{wyner2002sliding}%
  \BibitemOpen
  \bibfield  {author} {\bibinfo {author} {\bibfnamefont {A.~D.}\ \bibnamefont
  {Wyner}}\ and\ \bibinfo {author} {\bibfnamefont {J.}~\bibnamefont {Ziv}},\
  }\bibfield  {title} {\bibinfo {title} {The sliding-window lempel-ziv
  algorithm is asymptotically optimal},\ }\href@noop {} {\bibfield  {journal}
  {\bibinfo  {journal} {Proceedings of the IEEE}\ }\textbf {\bibinfo {volume}
  {82}},\ \bibinfo {pages} {872} (\bibinfo {year} {2002})}\BibitemShut
  {NoStop}%
\bibitem [{\citenamefont {Ornstein}\ and\ \citenamefont
  {Shields}(1990)}]{ornstein1990universal}%
  \BibitemOpen
  \bibfield  {author} {\bibinfo {author} {\bibfnamefont {D.~S.}\ \bibnamefont
  {Ornstein}}\ and\ \bibinfo {author} {\bibfnamefont {P.~C.}\ \bibnamefont
  {Shields}},\ }\bibfield  {title} {\bibinfo {title} {Universal almost sure
  data compression},\ }\href@noop {} {\bibfield  {journal} {\bibinfo  {journal}
  {The Annals of Probability}\ ,\ \bibinfo {pages} {441}} (\bibinfo {year}
  {1990})}\BibitemShut {NoStop}%
\bibitem [{\citenamefont {{Di Francesco}}\ \emph {et~al.}(1997)\citenamefont
  {{Di Francesco}}, \citenamefont {Mathieu},\ and\ \citenamefont
  {S{\'e}n{\'e}chal}}]{CFT1997}%
  \BibitemOpen
  \bibfield  {author} {\bibinfo {author} {\bibfnamefont {P.}~\bibnamefont {{Di
  Francesco}}}, \bibinfo {author} {\bibfnamefont {P.}~\bibnamefont {Mathieu}},\
  and\ \bibinfo {author} {\bibfnamefont {D.}~\bibnamefont {S{\'e}n{\'e}chal}},\
  }\href {https://doi.org/10.1007/978-1-4612-2256-9} {\emph {\bibinfo {title}
  {Conformal field theory}}},\ Graduate Texts in Contemporary Physics\
  (\bibinfo  {publisher} {Springer},\ \bibinfo {address} {Germany},\ \bibinfo
  {year} {1997})\BibitemShut {NoStop}%
\bibitem [{\citenamefont {Chen}\ and\ \citenamefont
  {Grover}(2024{\natexlab{c}})}]{chen2024unconventional}%
  \BibitemOpen
  \bibfield  {author} {\bibinfo {author} {\bibfnamefont {Y.-H.}\ \bibnamefont
  {Chen}}\ and\ \bibinfo {author} {\bibfnamefont {T.}~\bibnamefont {Grover}},\
  }\bibfield  {title} {\bibinfo {title} {Unconventional topological mixed-state
  transition and critical phase induced by self-dual coherent errors},\ }\href
  {https://doi.org/10.1103/PhysRevB.110.125152} {\bibfield  {journal} {\bibinfo
   {journal} {Phys. Rev. B}\ }\textbf {\bibinfo {volume} {110}},\ \bibinfo
  {pages} {125152} (\bibinfo {year} {2024}{\natexlab{c}})}\BibitemShut
  {NoStop}%
\bibitem [{\citenamefont {Rokhsar}\ and\ \citenamefont
  {Kivelson}(1988)}]{rokhsar1988superconductivity}%
  \BibitemOpen
  \bibfield  {author} {\bibinfo {author} {\bibfnamefont {D.~S.}\ \bibnamefont
  {Rokhsar}}\ and\ \bibinfo {author} {\bibfnamefont {S.~A.}\ \bibnamefont
  {Kivelson}},\ }\bibfield  {title} {\bibinfo {title} {Superconductivity and
  the quantum hard-core dimer gas},\ }\href@noop {} {\bibfield  {journal}
  {\bibinfo  {journal} {Physical review letters}\ }\textbf {\bibinfo {volume}
  {61}},\ \bibinfo {pages} {2376} (\bibinfo {year} {1988})}\BibitemShut
  {NoStop}%
\bibitem [{\citenamefont {Ardonne}\ \emph {et~al.}(2004)\citenamefont
  {Ardonne}, \citenamefont {Fendley},\ and\ \citenamefont
  {Fradkin}}]{ardonne2004topological}%
  \BibitemOpen
  \bibfield  {author} {\bibinfo {author} {\bibfnamefont {E.}~\bibnamefont
  {Ardonne}}, \bibinfo {author} {\bibfnamefont {P.}~\bibnamefont {Fendley}},\
  and\ \bibinfo {author} {\bibfnamefont {E.}~\bibnamefont {Fradkin}},\
  }\bibfield  {title} {\bibinfo {title} {Topological order and conformal
  quantum critical points},\ }\href@noop {} {\bibfield  {journal} {\bibinfo
  {journal} {Annals of Physics}\ }\textbf {\bibinfo {volume} {310}},\ \bibinfo
  {pages} {493} (\bibinfo {year} {2004})}\BibitemShut {NoStop}%
\bibitem [{\citenamefont {Cardy}(1996)}]{Cardy_1996_ch8}%
  \BibitemOpen
  \bibfield  {author} {\bibinfo {author} {\bibfnamefont {J.}~\bibnamefont
  {Cardy}},\ }\bibinfo {title} {Random systems},\ in\ \href@noop {} {\emph
  {\bibinfo {booktitle} {Scaling and Renormalization in Statistical
  Physics}}},\ \bibinfo {series and number} {Cambridge Lecture Notes in
  Physics}\ (\bibinfo  {publisher} {Cambridge University Press},\ \bibinfo
  {year} {1996})\ p.\ \bibinfo {pages} {145–168}\BibitemShut {NoStop}%
\bibitem [{\citenamefont {Melchert}(2009)}]{melchert2009autoscale}%
  \BibitemOpen
  \bibfield  {author} {\bibinfo {author} {\bibfnamefont {O.}~\bibnamefont
  {Melchert}},\ }\bibfield  {title} {\bibinfo {title} {autoscale. py-a program
  for automatic finite-size scaling analyses: A user's guide},\ }\href
  {https://arxiv.org/abs/0910.5403} {\bibfield  {journal} {\bibinfo  {journal}
  {arXiv preprint arXiv:0910.5403}\ } (\bibinfo {year} {2009})}\BibitemShut
  {NoStop}%
\bibitem [{\citenamefont {Le~Doussal}\ and\ \citenamefont
  {Harris}(1988)}]{le1988location}%
  \BibitemOpen
  \bibfield  {author} {\bibinfo {author} {\bibfnamefont {P.}~\bibnamefont
  {Le~Doussal}}\ and\ \bibinfo {author} {\bibfnamefont {A.~B.}\ \bibnamefont
  {Harris}},\ }\bibfield  {title} {\bibinfo {title} {Location of the ising
  spin-glass multicritical point on nishimori's line},\ }\href
  {https://doi.org/10.1103/PhysRevLett.61.625} {\bibfield  {journal} {\bibinfo
  {journal} {Phys. Rev. Lett.}\ }\textbf {\bibinfo {volume} {61}},\ \bibinfo
  {pages} {625} (\bibinfo {year} {1988})}\BibitemShut {NoStop}%
\bibitem [{\citenamefont {Le~Doussal}\ and\ \citenamefont
  {Harris}(1989)}]{le1989varepsilon}%
  \BibitemOpen
  \bibfield  {author} {\bibinfo {author} {\bibfnamefont {P.}~\bibnamefont
  {Le~Doussal}}\ and\ \bibinfo {author} {\bibfnamefont {A.~B.}\ \bibnamefont
  {Harris}},\ }\bibfield  {title} {\bibinfo {title} {\ensuremath{\epsilon}
  expansion for the nishimori multicritical point of spin glasses},\ }\href
  {https://doi.org/10.1103/PhysRevB.40.9249} {\bibfield  {journal} {\bibinfo
  {journal} {Phys. Rev. B}\ }\textbf {\bibinfo {volume} {40}},\ \bibinfo
  {pages} {9249} (\bibinfo {year} {1989})}\BibitemShut {NoStop}%
\bibitem [{\citenamefont {Hasenbusch}\ \emph {et~al.}(2008)\citenamefont
  {Hasenbusch}, \citenamefont {Toldin}, \citenamefont {Pelissetto},\ and\
  \citenamefont {Vicari}}]{hasenbusch2008multicritical}%
  \BibitemOpen
  \bibfield  {author} {\bibinfo {author} {\bibfnamefont {M.}~\bibnamefont
  {Hasenbusch}}, \bibinfo {author} {\bibfnamefont {F.~P.}\ \bibnamefont
  {Toldin}}, \bibinfo {author} {\bibfnamefont {A.}~\bibnamefont {Pelissetto}},\
  and\ \bibinfo {author} {\bibfnamefont {E.}~\bibnamefont {Vicari}},\
  }\bibfield  {title} {\bibinfo {title} {Multicritical nishimori point in the
  phase diagram of the $\ifmmode\pm\else\textpm\fi{}j$ ising model on a square
  lattice},\ }\href {https://doi.org/10.1103/PhysRevE.77.051115} {\bibfield
  {journal} {\bibinfo  {journal} {Phys. Rev. E}\ }\textbf {\bibinfo {volume}
  {77}},\ \bibinfo {pages} {051115} (\bibinfo {year} {2008})}\BibitemShut
  {NoStop}%
\bibitem [{\citenamefont {Honecker}\ \emph {et~al.}(2001)\citenamefont
  {Honecker}, \citenamefont {Picco},\ and\ \citenamefont
  {Pujol}}]{honecker2001universality}%
  \BibitemOpen
  \bibfield  {author} {\bibinfo {author} {\bibfnamefont {A.}~\bibnamefont
  {Honecker}}, \bibinfo {author} {\bibfnamefont {M.}~\bibnamefont {Picco}},\
  and\ \bibinfo {author} {\bibfnamefont {P.}~\bibnamefont {Pujol}},\ }\bibfield
   {title} {\bibinfo {title} {Universality class of the nishimori point in the
  2d $\ifmmode\pm\else\textpm\fi{}\mathit{J}$ random-bond ising model},\ }\href
  {https://doi.org/10.1103/PhysRevLett.87.047201} {\bibfield  {journal}
  {\bibinfo  {journal} {Phys. Rev. Lett.}\ }\textbf {\bibinfo {volume} {87}},\
  \bibinfo {pages} {047201} (\bibinfo {year} {2001})}\BibitemShut {NoStop}%
\bibitem [{\citenamefont {Cho}\ and\ \citenamefont
  {Fisher}(1997)}]{cho1997criticality}%
  \BibitemOpen
  \bibfield  {author} {\bibinfo {author} {\bibfnamefont {S.}~\bibnamefont
  {Cho}}\ and\ \bibinfo {author} {\bibfnamefont {M.~P.~A.}\ \bibnamefont
  {Fisher}},\ }\bibfield  {title} {\bibinfo {title} {Criticality in the
  two-dimensional random-bond ising model},\ }\href
  {https://doi.org/10.1103/PhysRevB.55.1025} {\bibfield  {journal} {\bibinfo
  {journal} {Phys. Rev. B}\ }\textbf {\bibinfo {volume} {55}},\ \bibinfo
  {pages} {1025} (\bibinfo {year} {1997})}\BibitemShut {NoStop}%
\bibitem [{\citenamefont {Venn}\ \emph {et~al.}(2023)\citenamefont {Venn},
  \citenamefont {Behrends},\ and\ \citenamefont {B{\'e}ri}}]{venn2023coherent}%
  \BibitemOpen
  \bibfield  {author} {\bibinfo {author} {\bibfnamefont {F.}~\bibnamefont
  {Venn}}, \bibinfo {author} {\bibfnamefont {J.}~\bibnamefont {Behrends}},\
  and\ \bibinfo {author} {\bibfnamefont {B.}~\bibnamefont {B{\'e}ri}},\
  }\bibfield  {title} {\bibinfo {title} {Coherent-error threshold for surface
  codes from majorana delocalization},\ }\href
  {https://doi.org/10.1103/PhysRevLett.131.060603} {\bibfield  {journal}
  {\bibinfo  {journal} {Physical Review Letters}\ }\textbf {\bibinfo {volume}
  {131}},\ \bibinfo {pages} {060603} (\bibinfo {year} {2023})}\BibitemShut
  {NoStop}%
\bibitem [{\citenamefont {Levine}(2022)}]{levine_rutgers_talk}%
  \BibitemOpen
  \bibfield  {author} {\bibinfo {author} {\bibfnamefont {D.}~\bibnamefont
  {Levine}},\ }\bibfield  {title} {\bibinfo {title} {Order and information out
  of equilibrium}} (\bibinfo {year} {2022}),\ \bibinfo {note} {talk at Rutgers
  Center for Mathematical Sciences Research}\BibitemShut {NoStop}%
\bibitem [{\citenamefont {Troyer}\ and\ \citenamefont
  {Wiese}(2005)}]{troyer2005computational}%
  \BibitemOpen
  \bibfield  {author} {\bibinfo {author} {\bibfnamefont {M.}~\bibnamefont
  {Troyer}}\ and\ \bibinfo {author} {\bibfnamefont {U.-J.}\ \bibnamefont
  {Wiese}},\ }\bibfield  {title} {\bibinfo {title} {Computational complexity
  and fundamental limitations to fermionic quantum monte carlo simulations},\
  }\href@noop {} {\bibfield  {journal} {\bibinfo  {journal} {Physical review
  letters}\ }\textbf {\bibinfo {volume} {94}},\ \bibinfo {pages} {170201}
  (\bibinfo {year} {2005})}\BibitemShut {NoStop}%
\bibitem [{\citenamefont {Pan}\ and\ \citenamefont {Meng}(2022)}]{pan2022sign}%
  \BibitemOpen
  \bibfield  {author} {\bibinfo {author} {\bibfnamefont {G.}~\bibnamefont
  {Pan}}\ and\ \bibinfo {author} {\bibfnamefont {Z.~Y.}\ \bibnamefont {Meng}},\
  }\bibfield  {title} {\bibinfo {title} {Sign problem in quantum monte carlo
  simulation},\ }\href@noop {} {\bibfield  {journal} {\bibinfo  {journal}
  {arXiv preprint arXiv:2204.08777}\ } (\bibinfo {year} {2022})}\BibitemShut
  {NoStop}%
\bibitem [{\citenamefont {Hastings}\ and\ \citenamefont
  {Freedman}(2013)}]{hastings2013obstructions}%
  \BibitemOpen
  \bibfield  {author} {\bibinfo {author} {\bibfnamefont {M.~B.}\ \bibnamefont
  {Hastings}}\ and\ \bibinfo {author} {\bibfnamefont {M.~H.}\ \bibnamefont
  {Freedman}},\ }\bibfield  {title} {\bibinfo {title} {Obstructions to
  classically simulating the quantum adiabatic algorithm},\ }\href@noop {}
  {\bibfield  {journal} {\bibinfo  {journal} {arXiv preprint arXiv:1302.5733}\
  } (\bibinfo {year} {2013})}\BibitemShut {NoStop}%
\bibitem [{\citenamefont {Hastings}(2021)}]{hastings2021power}%
  \BibitemOpen
  \bibfield  {author} {\bibinfo {author} {\bibfnamefont {M.~B.}\ \bibnamefont
  {Hastings}},\ }\bibfield  {title} {\bibinfo {title} {The power of adiabatic
  quantum computation with no sign problem},\ }\href@noop {} {\bibfield
  {journal} {\bibinfo  {journal} {Quantum}\ }\textbf {\bibinfo {volume} {5}},\
  \bibinfo {pages} {597} (\bibinfo {year} {2021})}\BibitemShut {NoStop}%
\bibitem [{\citenamefont {Gily{\'e}n}\ \emph {et~al.}(2021)\citenamefont
  {Gily{\'e}n}, \citenamefont {Hastings},\ and\ \citenamefont
  {Vazirani}}]{gilyen2021sub}%
  \BibitemOpen
  \bibfield  {author} {\bibinfo {author} {\bibfnamefont {A.}~\bibnamefont
  {Gily{\'e}n}}, \bibinfo {author} {\bibfnamefont {M.~B.}\ \bibnamefont
  {Hastings}},\ and\ \bibinfo {author} {\bibfnamefont {U.}~\bibnamefont
  {Vazirani}},\ }\bibfield  {title} {\bibinfo {title} {(sub) exponential
  advantage of adiabatic quantum computation with no sign problem},\
  }\href@noop {} {\bibfield  {journal} {\bibinfo  {journal} {Proceedings of the
  53rd Annual ACM SIGACT Symposium on Theory of Computing}\ ,\ \bibinfo {pages}
  {1357}} (\bibinfo {year} {2021})}\BibitemShut {NoStop}%
\bibitem [{\citenamefont {Lavasani}\ and\ \citenamefont
  {Vijay}(2024)}]{lavasani2024stability}%
  \BibitemOpen
  \bibfield  {author} {\bibinfo {author} {\bibfnamefont {A.}~\bibnamefont
  {Lavasani}}\ and\ \bibinfo {author} {\bibfnamefont {S.}~\bibnamefont
  {Vijay}},\ }\bibfield  {title} {\bibinfo {title} {The stability of gapped
  quantum matter and error-correction with adiabatic noise},\ }\href
  {https://arxiv.org/abs/2402.14906} {\bibfield  {journal} {\bibinfo  {journal}
  {arXiv preprint arXiv:2402.14906}\ } (\bibinfo {year} {2024})}\BibitemShut
  {NoStop}%
\bibitem [{\citenamefont {Sang}\ \emph
  {et~al.}(2024{\natexlab{b}})\citenamefont {Sang}, \citenamefont {Hsieh},\
  and\ \citenamefont {Zou}}]{sang2024approximate}%
  \BibitemOpen
  \bibfield  {author} {\bibinfo {author} {\bibfnamefont {S.}~\bibnamefont
  {Sang}}, \bibinfo {author} {\bibfnamefont {T.~H.}\ \bibnamefont {Hsieh}},\
  and\ \bibinfo {author} {\bibfnamefont {Y.}~\bibnamefont {Zou}},\ }\bibfield
  {title} {\bibinfo {title} {Approximate quantum error correcting codes from
  conformal field theory},\ }\href@noop {} {\bibfield  {journal} {\bibinfo
  {journal} {Physical Review Letters}\ }\textbf {\bibinfo {volume} {133}},\
  \bibinfo {pages} {210601} (\bibinfo {year} {2024}{\natexlab{b}})}\BibitemShut
  {NoStop}%
\bibitem [{\citenamefont {Liao}\ \emph {et~al.}(2019)\citenamefont {Liao},
  \citenamefont {Liu}, \citenamefont {Wang},\ and\ \citenamefont
  {Xiang}}]{liao2019differentiable}%
  \BibitemOpen
  \bibfield  {author} {\bibinfo {author} {\bibfnamefont {H.-J.}\ \bibnamefont
  {Liao}}, \bibinfo {author} {\bibfnamefont {J.-G.}\ \bibnamefont {Liu}},
  \bibinfo {author} {\bibfnamefont {L.}~\bibnamefont {Wang}},\ and\ \bibinfo
  {author} {\bibfnamefont {T.}~\bibnamefont {Xiang}},\ }\bibfield  {title}
  {\bibinfo {title} {Differentiable programming tensor networks},\ }\href
  {https://doi.org/10.1103/PhysRevX.9.031041} {\bibfield  {journal} {\bibinfo
  {journal} {Phys. Rev. X}\ }\textbf {\bibinfo {volume} {9}},\ \bibinfo {pages}
  {031041} (\bibinfo {year} {2019})}\BibitemShut {NoStop}%
\bibitem [{\citenamefont {Shields}(1999)}]{shields1999performance}%
  \BibitemOpen
  \bibfield  {author} {\bibinfo {author} {\bibfnamefont {P.~C.}\ \bibnamefont
  {Shields}},\ }\bibfield  {title} {\bibinfo {title} {Performance of lz
  algorithms on individual sequences},\ }\href
  {https://doi.org/10.1109/18.761286} {\bibfield  {journal} {\bibinfo
  {journal} {IEEE Transactions on Information Theory}\ }\textbf {\bibinfo
  {volume} {45}},\ \bibinfo {pages} {1283} (\bibinfo {year}
  {1999})}\BibitemShut {NoStop}%
\bibitem [{\citenamefont {Bravyi}(2005)}]{bravyi2005}%
  \BibitemOpen
  \bibfield  {author} {\bibinfo {author} {\bibfnamefont {S.}~\bibnamefont
  {Bravyi}},\ }\bibfield  {title} {\bibinfo {title} {Lagrangian representation
  for fermionic linear optics},\ }\href@noop {} {\bibfield  {journal} {\bibinfo
   {journal} {Quantum Info. Comput.}\ }\textbf {\bibinfo {volume} {5}},\
  \bibinfo {pages} {216–238} (\bibinfo {year} {2005})}\BibitemShut {NoStop}%
\bibitem [{\citenamefont {Or{\'u}s}(2014)}]{orus2014practical}%
  \BibitemOpen
  \bibfield  {author} {\bibinfo {author} {\bibfnamefont {R.}~\bibnamefont
  {Or{\'u}s}},\ }\bibfield  {title} {\bibinfo {title} {A practical introduction
  to tensor networks: Matrix product states and projected entangled pair
  states},\ }\href@noop {} {\bibfield  {journal} {\bibinfo  {journal} {Annals
  of physics}\ }\textbf {\bibinfo {volume} {349}},\ \bibinfo {pages} {117}
  (\bibinfo {year} {2014})}\BibitemShut {NoStop}%
\bibitem [{\citenamefont {Gray}(2018)}]{gray2018quimb}%
  \BibitemOpen
  \bibfield  {author} {\bibinfo {author} {\bibfnamefont {J.}~\bibnamefont
  {Gray}},\ }\bibfield  {title} {\bibinfo {title} {quimb: a python library for
  quantum information and many-body calculations},\ }\href
  {https://doi.org/10.21105/joss.00819} {\bibfield  {journal} {\bibinfo
  {journal} {Journal of Open Source Software}\ }\textbf {\bibinfo {volume}
  {3}},\ \bibinfo {pages} {819} (\bibinfo {year} {2018})}\BibitemShut {NoStop}%
\end{thebibliography}

%

\appendix

\section{Brief introduction to Lempel-Ziv compression}	
\label{sec:LZ77}

This appendix aims to introduce the Lempel-Ziv 77 (LZ77) lossless data compression algorithm and explain how to calculate the compressed code length from the compressed data.
In our context, the LZ77 algorithm compresses a sequence of spins \( x_\mathbf{j} = (x_1, x_2, \cdots, x_N),\ x_j \pm 1 \) by replacing repeated occurrences of spins with references to a single copy that appeared earlier in \( x_\mathbf{j} \). The result is the compressed sequence consisting of $|C(x_\mathbf{j})|$ tuples: \( C(x_\mathbf{j}) = [(d_1, l_1, b_1), (d_2, l_2, b_2), \cdots, (d_{|C(x_\mathbf{j})|}, l_{|C(x_\mathbf{j})|}, b_{|C(x_\mathbf{j})|})]\).
In this notation, \((d, l, b)\) indicates that each of the next \( l \) bits is equal to the spin that is \( d \) positions behind the current location. Here, \( b \) can be 0, 1, or -1. If \( b = \pm 1 \) , it denotes the spins that follows the \( l \) spins, while \( b = 0 \) means there is no following spin.
We note that \( l \) can be larger than \( d \) since the `current' location changes dynamically as previously occurred spins are copied and pasted.

The algorithm can be most easily explained through examples and from the decoder's point of view.
If $x_\mathbf{j} = [1,1,1,1,1,1,1,1]$, LZ77 outputs a sequence of $2$ tuples $C(x_\mathbf{j}) = [(0,0,1), (1,7,0)]$.
Here,
$d_1 = l_1 =0$ means the next bit cannot be determined from the previously occurred bits (which is empty) and is specificed by $b_1 = 1$.
$d_2 = 1,\ l_2 = 7$ means each of the next 7 bits is equal to the bit that is 1 position behind the current location, which is always $1$.
$b_2 = 0$ means there is no bit following after copying and pasting the next 7 bits.
If 
$x_\mathbf{j} = [1,-1,1,-1,1,-1,1,-1]$, LZ77 outputs $C(x_\mathbf{j}) = [(0, 0, 1), (0, 0, -1), (2, 6, 0)]$.
Here, $(d_1, l_1, b_1) = (0,0,1)$ means the next bit cannot be determined by the previous bit (which is empty) and is $1$.
$(d_2, l_2, b_2) = (0,0,-1)$ means the next bit cannot be determined by the previous bit (which is $1$) and is $-1$.
$(d_3, l_3, b_3) = (2,6,0)$ means each of the next $6$ bits is equal to the bit that is $2$ positions behind the current location, and there is no bit following.
As a final example, LZ77 outputs $C(x_\mathbf{j}) = [(0, 0, -1), (1, 3, 1), (1, 2, -1)]$ if $x_\mathbf{j} = [-1, -1,-1,-1,1,1,1,-1]$, and the compressed sequence $C(x_\mathbf{j})$ can be decoded following the same logic as the previous two examples.

Given the number $|C(x_\mathbf{j})|$ of tuples in $C(x_\mathbf{j})$, the compressed code length $\mathcal{N}(x_\mathbf{j})$ can be upper bounded through  $\mathcal{N}(x_\mathbf{j}) \leq |C(x_\mathbf{j})|\log |C(x_\mathbf{j})| + 2M \log(N/|C(x_\mathbf{j})|)+ O(N \log \log(N/|C(x_\mathbf{j})|))$ \cite{shields1999performance}.
Following Refs.\cite{stefano2019quantifying,martiniani2020correlation}, we approximate $\mathcal{N}(x_\mathbf{j})$ by
\begin{equation}
    \mathcal{N}(x_\mathbf{j}) \approx |C(x_\mathbf{j})|\log |C(x_\mathbf{j})| + 2|C(x_\mathbf{j})| \log(\frac{N}{|C(x_\mathbf{j})|}).
\end{equation}
The computable information density (CID) can then be computed through Eq.\eqref{Eq:CID} in the main text.

\section{Numerical estimation of diagonal entropy and computatble information density}	
\label{sec:app_numerics}

This appendix provides additional details on the numerical estimation of diagonal entropy and computable information density (CID). First, we briefly discuss the general scheme for generating samples based on the Born probability $\rho_{x_{\mathbf{j}}} = \langle x_{\mathbf{j}} |\rho |x_{\mathbf{j}}\rangle$ for mixed states $\rho$ in Sec.~\ref{sec:tfim}, Sec.~\ref{sec:ground_state_2d}, and Sec.~\ref{sec:coherent}, using the standard Metropolis-Hastings algorithm. Later, we explain how $\rho_{x_{\mathbf{j}}}$ is computed for all examples in this paper.
Note that the samples of RBIM along the Nishimori line studied in Sec.~\ref{sec:Nishimori} can be generated much more easily without the process discussed in the next paragraph. Instead, as discussed in the main text, the samples can be obtained by independently sampling the bond configurations $J_{\langle \tilde{i}, \tilde{j} \rangle}$ on the dual lattice according to the probability distribution $P(J_{\langle \tilde{i}, \tilde{j} \rangle} =1) = 1-p$, $P(J_{\langle \tilde{i}, \tilde{j} \rangle} =-1) = p$, and then projecting them onto the vortex configurations.

Generating typical measurement outcomes for a mixed state $\rho$ in the Pauli-$X$ basis involves the following steps. 
(i) Initialize the configuration $x_{\mathbf{j}} = (x_1, \cdots, x_j,\cdots, x_N)$ with $x_j = \pm 1$. In all examples considered, we choose the uniform configuration $x_{\mathbf{j}} = 1, \forall j$. 
(ii) Generate a candidate state $x'_{\mathbf{j}}$ by randomly picking a bond $\langle i,j\rangle$ and flipping $x_i \rightarrow -x_i,\ x_j \rightarrow -x_j$ in the previous configuration $x_{\mathbf{j}}$. 
(iii) Compute the acceptance probability $A(x'_\mathbf{j}|x_\mathbf{j}) = \min(1, \rho_{x_{\mathbf{j}}}/\rho_{x'_{\mathbf{j}}})$. 
(iv) Generate a uniform random number $u \in [0,1]$ and accept the new state if $u < A(x'_\mathbf{j}|x_\mathbf{j})$. 
(v) Thermalize the samples by performing the above iterations for at least $5N$ Monte Carlo steps, where $N$ is the total system size. 
(vi) We then perform an additional $N \times N_s$ Monte Carlo steps. After every $N$ steps, we store the sample $x_{\mathbf{j}}$ and compute $-\log(\rho_{x_\mathbf{j}})$ and $\text{CID}(x_\mathbf{j})$. Thus, we obtain $N_s$ samples. 
(vii) Compute the diagonal entropy and CID as $s_d = -\sum_{x_{\mathbf{j}} \in \text{samples}} \log(\rho_{x_{\mathbf{j}}})/(N N_s)$ and $\mathbb{E}[\text{CID}] = \sum_{x_{\mathbf{j}} \in \text{samples}} \text{CID}(x_\mathbf{j})/N_s$, respectively.

We now discuss the computation of the Born probability $\rho_{x_{\mathbf{j}}} = \langle x_{\mathbf{j}}|\rho|x_{\mathbf{j}}\rangle$. For the $(1+1)$-D Ising model in Sec.~\ref{sec:tfim}, where $\rho = |\Psi(J)\rangle \langle \Psi(J)|$, the Born probability $\rho_{x_{\mathbf{j}}}$ can be estimated efficiently by mapping the system to free fermions. The key idea is that both $|x_\mathbf{j}\rangle$ and $|\Psi (J)\rangle$ can be mapped to fermionic Gaussian states $|\psi_1\rangle$ and $|\psi _2\rangle$, respectively, via the Jordan-Wigner transformation. A fermionic Gaussian state $|\psi\rangle$ is completely specified by the covariance matrix $M_{j,k}= -i  \langle\psi| \gamma_j \gamma_k - \delta_{j,l}|\psi\rangle$ (up to a phase), and the square of the overlap between two fermionic Gaussian states can be computed as $|\langle \psi_1 |\psi_2\rangle|^2 = \sqrt{\det(\frac{I+M_1M_2}{2})}$. Thus, $\rho_{x_{\mathbf{j}}} = \langle x_{\mathbf{j}}|\rho|x_{\mathbf{j}}\rangle = |\langle x_{\mathbf{j}}|\Psi(J)\rangle|^2$ for a system of size $L$ can be efficiently computed by evaluating the determinant of a $2L \times 2L$ matrix \cite{bravyi2005}.

On the other hand, $\rho_{x_{\mathbf{j}}}$ for the models studied in Sec.~\ref{sec:ground_state_2d} and Sec.~\ref{sec:coherent} is computed using the standard tensor-network-based approach (see Ref.~\cite{orus2014practical} for a review). The key idea is as follows. As shown in Eq.~\eqref{Eq:rho2_RBIM} and Eq.~\eqref{Eq:complex_RBIM}, $\rho_{x_\mathbf{j}}$ can be mapped to the squared norm of the partition function of the disordered real and complex Ising models on the dual lattice. These models can naturally be represented as a tensor network (TN) with bond dimension $\chi = 2$. Such a TN can always be contracted row by row, starting from the bottom to the top. However, the bond dimension required for exact contraction scales exponentially with the system size $L$, necessitating truncation to make the numerical computation feasible.
We implement the truncations and contractions of the TN using the open-source Python library Quimb \cite{gray2018quimb}, and the key idea is as follows. We treat the tensor network being contracted from the bottom as a matrix product state (MPS). When contracting the MPS to the next row of tensors, the bond dimension of the MPS is truncated by minimizing the error with respect to the exact MPS using the standard singular value decomposition (throughout the calculations, we ensure that the tolerance remains below $10^{-8}$). This process is performed row by row during the TN contraction until we reach the top row, which yields the desired quantity $\rho_{x_{\mathbf{j}}}$.
Note that in the main text, we also compute the disorder-averaged of the spin-spin correlation function $ [\langle s_{\tilde{i}} s_{\tilde{j}} \rangle ]  \approx \sum_{J_{\langle \mathbf{\tilde{k}, \tilde{l} }\rangle} \in \text{samples}} \langle s_i s_j \rangle_{J_{\langle \mathbf{\tilde{k}, \tilde{l}} \rangle}}/N_s $ and the the disorder-averaged of the free energy cost of a single Ising vortex  $[ \langle e^{-\Delta F_l /2} \rangle ]   \approx \sum_{J_{\langle \mathbf{\tilde{k}, \tilde{l} }\rangle} \in \text{samples}} \sqrt{ \mathcal{Z}_{J_{\langle \tilde{i}, \tilde{j} \rangle, l}}/ \mathcal{Z}_{J_{\langle \tilde{i}, \tilde{j} \rangle}}}/N_s$, where the quantities in the summand are all computed using the same method.

\end{document}